\documentclass[a4paper,11pt]{article}
\pdfoutput=1

\usepackage{jheppub} 
\usepackage[T1]{fontenc}
\usepackage{multirow}
\usepackage{units}
\usepackage{placeins}

\newcommand{\secref}[1]{section~\ref{#1}}
\newcommand{\Secref}[1]{Section~\ref{#1}}
\newcommand{\appref}[1]{appendix~\ref{#1}}

\newcommand{\figref}[1]{figure~\ref{#1}}
\newcommand{\Figref}[1]{Figure~\ref{#1}}
\renewcommand{\eqref}[1]{equation~\ref{#1}}

\renewcommand{\d}{\text{d}}

\title{A dynamical implementation of colour coherence for quenched jets in JEWEL}

\author{Korinna Zapp}
\affiliation{Department of Physics, Lund University, SE-22362 Lund, Sweden}

\emailAdd{korinna.zapp@fysik.lu.se}

\abstract{Colour coherence affects the radiation pattern of hard partons both in vacuum and in a dense coloured background formed in heavy ion collisions. In vacuum evolution it leads to the well-known phenomenon of angular ordering, and in heavy ion collisions the appearance of a medium resolution scale strongly affects the way in which a fragmenting hard parton interacts with the background medium. In this paper I present the implementation of colour coherence in the \textsc{Jewel} event generator for jet evolution in a dense medium. In each  interaction between a hard parton and the medium it is checked whether the momentum transfer of the scattering is sufficient to resolve the colour dipole. In this way it is dynamically decided which structures stay coherent. Importantly, scatterings that resolve an individual parton disrupt the colour coherence, which affects the next splitting via the loss of angular ordering. This leads to a suppression of hard radiation, and consequently a reduction in overall scattering rate, which is the dominant source of effects of colour coherence observable in reconstructed jets. I discuss these modifications using the examples of nuclear modification factor, jet fragmentation function and jet--hadron correlations. }

\begin{document} 
\maketitle
\flushbottom
	
\section{Introduction}
\label{sec:intro}

Colour coherence is a fundamental property of QCD and refers to the interference between partons carrying matching colour and anti-colour. The physical picture is that when the two partons are not resolved individually by a process, e.g. the emission of an additional gluon, they remain coherent and effectively act as one object. These interference effects lead to the string~\cite{Andersson:1980vk,Azimov:1985zta} or drag effect~\cite{Dokshitzer:1987nm} experimentally first observed in electron--positron collisions by the \textsc{Jade} experiment at \textsc{Petra}~\cite{JADE:1981ofk} and later by \textsc{Cdf} in proton--anti-proton collisions at the Tevatron~\cite{CDF:1994zkl}. Another consequence of soft gluon interference is the formation of the hump-backed plateau of the MLLA~\cite{Mueller:1982cq,Azimov:1985by,Bassetto:1983mvz} observed at \textsc{Lep} by the \textsc{Opal} experiment~\cite{OPAL:1990vmr}. Colour coherence is also responsible for the angular ordering property of subsequent gluon emissions in parton showers~\cite{Marchesini:1983bm}.

\smallskip

\begin{figure}
	\centering
	\resizebox{0.5\linewidth}{!}{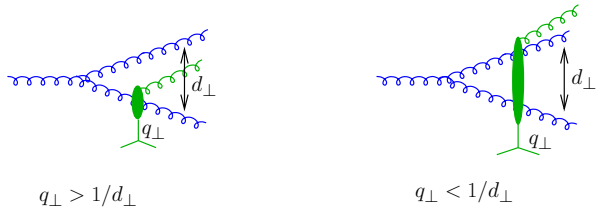}
	\caption{\textbf{Left:} Resolved re-scattering of a colour dipole, where the resolution set by the inverse transverse momentum transfer $q_\perp$ is sufficient to resolve the transverse separation between the partons and both legs radiate independently. \textbf{Right:} Unresolved scattering, where the momentum transfer does not resolve the partons individually and the dipole therefore scatters and radiates coherently.}
	\label{fig:medres}
\end{figure}

In the context of heavy ion collisions the main interest in colour coherence results from the theoretical expectation that the coloured background medium has an intrinsic resolution scale below which hard partonic structures are not resolved and stay coherent.	
The physical picture emerging from first calculations of induced gluon emissions off a colour dipole (antenna) is that the antenna radiates coherently when the medium does not resolve the separation between the two partons, and the partons radiate independently when they are resolved by the medium~\cite{Casalderrey-Solana:2011ule,Mehtar-Tani:2011hma,Mehtar-Tani:2011vlz,Mehtar-Tani:2011lic,Mehtar-Tani:2012mfa} (cf. \figref{fig:medres}). It is important to note that the resolution of the medium is set by the transverse momentum transfer between the hard parton/dipole and the medium, which is generically expected to be of order the screening mass (or inverse screening length). The antenna thus stays coherent in the medium until the two partons have separated by a transverse distance that is resolved by the medium. In these calculations there appears a critical opening angle below which a colour dipole stays coherent during its entire passage through a medium of a given length and density. A dipole that remains unresolved by the medium and therefore interacts with the medium as one effective object can still emit vacuum radiation that resolves its inner structure. The reason for this is that the resolution of the medium and the vacuum emissions are different. One therefore has to be careful to correctly separate the medium resolution determining which object the medium interacts with from the implications for different types of emissions. It is worth noting that while in the earliest calculations the critical angle is determined exclusively by medium properties, relaxing some of the assumptions transforms the critical angle into a dynamical quantity that depends also on the dipole kinematics and differs from splitting to splitting~\cite{Abreu:2024wka}. Furthermore, it was observed that the medium induced radiation off the coherent antenna follows an anti-angular ordering pattern~\cite{Mehtar-Tani:2010ebp}. It was then suggested that the experimental observation of the separation scale at which structures inside jets start to be modified in heavy ion collisions compared to proton--proton collisions could give access to the resolution scale of the medium~\cite{CasalderreySolana:2012ef}. Since these early works the theoretical modelling of antenna radiation in a medium has improved considerably~\cite{Armesto:2012qa,Apolinario:2014csa,Calvo:2014cba,Barata:2021byj,Barata:2024bqp,Abreu:2024wka,Kuzmin:2025fyu,Andres:2025prc,Andres:2026tas}. Although there is a first indication from the hybrid strong/weak model of jet quenching that jet sub-structure data seems to favour a finite resolution scale~\cite{Kudinoor:2025ilx,Kudinoor:2025gao} and also other works find improvements in certain observables due to colour coherence~\cite{Mehtar-Tani:2014yea,Chang:2017gkt,JETSCAPE:2023hqn,Duan:2026nvr}, compelling experimental evidence of colour coherence in quenched jets is presently lacking.

While many analytical and semi-analytical calculations include colour (de)coherence effects~\cite{Dominguez:2019ges,Mehtar-Tani:2016aco,Mehtar-Tani:2017ypq,Mehtar-Tani:2017web,Mehtar-Tani:2021fud,Takacs:2021bpv,Pablos:2022mrx,Cunqueiro:2023vxl,Mehtar-Tani:2024mvl,Apolinario:2026hff,Duan:2026nvr}, up to now the only Monte Carlo models for jet quenching that consider effects of colour coherence are the hybrid strong/weak model of jet quenching~\cite{Casalderrey-Solana:2014bpa} and JetMed~\cite{Caucal:2019uvr}. In the \textsc{Jetscape} framework it is possible to mimic colour coherence by a scale dependent reduction of the jet transport coefficient $\hat q$~\cite{JETSCAPE:2022jer}, while the approach to colour evolution taken in \textsc{Angantyr}~\cite{Lundberg:2025ecc} is also based on dipoles and has some similarities with the traditional picture of colour coherence.

The hybrid strong/weak model of jet quenching, aka the Hybrid Model, uses \textsc{Pythia} to  generate jet production and parton showers without medium effects. The partons in the parton shower (intermediate and final partons) then lose energy according to a holographic calculation in a hydrodynamic background medium. To include coherence effects it is stipulated that after a splitting the daughters will only start interacting independently after they have separated by a distance (in the lab frame) given by the resolution scale~\cite{Hulcher:2017cpt}. Until then the parent parton continues to lose energy. When a parton splits before becoming resolved from its colour partner, coherent systems consisting of more than two partons can form. The temperature dependent resolution scale, which is expected to be of the order of the screening length, is treated as a parameter of the model. The main effects of a finite resolution identified in~\cite{Hulcher:2017cpt} are a modest enhancement of low to intermediate energy fragments found at large angles from the jet axis, while the hard fragments close to the jet axis lose slightly more energy. The overall reduction in energy loss is compensated by re-tuning the coupling parameter, resulting in the same nuclear modification factor as without finite resolution scale. The resolution scale was later constrained using jet sub-structure data~\cite{Casalderrey-Solana:2019ubu,Kudinoor:2025ilx,Kudinoor:2025gao} and it was concluded that a finite resolution length yields the best agreement with the latest measurements.

The JetMed model is based on the observation that in the double logarithmic approximation hard vacuum like and medium induced emissions factorise due to a separation of formation times~\cite{Caucal:2018dla}. The JetMed implementation~\cite{Caucal:2019uvr} contains single logarithmic corrections, but the overall structure remains. The first stage of the evolution consists of vacuum like emissions with short formation times. Then medium induced emissions off the partons produced in the first stage are generated. Finally, after the partons have left the medium they continue to undergo vacuum like splittings. Colour coherence is incorporated in the model by requiring that only partons with a relative angle large enough to be resolved by the medium act as independent sources of medium induced emissions. Furthermore, the first emission outside of the medium does not have to satisfy angular ordering, because the medium will have washed out colour correlations. Structures with angles smaller than the critical angle remain coherent and energy loss is suppressed. 

\smallskip

Here I present the implementation of colour coherence in the \textsc{Jewel}~\cite{Zapp:2012ak} model. In contrast to the Hybrid Model, \textsc{Jewel} assumes a weak coupling scenario and simulates individual scattering processes between the hard partons and thermal partons forming the background medium. This allows a much more dynamical modelling of colour coherence at the level of individual interactions. The scattered object is either a coherent colour dipole (when the transverse momentum transfer is so small that it does not resolve the colour charges individually), or a parton (when the transverse momentum transfer is sufficient to resolve the dipole's structure or the parton has already lost its colour partner due an earlier resolved scattering). The notion of a critical angle is not helpful in this dynamical picture, since it now depends on the dipole and the momentum transfer in each re-scattering which structures are resolved. A resolved scattering involves the exchange of colour between one leg of the dipole and the background medium, which breaks the connection to the colour partner. From that point on the two partons interact independently, i.e.~all scatterings resolve the individual partons (irrespective of the momentum transfer). The loss of the colour partner also has implications for the next emission. When the re-scattering is too soft to affect the parton shower evolution, the next splitting is simply generated by the current parton shower. But after the breaking of the coherence of the colour dipole emissions off the two partons don't interfere any more. Therefore, the next emission off either of the partons is not subject to angular ordering. When the re-scattering is hard and re-sets the parton shower the first emission from the new parton shower is never constrained by angular ordering. Since \textsc{Jewel} has its own parton shower in which splittings are interleaved with scatterings, it is possible to dynamically turn angular ordering on and off depending on whether there was a resolved scattering between the splittings. Therefore, including colour coherence in \textsc{Jewel} leads to a significantly harder fragmentation pattern with fewer splittings. The number of scatterings per parton does not change much, but since fewer partons are produced the total number of scatterings and with it the medium response component is reduced. The consequences of this are visible in jet observables.

Already in previous \textsc{Jewel} versions angular ordering was by default turned off for emissions inside the medium and the first emission after a parton shower re-set due to a hard re-scattering was not required to satisfy angular ordering. What is new in \textsc{Jewel}\,2.6.0 is that angular ordering can be dynamically turner on and off for each emission instead of globally off for all emissions inside the medium. Also new is the introduction of the resolution for momentum transfers from the medium with the possible outcomes of scattering of a coherent colour dipole or an individual parton. A limitation imposed by the \textsc{Jewel} code structure is that only coherent states of two partons can be considered, and that scatterings of a coherent dipole can only be elastic. A benefit of this approach, on the other hand, is that it allows for fluctuations. 

\smallskip

The rest of this paper is organised as follows: the dynamical treatment of angular ordering is discussed in detail in section~\ref{sec:angord}. \Secref{sec:implementation} details how coherent and incoherent scattering is implemented and the impact of colour coherence on the nuclear modification factor, jet fragmentation function, and jet--hadron correlations is shown in \secref{sec:results}. Final discussion and conclusions are contained in \secref{sec:conclusions}. All Monte Carlo results shown in this paper have been generated with the new version 2.6.0 of  \textsc{Jewel}\footnote{The source code is available on \url{jewel.hepforge.org}.} and the paper  documents the new features of this version of the model.

\section{Angular ordering with and without medium modifications}
\label{sec:angord}

\begin{figure}
	\resizebox{\linewidth}{!}{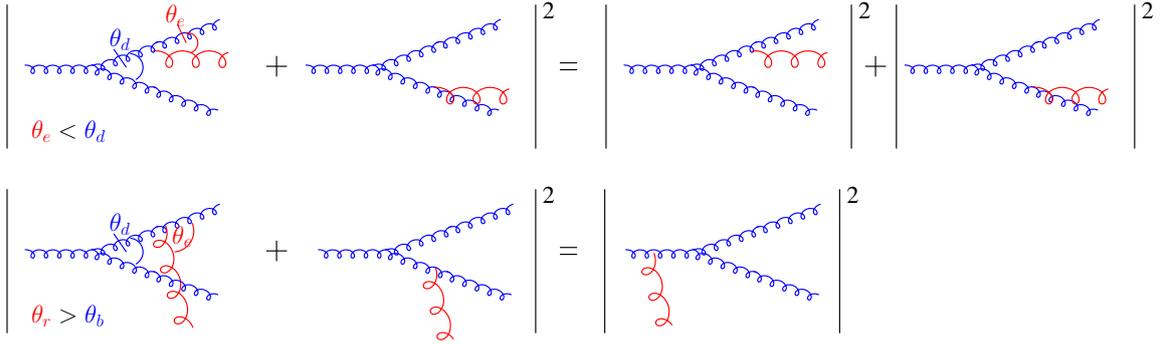}
	\caption{Angular ordering property of subsequent emissions in QCD: when the emission angle $\theta_e$ of the red gluon is smaller than the opening angle $\theta_d$ of the radiating dipole the two blue legs radiate independently (top row). When the emission angle is larger than the dipole opening angle the emissions off the two legs interfere destructively and it looks as if the emission was coming off the combined colour charge of the dipole (bottom row).}
	\label{fig:angord}
\end{figure}

Angular ordering is a consequence of colour coherence and occurs in emissions off a colour dipole~\cite{Marchesini:1983bm}.\footnote{Angular ordering is the QCD analogue of the Chudakov effect~\cite{Chudakov:1955} in QED.} A colour dipole is typically created by a previous emission as depicted in blue in \figref{fig:angord} for the case of a gluon emitting another gluon. The two resulting gluons have a matching colour and anti-colour charge and thus form a colour dipole. When this dipole emits a further gluon (shown in red in \figref{fig:angord}) the emissions off both legs interfere. The physical picture is that when the emission angle of the second gluon is smaller than the opening angle of the dipole (i.e.\ the emission angle of the first emission) the second gluon resolves the two colour charges and both legs of the dipole radiate independently. In the opposite case where the emission angle of the second gluon is larger than the dipole opening angle the two charges are not resolved and the dipole stays coherent. In this case there is destructive interference between the emissions off both legs of the dipole. The resulting radiation off the coherent dipole looks as if it was coming from the combined colour charge (which is the colour charge of the gluon before the first emission). The emission histories are thus ordered in emission angle. This qualitative picture is supported by calculations in QCD which confirm the angular ordering property of subsequent emissions (in the soft gluon limit and after integrating out the azimuthal angle).

\smallskip

Parton showers, which simulate the QCD scale evolution by recursive generation of QCD radiation, have to respect the restriction of the radiation phase space due to angular ordering. In parton showers that have the emission angle as ordering variable for subsequent emission~\cite{Bewick:2019rbu,Lee:2023hef} angular ordering is automatically satisfied. In antenna showers, in which the elementary splitting process is an emission off a colour dipole (antenna) leading to two new antennae, it can be included through an appropriate choice of evolution variable~\cite{Gustafson:1986db}. In other parton showers it can be implemented as an additional requirement by explicitly restricting the radiation phase space.

\subsection{Angular ordering implementations in JEWEL}

The \textsc{Jewel} parton shower is a virtuality ordered shower in off-shell kinematics. In this way it is ensured that the recoil from an emission is absorbed locally (except for the first emission, which will be discussed below) and there is no need to modify the kinematics of earlier splittings, which would lead to inconsistencies in the presence of re-scattering.

In the evolution of a parton $a$ the probability for no splitting to occur between a starting scale $Q_\text{max}^2$ and some lower scale $Q^2$ is given by the Sudakov form factor
\begin{equation}
	\label{eq:sudakov}
	\mathcal{S}_a(Q_\text{max}^2,Q^2) =
	\exp\left\{-
	\int\limits_{Q^2}^{Q_\text{max}^2}\frac{\d Q'^2}{Q'^2}
	\int\limits_{z_-(Q'^2)}^{z_+(Q'^2)}\d z\, \sum_b
	\frac{\alpha_s(k_\perp^2)}{2\pi}\hat 
	P_{ba}(z)
	\right\} \,,
\end{equation}
where $z$ is the energy sharing between the partons $b$ and $c$, $\hat P_{ba}$ is the unregularised Altarelli-Parisi splitting function for the splitting $a\to b+c$, and the strong coupling $\alpha_s$ is evaluated at the relative squared transverse momentum $k_\perp^2 \approx z(1-z)Q'^2$ of the splitting. In \textsc{Jewel} the limits $z_\pm(Q'^2)$ specifieing the allowed $z$-range are derived from the requirement of a minimum transverse momentum $k_\perp^2 \ge Q_0^2/4$ for a splitting of $a$ into massless daughters. This leads to 
\begin{equation}
	z_\pm(Q'^2) = \frac{1}{2} \pm \frac{1}{2}\sqrt{\left( 1 - \frac{Q_0^2}{Q'^2}\right)\left( 1 - \frac{Q'^2}{E_a^2}\right)} \,,
\end{equation}
where $E_a$ is the energy of parton $a$. The differential probability for splitting at a scale is obtained as the negative derivative of the Sudakov form factor with respect to $Q^2$
\begin{equation}
	\label{eq:splitprob}
	\frac{\d \mathcal{P}}{\d Q^2} = - \frac{\d \mathcal{S}_a(Q_\text{max}^2,Q^2)}{\d Q^2} \,.
\end{equation}
This determines the virtual mass $m_a = \sqrt{Q^2}$ of parton $a$. When the veto algorithm is employed to sample \eqref{eq:splitprob}, a value for the energy sharing $z_a$ at the splitting is obtained at the same time. 
\begin{figure}
	\centering
	\resizebox{0.3\linewidth}{!}{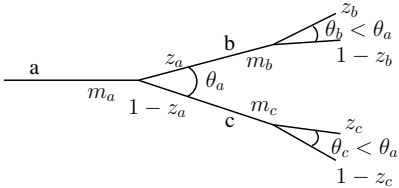}
	\caption{Splitting of parton $a$ with mass $m_a$ into $b$ and $c$ with masses $m_b$ and $m_c$, respectively, and the opening angles subject to angular ordering.}
	\label{fig:splitting}
\end{figure}
As mentioned earlier, \textsc{Jewel} generates the parton shower in off-shell kinematics in order to avoid having to adjust the kinematics of an earlier splitting \textit{a posteriori}. Therefore, when the splitting $a\to b+c$ depicted in \figref{fig:splitting} is generated the masses of the daughters $m_b$ and $m_c$ already have to be determined. The daughter masses are again obtained from \eqref{eq:splitprob} with $Q_\text{max}^2 = m_a^2$ as starting scale. However, $m_b$ and $m_c$ have to satisfy additional constraints, which means that the masses obtained from \eqref{eq:splitprob} can get rejected. Firstly, there are the kinematic constraints $k_\perp^2 > 0$ and $m_b+m_c < m_a$ at the $a\to b+c$ vertex, and secondly there are the angular ordering conditions $\theta_b < \theta_a$ and $\theta_c < \theta_a$. The splitting angle of a parton $i$ can be approximated by
\begin{equation}
	\label{eq:splitangle}
	\theta_i \approx \frac{m_i}{\sqrt{z_i(1-z_i)}E_i} \,.
\end{equation}
The opening angle $\theta_a$ of the splitting $a\to b+c$ thus depends only on the virtual mass $m_a$ of parton $a$ (i.e.\ the scale of the splitting), the energy sharing $z_a$ in the splitting, and the splitting parton's energy $E_a$. Similarly, the splitting angles of the daughters depend only on the daughters' virtual mass and energy and the energy sharing at the daughter splittings. The kinematic constraints thus involve both daughters while the angular ordering condition affects each daughter individually.\footnote{There is a correlation between the daughters because they share the common mother's energy, but as far as the angular ordering condition is concerned the two daughters are independent.} It is worth noting that angular ordering favours smaller masses and more symmetric splittings. It thus affects not only the angular distribution of fragments, but also the energy sharing among them. 

If the kinematic constraints or either of the angular ordering conditions are violated the one or both values for the virtual masses of daughters are rejected. 
When this happens, a new value for $m_b$ and/or $m_c$ (together with the corresponding values for $z_{b,c}$) are generated from \eqref{eq:splitprob} with the scale of the rejected splitting as starting scale $Q_\text{max}$.
 
\begin{figure}
	\begin{tabular}{ccc}
		\multirow{2}{0.3\textwidth}[1.75cm]{
		\resizebox{0.3\textwidth}{!}{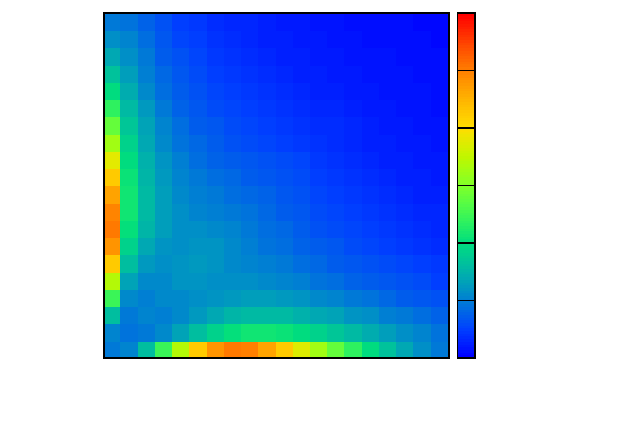}} &
		\resizebox{0.3\textwidth}{!}{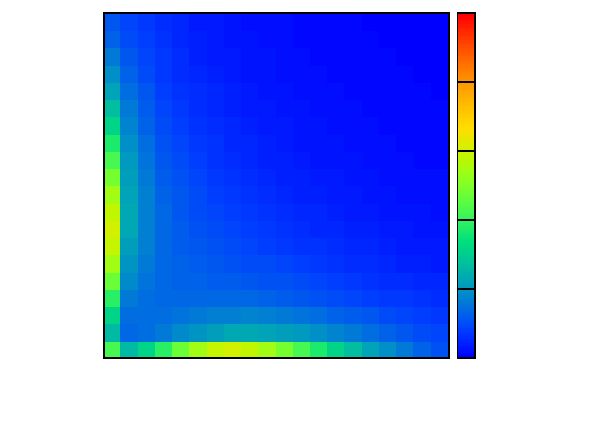} & 
		\resizebox{0.3\textwidth}{!}{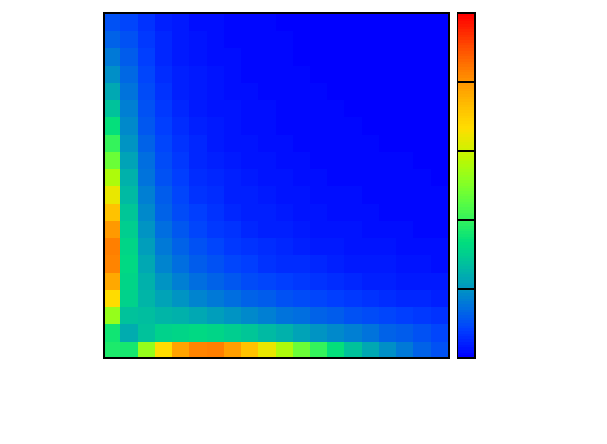} \\
		& \resizebox{0.3\textwidth}{!}{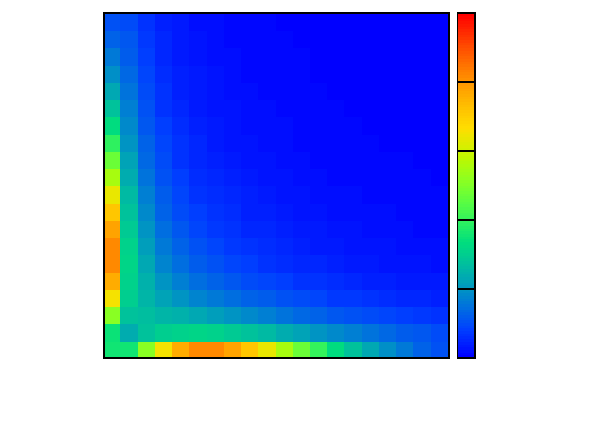} & 
	    \resizebox{0.3\textwidth}{!}{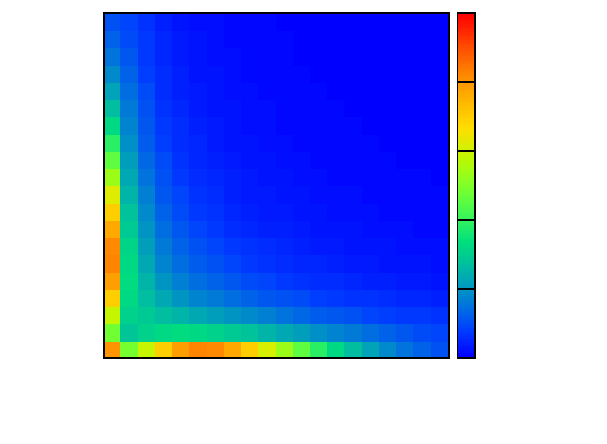}
	\end{tabular}
	\caption{Values of $m_b$ and $m_c$ obtained for the splitting $a\to b+c$ with $E_a = \unit[200]{GeV}$. The splitting variables $m_a$ and $z_a$ are generated from \eqref{eq:splitprob} with $Q_\text{max} = E_a$ without further constraints. The plot in the left column shows the values of $m_b$ and $m_c$ obtained from \eqref{eq:splitprob} with $Q_\text{max} = m_a$. The middle column shows the result of imposing either the kinematic constraints (top) or angular ordering (bottom). Finally, the right column shows the distributions after imposing both constraints in different orders: kinematics first and angular ordering afterwards (top) or angular ordering first and kinematic constraints afterwards (bottom).}
	\label{fig:ppoptions}
\end{figure}

As mentioned earlier, in the approximation of \eqref{eq:splitangle} the opening angle depends only on the splitting variables and properties of the splitting parton. One can thus choose in which order kinematic and angular ordering constraints should be imposed (this choice is beyond the formal LL accuracy of the parton shower). Since the two constraints prefer different regions in the $(m_b,m_c)$ plane the two options lead to slightly different outcomes, as illustrated in \figref{fig:ppoptions} for the splitting of \figref{fig:splitting}. Both the probability for further splitting of the daughters and the distribution of splitting scales of the daughters differ. These differences are also visible in jet observables, as will be discussed in the next sub-section.

\smallskip

Both options for the order in which kinematic and angular ordering constraints are imposed are implemented in \textsc{Jewel}\,2.6.\footnote{In earlier versions only the option of enforcing angular ordering first and the kinematic constraints afterwards was available (cf.\ \appref{sec:jewelchanges} for more details on changes in \textsc{Jewel}\,2.6.0).} The default choice is to use the approximated splitting angle of \eqref{eq:splitangle}. However, when a kinematically valid configuration is enforced first one can work with the exact expression for the splitting angle instead. Angular ordering can also be turned off for comparison.

An important advantage of the \textit{kinematics first} option is that the check whether angular ordering is satisfied can be postponed until potential scatterings of the daughters have taken place. Angular ordering can thus be dynamically turned off when the parton shower is evolving in a background medium. When either of the partons $b$ or $c$ coming from a given splitting $a\to b+c$ experiences a re-scattering before splitting the colour exchange with the background breaks the colour coherence of the pair. The subsequent splittings of $b$ and $c$ are then not subject to angular ordering. The virtual masses of $b$ and $c$ will then never be rejected and there is thus no need to modify the past. When there was no scattering and a virtual mass gets rejected, nothing has happened since the last splitting and the rejection does not lead to problems. 

In the new implementation in \textsc{Jewel}\,2.6 it is therefore checked whether parton $b$ or $c$ underwent a re-scattering before splitting. If this is the case angular ordering is not required when $b$ and $c$ split. If neither of them scattered angular ordering is imposed. This can lead to the rejection of the splitting of either one or both of the partons. The new splitting scale is lower than the rejected one by construction and therefore the formation time increases. During this additional time the parton can undergo a re-scattering with corresponding implications for the following splitting. In addition to this dynamical procedure it is also possible to always enforce angular ordering (irrespective of whether or not a re-scattering took place between the splittings).

With the \textit{angular ordering first} option the decision whether or not to require angular ordering for the next splitting has to be taken before it is known whether there is going to be a re-scattering. The solution implemented in \textsc{Jewel} is to require angular ordering only for splittings that take place outside the background medium while splittings in a region with $T(\mathbf{r},t) > T_c$ are not required to obey angular ordering. Again, it is also possible to always enforce angular ordering.

\smallskip

The first splitting after the matrix element is different from the subsequent ones. To determine the largest splitting angle allowed by angular ordering for the first splitting the colour partner in the final state of the respective parton is found. The angle between the splitting parton and the colour partner is taken as the largest possible splitting angle (in case of gluons it is taken to be the smaller of the angles between the gluon and its two colour partners). When the colour partner is in the initial state the angle of the first emission is unconstrained. This prescription only considers interference between emissions off final state partons, the interference between final and initial state partons can in \textsc{Jewel} not be taken into account due to the separation of initial and final state parton shower.\footnote{For the same reason the angle of the first final state emission of a parton radiated off an initial state parton is unconstrained.} The two hard partons coming from the matrix element are coupled together by kinematics because the recoil from the first emission off one of them has to be transferred to the other one. Since the recoil has an impact on the radiation phase space of the parton absorbing the recoil the only consistent way of imposing angular ordering is to first find a kinematically allowed configuration and impose angular ordering afterwards. The recoil handling is also the reason why angular ordering for the first emission can not be turned on and off dynamically. It is either always on or always off. The default choice is to have it always on since the first splitting typically has a very short formation time and the scattering rate at very early times is not high (cf.~\figref{fig:thats}	).

\subsection{Comparison to experimental data}

\begin{figure}
	\includegraphics[width=0.5\textwidth]{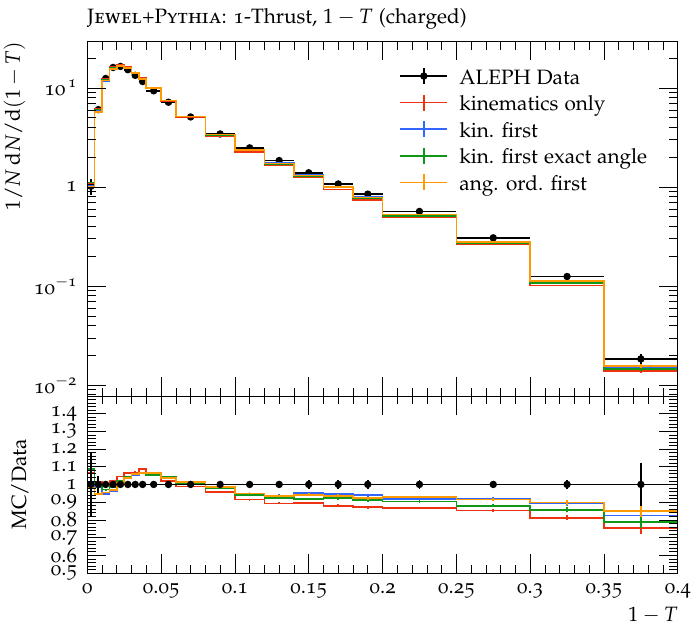}
	\includegraphics[width=0.5\textwidth]{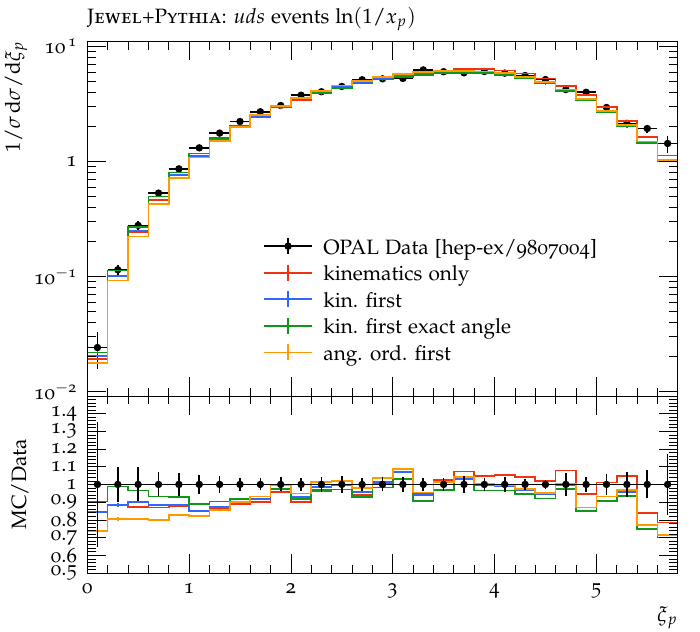}
	\caption{\textsc{Jewel}\,2.6.0+\textsc{Pythia} results with different angular ordering options compared to \textsc{Aleph} data~\cite{ALEPH:1996oqp} for thrust (left) and \textsc{Opal} data~\cite{OPAL:1998arz} for the charged particle fragmentation function (right) in $e^++e^-$ collisions at $\sqrt{s} = \unit[91.2]{GeV}$.}
	\label{fig:lep1}
\end{figure}

\begin{figure}
	\includegraphics[width=0.5\textwidth]{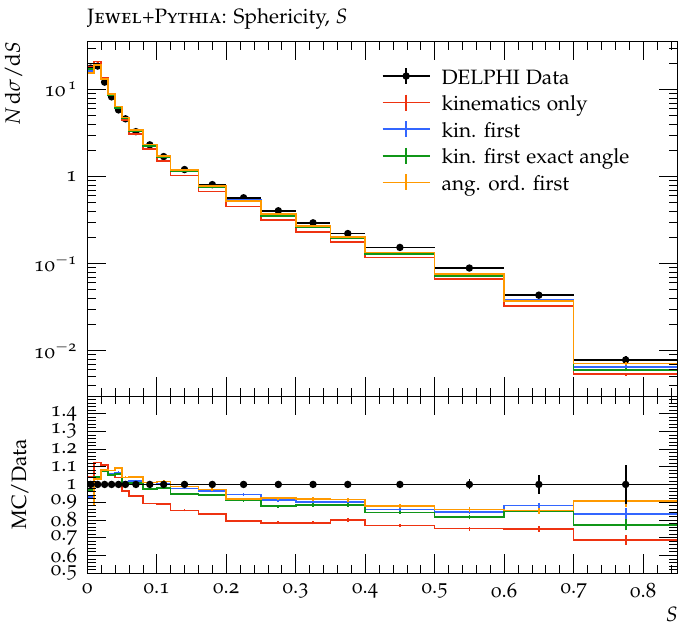}
	\includegraphics[width=0.5\textwidth]{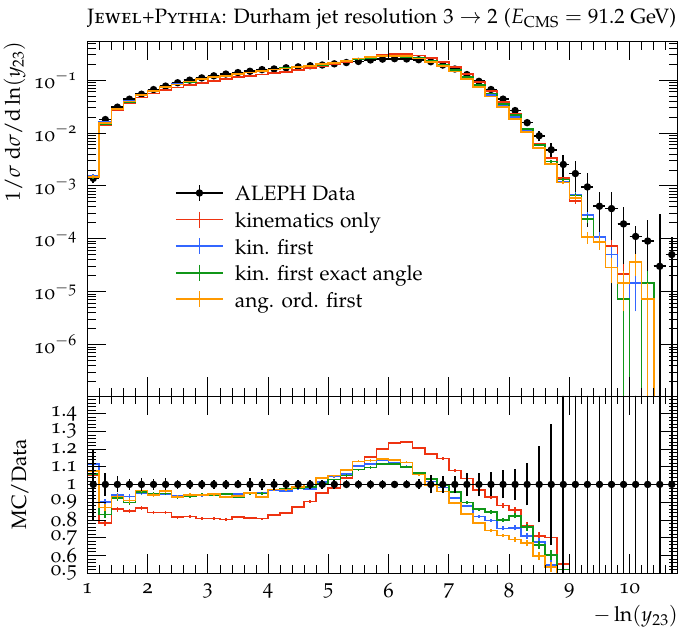}
	\caption{\textsc{Jewel}\,2.6.0+\textsc{Pythia} results with different angular ordering options compared to \textsc{Delphi} data~\cite{DELPHI:1996sen} for sphericity (left) and \textsc{Aleph} data~\cite{ALEPH:2003obs} for the $3 \to 2$ Durham jet resolution (right) in $e^++e^-$ collisions at $\sqrt{s} = \unit[91.2]{GeV}$.}
	\label{fig:lep2}
\end{figure}

All results shown here were generated with \textsc{Jewel}\,2.6.0 using the \textsc{Ct14nlo} PDF set~\cite{Dulat:2015mca} for p+p and the \textsc{Epps16nlo} nuclear PDF set~\cite{Eskola:2016oht} for Pb+Pb collisions, all PDF sets are provided by \textsc{Lhapdf\,6}~\cite{Buckley:2014ana}. The Monte Carlo events are analysed using the \textsc{Rivet}~\cite{Bierlich:2019rhm}, HepMC~\cite{Buckley:2019xhk}, FastJet~\cite{Cacciari:2011ma} and \textsc{Yoda} packages. Medium response is included for simulations in heavy ion collisions and the uncorrelated background contribution is subtracted using the constituent subtraction method~\cite{Milhano:2022kzx}. Since the focus is here on understanding the effects of colour coherence, which does not benefit from a sophisticated model for the background medium, the fast and lightweight simplified model shipped with \textsc{Jewel} is used. The emphasis is on measurements in p+p and central Pb+Pb collisions at $\sqrt{s_{NN}} = \unit[5.02]{TeV}$. Further parameter settings are specified in \appref{sec:jewelsettings}. 

\smallskip

The first part of this sub-section is devoted to comparing the different angular ordering options in vacuum, followed by a selection of results for heavy ion collisions. The options that are compared for jets evolving in vacuum are \textit{kinematics only} (i.e. no angular ordering), \textit{kinematics first}, \textit{kinematics first with true angle}, and \textit{angular ordering first}. In electron--positron collisions at \textsc{Lep} the three options with angular ordering yield almost identical results for most event shape variables, while the two \textit{kinematics first} versions describe the hard tail of the fragmentation functions slightly better. As an example \figref{fig:lep1} shows the comparison of \textsc{Jewel} to an \textsc{Aleph} measurement of thrust~\cite{ALEPH:1996oqp} and a \textsc{Opal} measurement of the charged particle fragmentation function~\cite{OPAL:1998arz}. The event shapes and jet rates generally disfavour the option without angular ordering. \Figref{fig:lep2} shows as an example the comparison to \textsc{Delphi} data for sphericity~\cite{DELPHI:1996sen} and to \textsc{Aleph} data for the $3 \to 2$ Durham jet resolution~\cite{ALEPH:2003obs}.

\begin{figure}
	\includegraphics[width=0.5\textwidth]{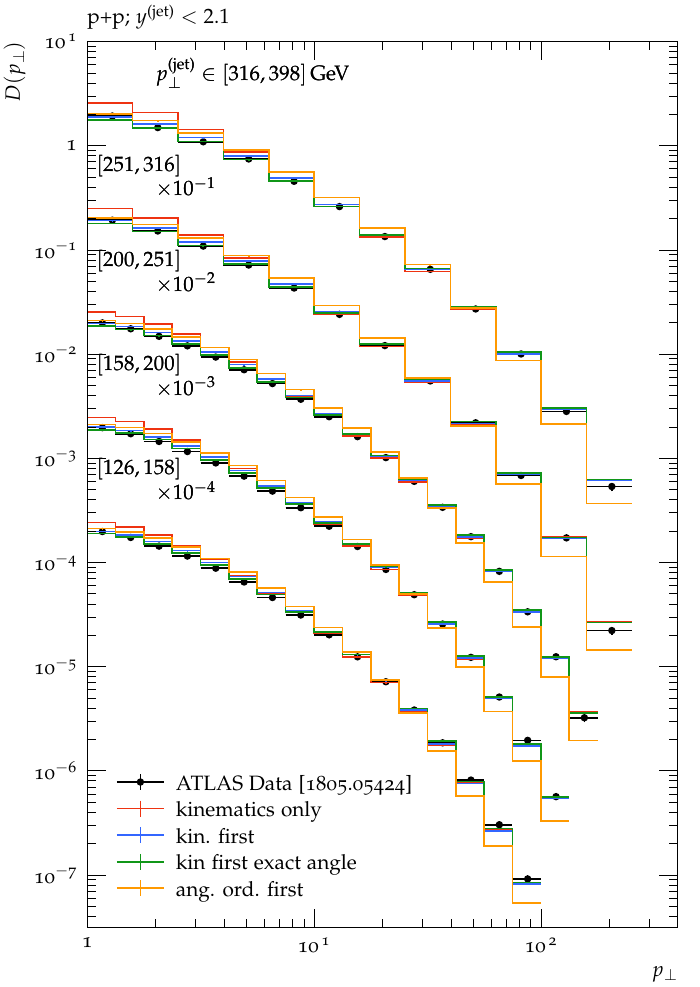}
	\includegraphics[width=0.5\textwidth]{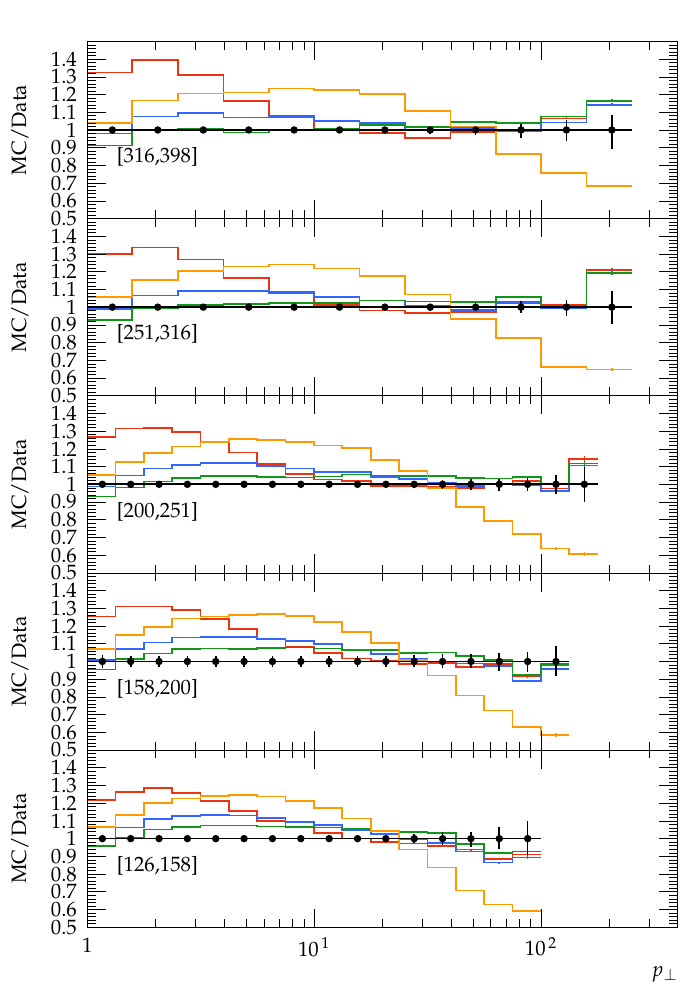}
	\caption{\textsc{Jewel}\,2.6.0+\textsc{Pythia} results with different angular ordering options compared to jet fragmentation functions in p+p collisions at $\sqrt{s_{NN}} = \unit[5.02]{TeV}$ measured by \textsc{Atlas}~\cite{ATLAS:2018bvp} for $R=0.4$ jets in bins of jet transverse momentum.}
	\label{fig:FF_pp}
\end{figure}

\begin{figure}
	\includegraphics[width=0.5\textwidth]{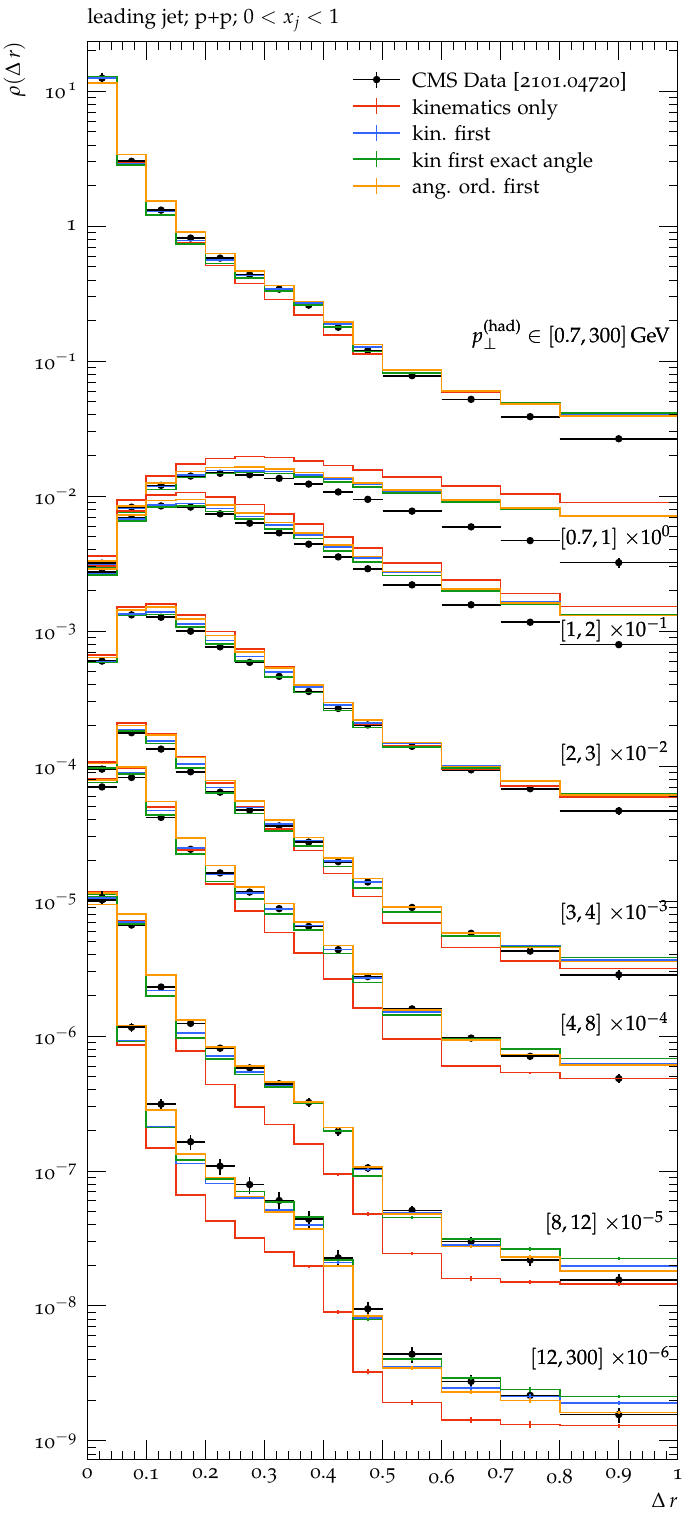}
	\includegraphics[width=0.5\textwidth]{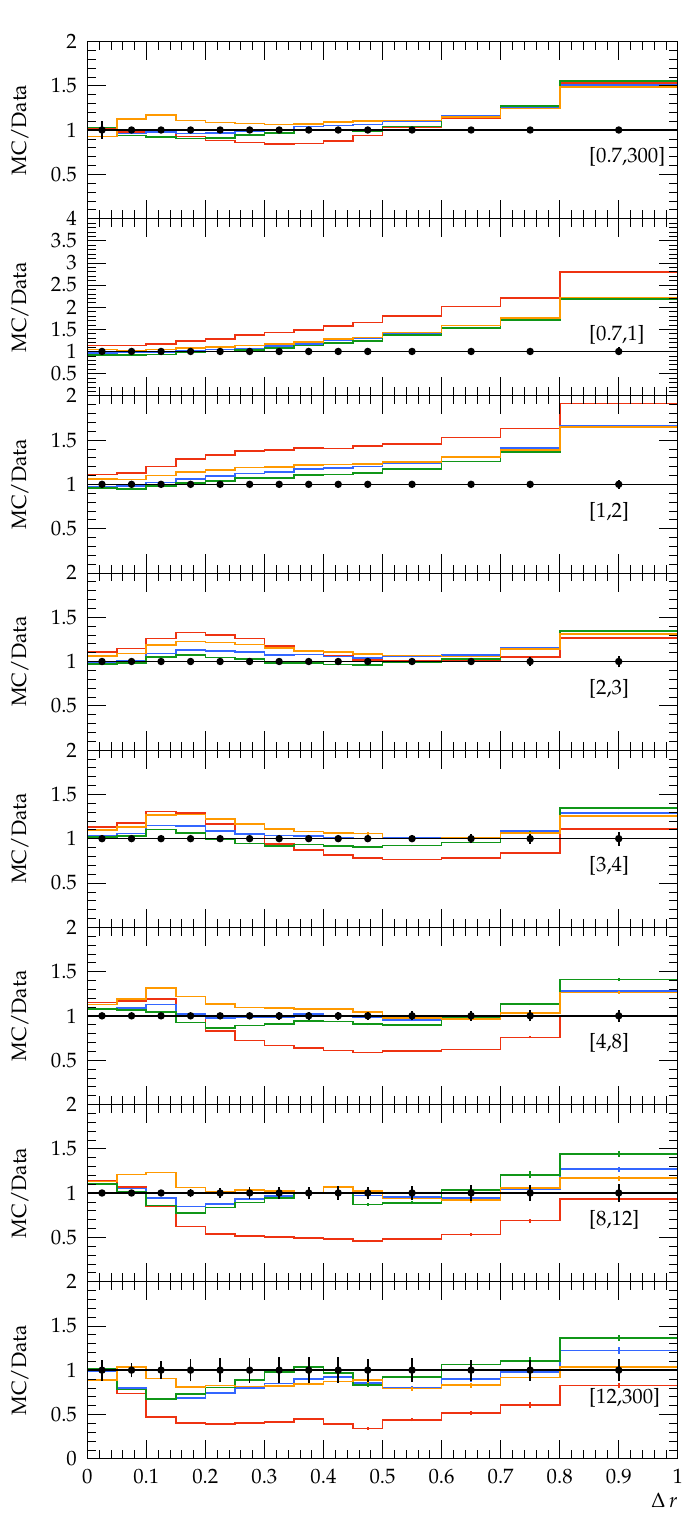}
	\caption{\textsc{Jewel}\,2.6.0+\textsc{Pythia} results with different angular ordering options compared to jet--hadron correlations in p+p collisions at $\sqrt{s_{NN}} = \unit[5.02]{TeV}$ measured by \textsc{Cms}~\cite{CMS:2021nhn} for $R=0.4$ jets in bins of hadron transverse momentum. Shown here are the correlations for the leading jet in di-jet events. The leading jet is required to have $p_\perp^\text{(jet)} > \unit[120]{GeV}$ and the sub-leading jet $p_\perp^\text{(jet)} > \unit[50]{GeV}$, the azimuthal angle between the two jets has to satisfy $\Delta \phi_{jj} > 5\pi/6$ and both jets have to be within $|\eta^\text{(jet)}| < 1.6$.}
	\label{fig:rho_lead_pp}
\end{figure}

\begin{figure}
	\includegraphics[width=0.5\textwidth]{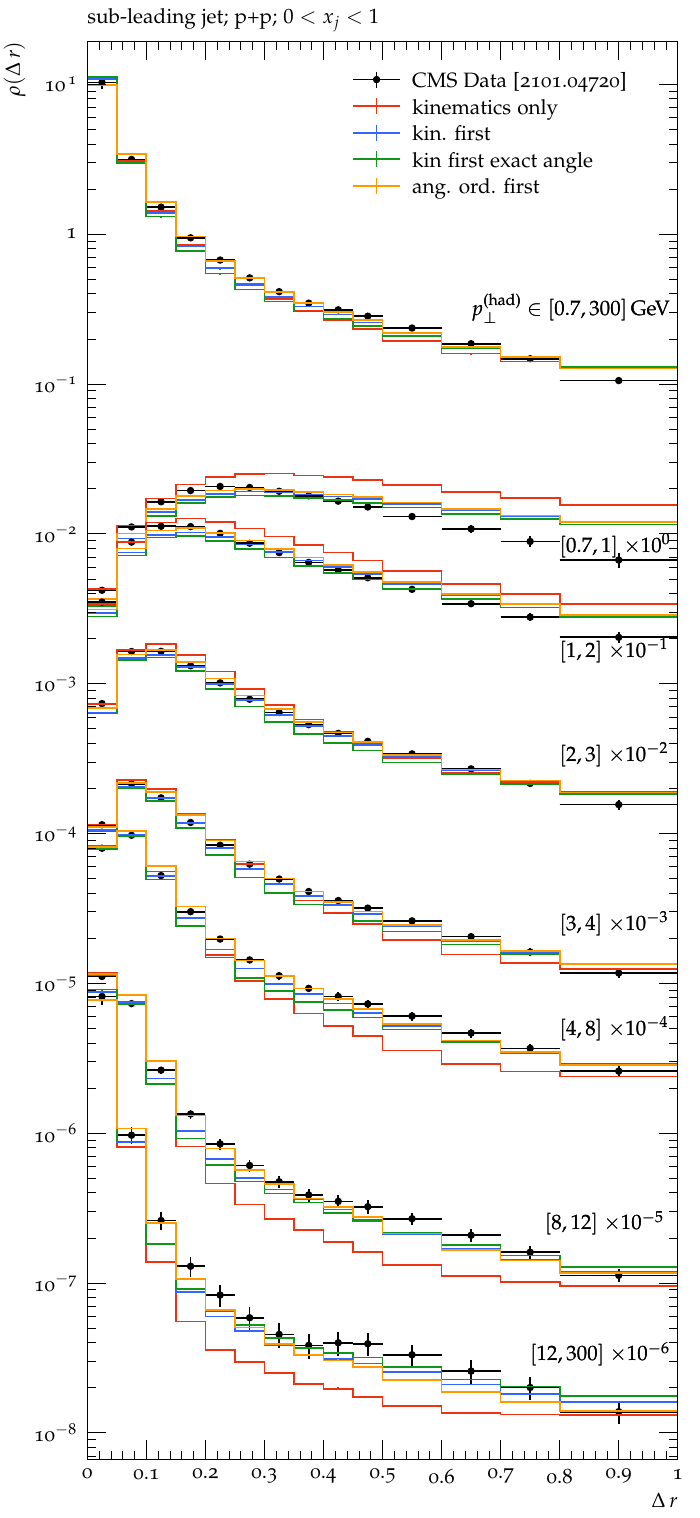}
	\includegraphics[width=0.5\textwidth]{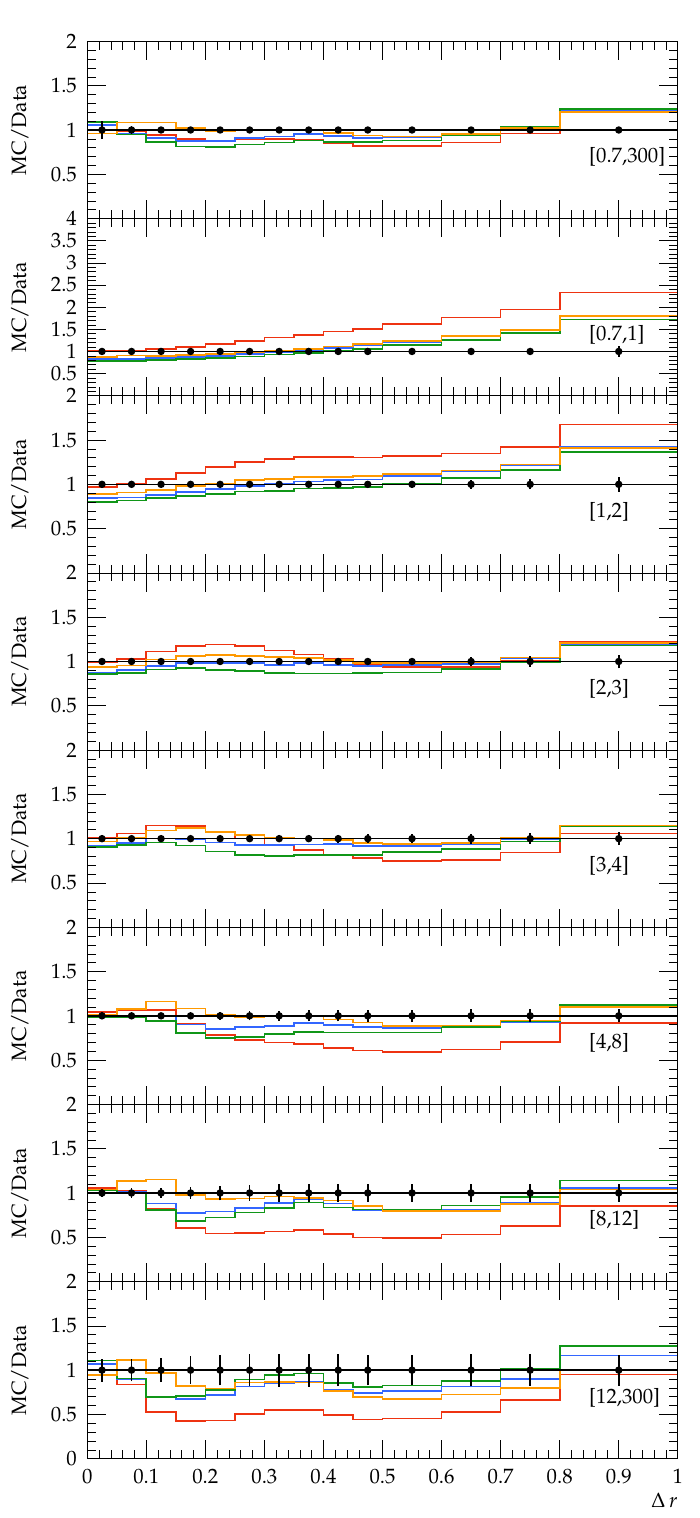}
	\caption{\textsc{Jewel}\,2.6.0+\textsc{Pythia} results with different angular ordering options compared to jet--hadron correlations in p+p collisions at $\sqrt{s_{NN}} = \unit[5.02]{TeV}$ measured by \textsc{Cms}~\cite{CMS:2021nhn} for $R=0.4$ jets in bins of hadron transverse momentum. Shown here are the correlations for the sub-leading jet in di-jet events. The leading jet is required to have $p_\perp^\text{(jet)} > \unit[120]{GeV}$ and the sub-leading jet $p_\perp^\text{(jet)} > \unit[50]{GeV}$, the azimuthal angle between the two jets has to satisfy $\Delta \phi_{jj} > 5\pi/6$ and both jets have to be within $|\eta^\text{(jet)}| < 1.6$.}
	\label{fig:rho_subl_pp}
\end{figure}

Moving to proton--proton collisions at the \textsc{Lhc}, \figref{fig:FF_pp} shows the comparison to the \textsc{Atlas} measurement of the jet fragmentation function~\cite{ATLAS:2018bvp} at $\sqrt{s_{NN}} = \unit[5.02]{TeV}$. Without angular ordering the jet fragmentation functions in \textsc{Jewel} are clearly too soft. With angular ordering turned on the soft and hard regions get suppressed while hadrons with intermediate $p_\perp$ are enhanced. This effect, which is a consequence of splittings getting more symmetric, is more pronounced in the \textit{angular ordering first} option. The \textit{kinematics first} options, on the other hand, describe the data significantly better. The difference between the approximate and exact splitting angle is not very pronounced, the latter is in slightly better agreement with the data. A similar picture is revealed by the jet--hadron correlations measured by \textsc{Cms}~\cite{CMS:2021nhn} and shown in \figref{fig:rho_lead_pp} for the leading jet and in \figref{fig:rho_subl_pp} for sub-leading jet in di-jet pairs. One might naively expect to see a narrowing of the distribution when angular ordering is turned on, but this is not seen in the \textsc{Jewel} results. The reason is that this measurement measures the correlation with respect to the WTA axis, which can change when the energy sharing among the hadrons changes due to angular ordering. Again, the data disfavour the \textit{kinematics only} option and the two \textit{kinematics first} versions are similar. The leading jet correlations are slightly better described by the \textit{kinematics first} option, but for the sub-leading it is not clear which option is preferred. 

As seen in table~\ref{tab:nsplit_incoh} angular ordering severely reduces the number of splittings generated by the parton shower. The different angular ordering options differ significantly in how restrictive they are. \textit{Angular ordering first} allows for more splittings than the \textit{kinematics first} versions, and of the latter the one using the exact angle is more restrictive.

\begin{table}
	\centering
	\begin{tabular}{|l|c|c|c|}
	  \hline
	  & p+p & \multicolumn{2}{c|}{\unit[0 - 10]{\%} Pb+Pb} \\
	  \hline
	  & $\langle N_\text{split} \rangle$ & $\langle N_\text{split} \rangle$ & $\langle f_\text{AO} \rangle$ \ [\unit{\%}]  \\ \hline
	  kinematics only & $26.00 \pm 0.05$  & $33.85 \pm 0.07$ & 0 \\
	  kin. first dyn. & $15.03 \pm 0.03$  & $26.93 \pm 0.05$ & $15.51 \pm 0.05$ \\
	  kin. first exact angle dyn. & $11.72 \pm 0.02$  & $25.11 \pm 0.05$ & $9.60 \pm 0.04$ \\
 	  kin. first always ang. ord. & $15.03 \pm 0.03$  & $21.93 \pm 0.04$ & $72.75 \pm 0.06$ \\
	  ang. ord. first dyn. & $18.53 \pm 0.03$  & $26.69 \pm 0.05$ & $32.47 \pm 0.07$ \\
	  ang. ord. first always ang. ord. &  $18.53 \pm 0.03$  & $26.48 \pm 0.05$ & $75.52 \pm 0.05$ \\
	  \hline
    \end{tabular}
	\caption{Mean number of splittings generated by \textsc{Jewel}'s final state parton shower per event in events producing at least two jets with $p_\perp > \unit[100]{GeV}$ within $|\eta| < 2.8$ (including initial state emissions associated with hard re-scatterings), and average fraction of splittings that have to satisfy angular ordering (denoted $f_\text{AO}$). In the former the first splitting in the final state parton shower is included, while in the latter the first splitting is not included because angular ordering for the first splitting is treated differently from subsequent splittings.}
	\label{tab:nsplit_incoh}
\end{table}

\begin{figure}
	\includegraphics[width=0.5\textwidth]{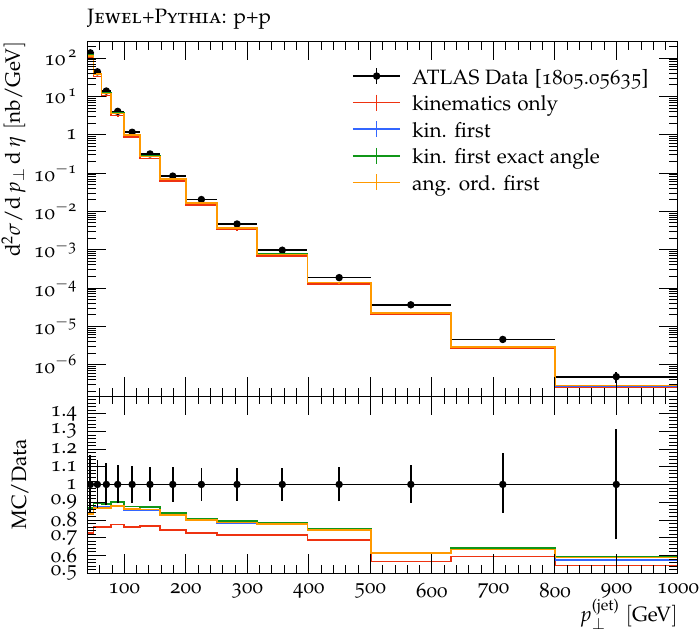}
	\includegraphics[width=0.5\textwidth]{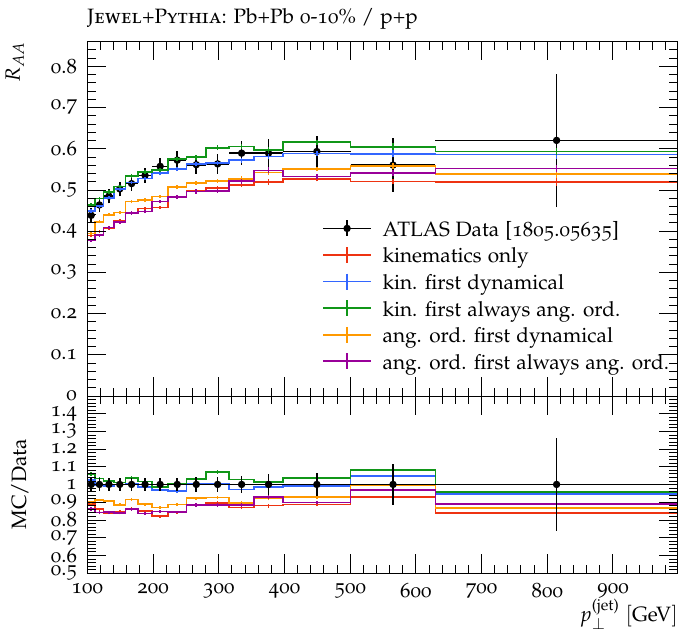}
	\caption{\textbf{Left:} Jet spectrum in p+p collisions at $\sqrt{s_{NN}} = \unit[5.02]{TeV}$ measured by \textsc{Atlas}~\cite{ATLAS:2018gwx} for $R=0.4$ jets within $|\eta^\text{(jet)}| < 2.8$ compared to \textsc{Jewel}\,2.6.0+\textsc{Pythia} with different angular ordering options. \textbf{Right:} Nuclear modification factor $R_\text{AA}$ for $0-10\%$ Pb+Pb collisions at $\sqrt{s_{NN}} = \unit[5.02]{TeV}$ measured by \textsc{Atlas}~\cite{ATLAS:2018gwx} compared to \textsc{Jewel}\,2.6.0+\textsc{Pythia} with different angular ordering options. The p+p reference for the different angular ordering options in medium is the corresponding angular ordering option in p+p, with the exception of the \textit{kinematics only} setting, for which a p+p reference with angular ordering (\textit{kinematics first}) was used.}
	\label{fig:RAA_PbPbm01234}
\end{figure}

\begin{figure}
	\includegraphics[width=0.5\textwidth]{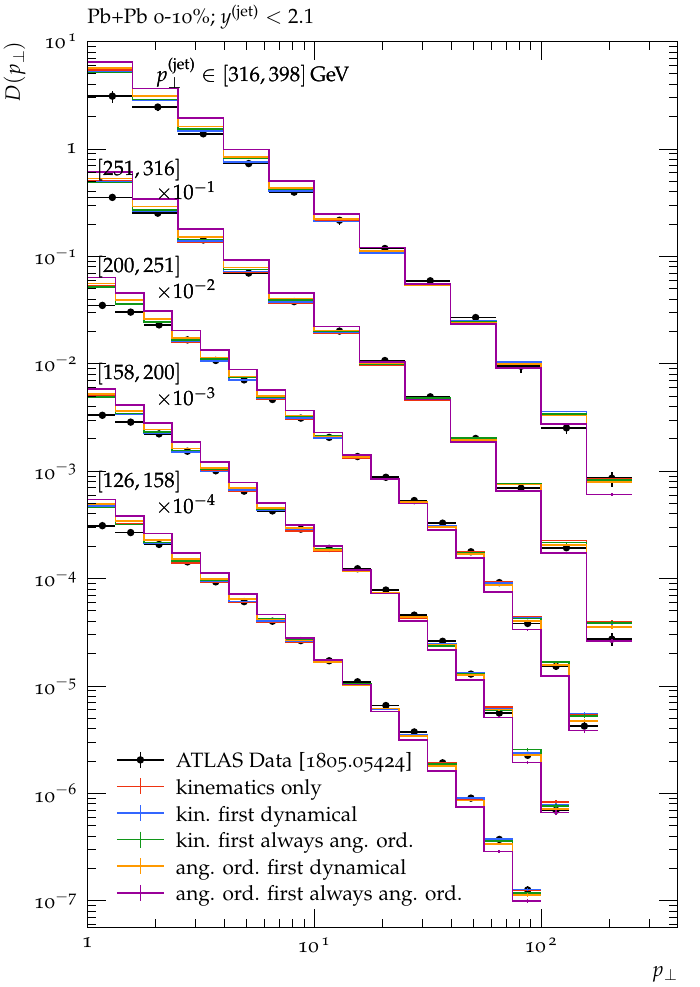}
	\includegraphics[width=0.5\textwidth]{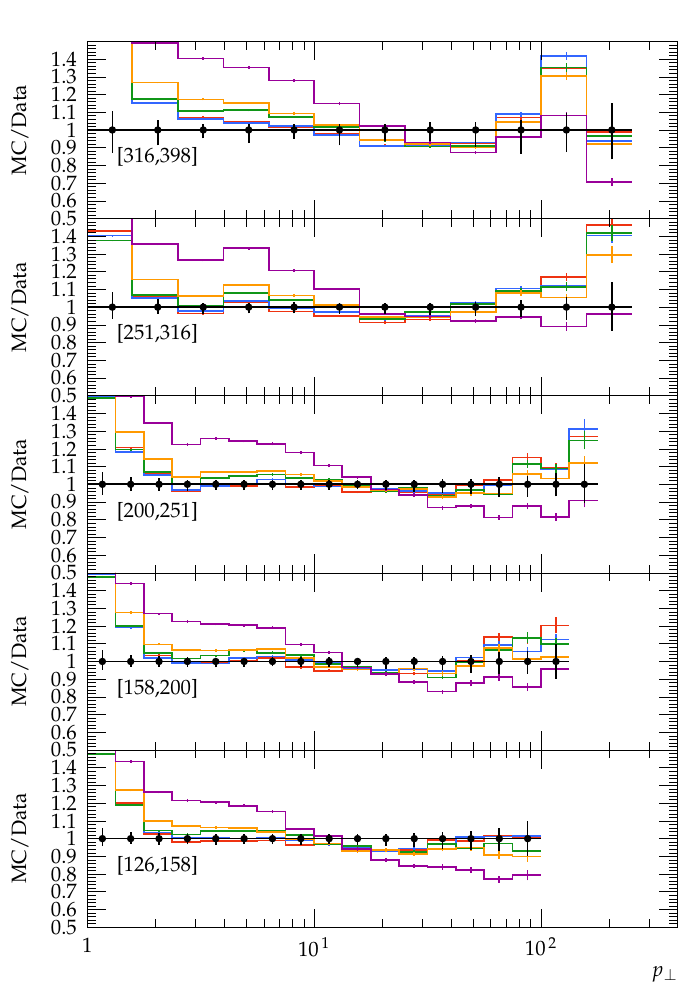}
	\caption{\textsc{Jewel}\,2.6.0+\textsc{Pythia} results with different angular ordering options compared to jet fragmentation functions in $0-10\%$ central Pb+Pb collisions at $\sqrt{s_{NN}} = \unit[5.02]{TeV}$ measured by \textsc{Atlas}~\cite{ATLAS:2018bvp} for $R=0.4$ jets in bins of jet transverse momentum.}
	\label{fig:FF_PbPbm01234}
\end{figure}

\begin{figure}
	\includegraphics[width=0.5\textwidth]{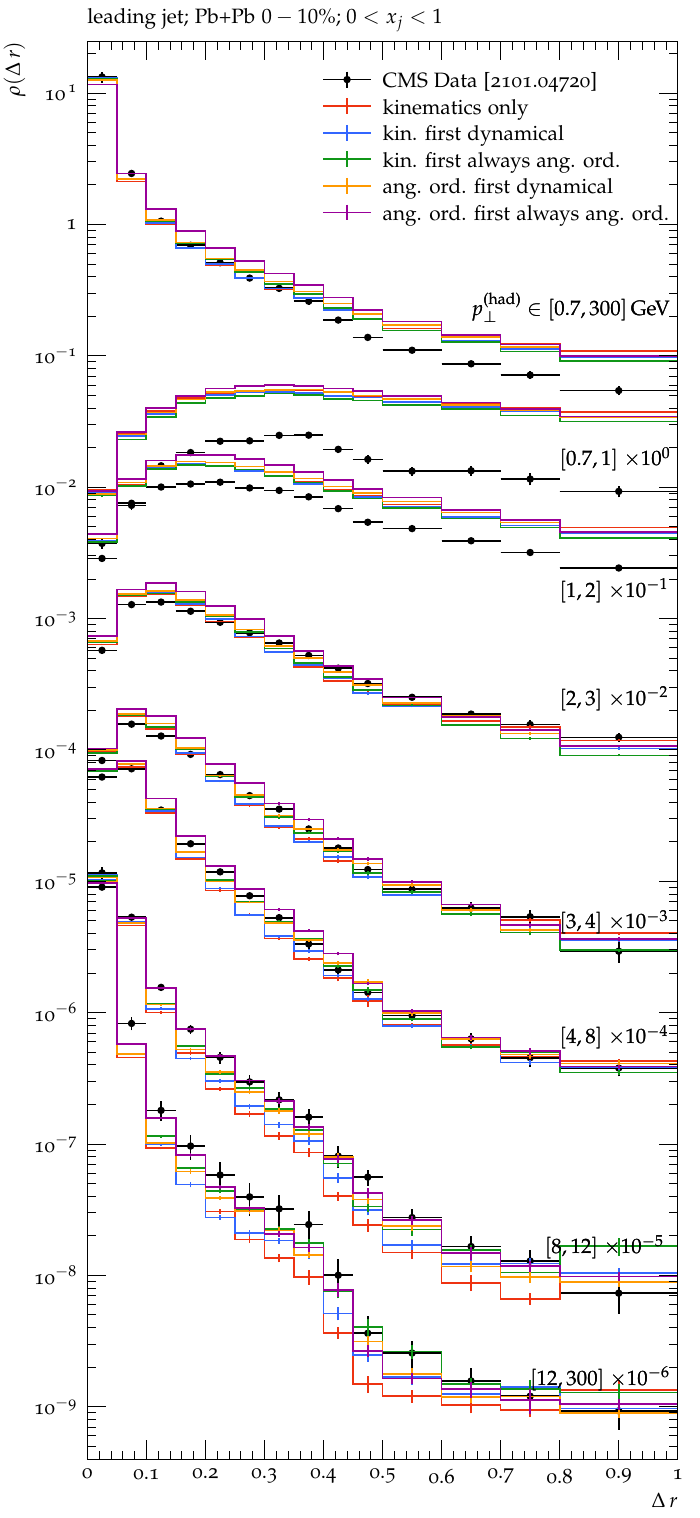}
	\includegraphics[width=0.5\textwidth]{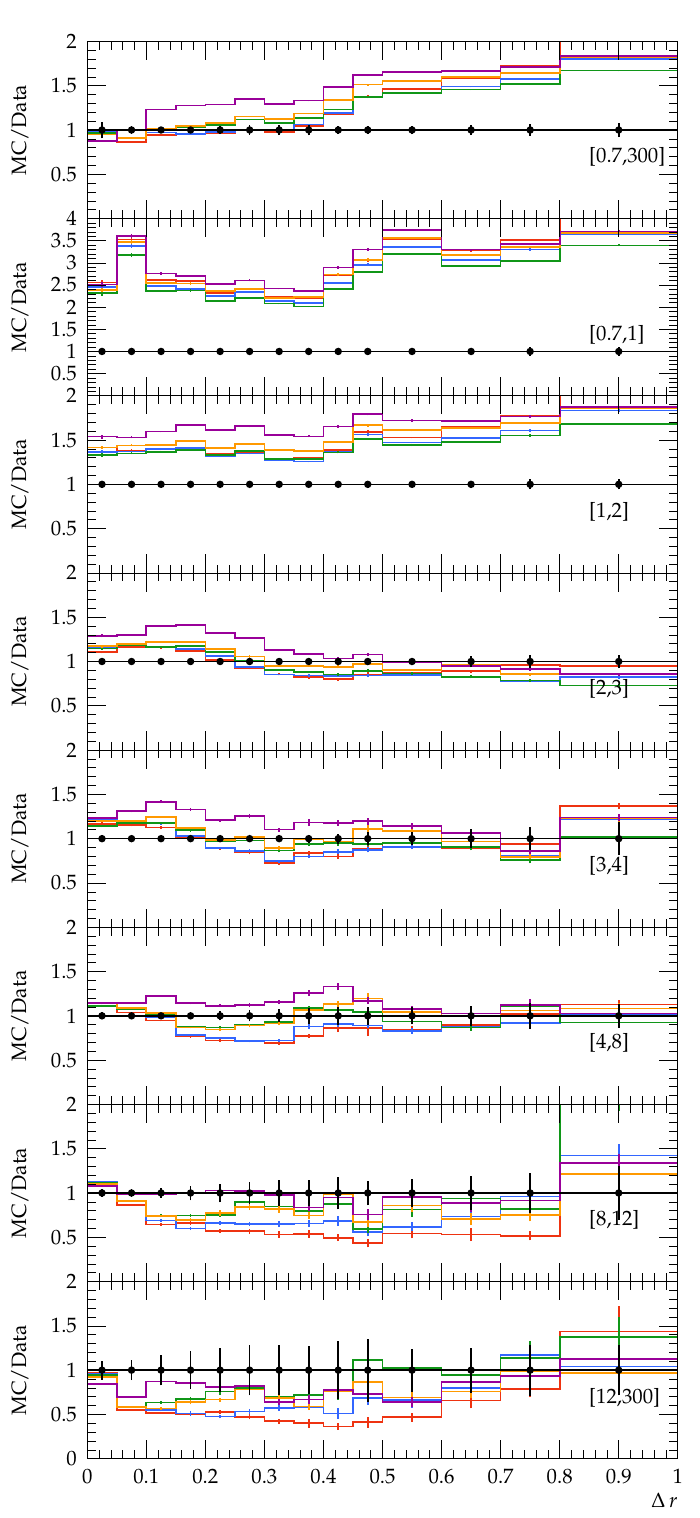}
	\caption{\textsc{Jewel}\,2.6.0+\textsc{Pythia} results with different angular ordering options compared to jet--hadron correlations in $0-10\%$ central Pb+Pb collisions at $\sqrt{s_{NN}} = \unit[5.02]{TeV}$ measured by \textsc{Cms}~\cite{CMS:2021nhn} for $R=0.4$ jets in bins of hadron transverse momentum for $0 < x_\text{j} < 1$. Shown here are the correlations for the leading jet in di-jet events. The leading jet is required to have $p_\perp^\text{(jet)} > \unit[120]{GeV}$ and the sub-leading jet $p_\perp^\text{(jet)} > \unit[50]{GeV}$, the azimuthal angle between the two jets has to satisfy $\Delta \phi_{jj} > 5\pi/6$ and both jets have to be within $|\eta^\text{(jet)}| < 1.6$.}
	\label{fig:rho_lead_PbPbm01234}
\end{figure}

\begin{figure}
	\includegraphics[width=0.5\textwidth]{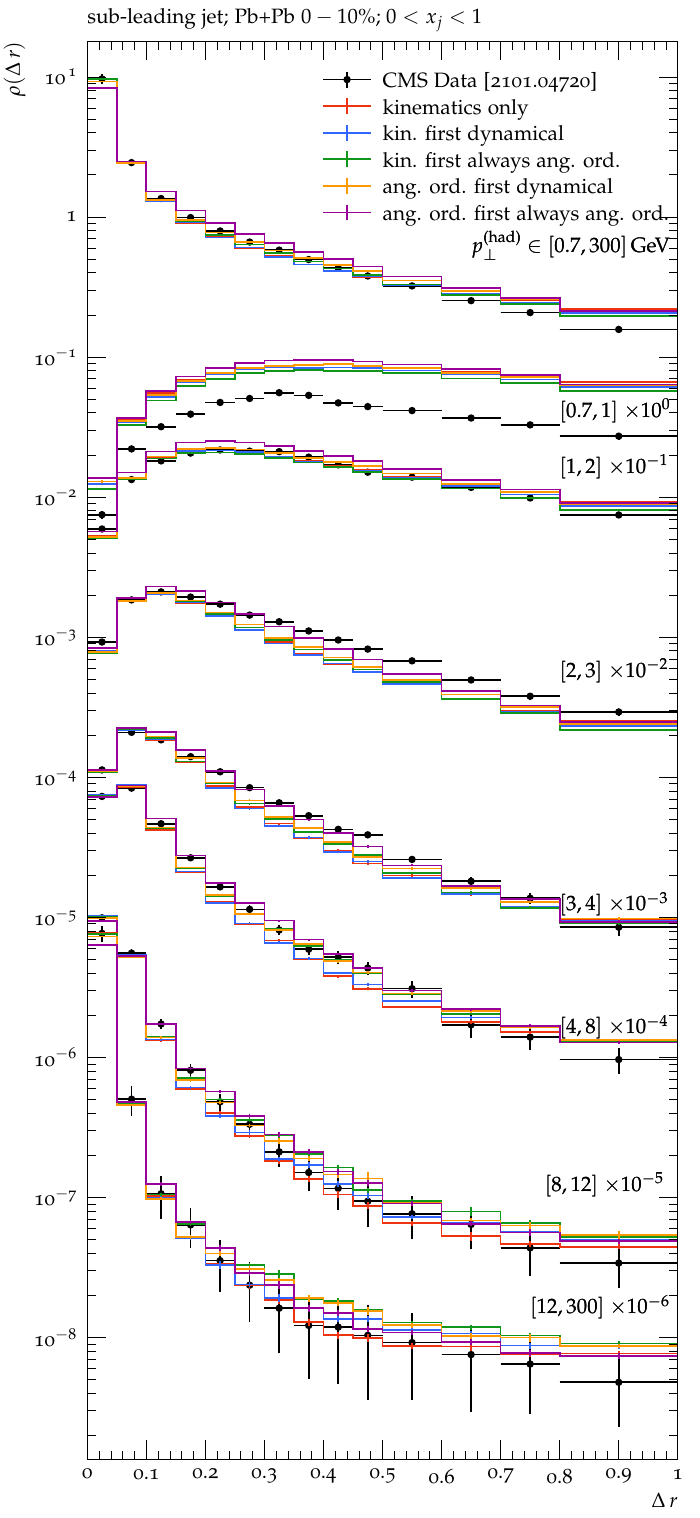}
	\includegraphics[width=0.5\textwidth]{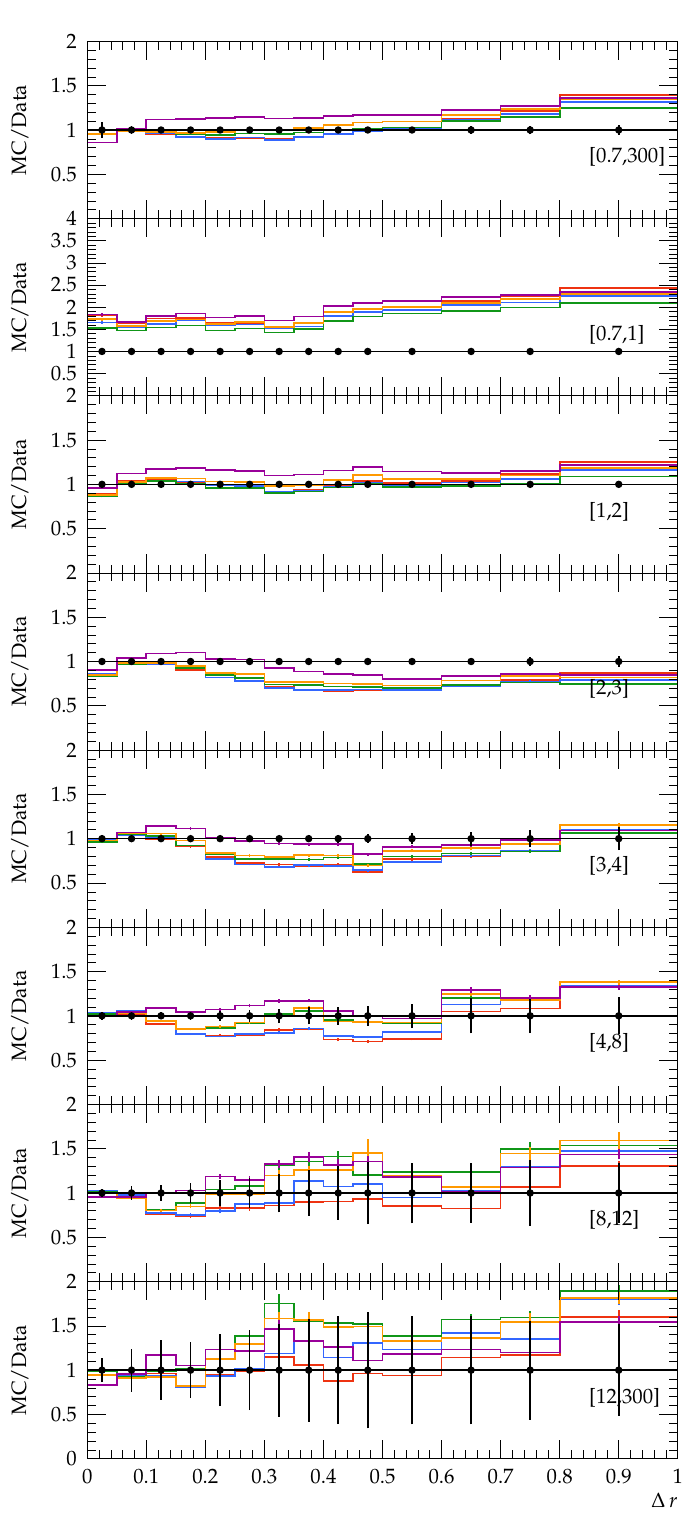}
	\caption{\textsc{Jewel}\,2.6.0+\textsc{Pythia} results with different angular ordering options compared to jet--hadron correlations in $0-10\%$ central Pb+Pb collisions at $\sqrt{s_{NN}} = \unit[5.02]{TeV}$ measured by \textsc{Cms}~\cite{CMS:2021nhn} for $R=0.4$ jets in bins of hadron transverse momentum for $0 < x_\text{j} < 1$. Shown here are the correlations for the sub-leading jet in di-jet events. The leading jet is required to have $p_\perp^\text{(jet)} > \unit[120]{GeV}$ and the sub-leading jet $p_\perp^\text{(jet)} > \unit[50]{GeV}$, the azimuthal angle between the two jets has to satisfy $\Delta \phi_{jj} > 5\pi/6$ and both jets have to be within $|\eta^\text{(jet)}| < 1.6$.}
	\label{fig:rho_subl_PbPbm01234}
\end{figure}

\medskip

The jet spectrum in p+p collisions shown in \figref{fig:RAA_PbPbm01234} is almost the same with all versions of angular ordering. The differences in the nuclear modification factor for central Pb+Pb collisions thus arise from differences in the Pb+Pb spectra. As can be seen in \figref{fig:RAA_PbPbm01234} the nuclear modification factor is significantly different for the \textit{kinematics first} and \textit{angular ordering first} options even when angular ordering is always required also for jet evolution in medium. A number of factors are at play here. For one thing, the different angular ordering options yield different numbers of splittings already in vacuum, as can be seen from Table~\ref{tab:nsplit_incoh}. The same is true for the two cases where angular ordering is always required also in Pb+Pb collisions. Without colour coherence each parton interacts and loses energy independently. Therefore, the overall amount of energy loss is strongly correlated with the number of partons. Consequently, the \textit{angular ordering first} option with its softer fragmentation is more strongly suppressed.  Table~\ref{tab:nsplit_incoh} also lists the fractions of splittings that are required to satisfy angular ordering. Two observations may be surprising at first sight: the fraction does not go to \unit[100]{\%} when angular ordering is always required, and the dynamical version of the \textit{angular ordering first} option has a higher fraction than the dynamical \textit{kinematics first} options. The reason for the former is that medium induced emissions, which here refers to the first emission after the parton shower has been reset to a higher scale and possible initial state emissions associated with this, are never angular ordered. The second point may be surprising because the condition that angular ordering is not required when a splitting happens inside the medium looks like an overestimate of medium effects. But this is not the case when the temperature is increasing, and most splittings happen before the initial time $\tau_i$ up to which the temperature is increasing in the simplified medium model. It is then not unlikely that a parton $a$ splits in vacuum, but before the daughters split the temperature comes above $T_\text{c}$. In this case, in the \textit{angular ordering first} option angular ordering is enforced for the splittings of the daughters, while in the \textit{kinematics first} option it will not be required if the daughters experienced a scattering before splitting. The lower fraction of splittings that have to obey angular ordering in the \textit{kinematics first} option thus partly explains why the number of splittings in central Pb+Pb collisions happens to be the same for the dynamical \textit{kinematics first} and \textit{angular ordering first} options.

\smallskip

Because the exact angle yields similar results as the approximated one from now on only results obtained with the (simpler) approximate angle will be shown. 
The jet fragmentation functions in central Pb+Pb collisions are shown in \figref{fig:FF_PbPbm01234}, and the jet--hadron correlations in \figref{fig:rho_lead_PbPbm01234} for the leading in \figref{fig:rho_subl_PbPbm01234} for the sub-leading jet (ratios of jet fragmentation functions and jet--hadron correlations can be found in \appref{sec:ratios}). 
As in p+p collisions, the \textit{kinematics first} option agrees generally better with the data, particularly for the jet fragmentation functions. The \textit{angular ordering first} again leads to a clearly too soft fragmentation when angular ordering is always enforced. With dynamical angular ordering the results are similar to the \textit{kinematics first} option. For the latter no large differences between the cases with dynamical and enforced angular ordering are observed. Only the correlations for the hardest particles show a significant difference. This could to some extent be a coincidence given that the overall suppression is larger without angular ordering, which implies larger quenching effects, but also a stronger survivor bias. One reason for the smaller differences between the angular ordering options is also the comparatively low percentage of splittings affected by angular ordering (cf.~table~\ref{tab:nsplit_incoh}). Furthermore, the effects of angular ordering on the internal structure of jets are also washed out by energy loss, which also restricts the radiation phase space and smears out angular structures as well as energy sharing patterns.

\section{Implementing colour coherence for re-scattering in \textsc{Jewel}}
\label{sec:implementation}

According to the calculations of medium induced emissions off an antenna~\cite{Casalderrey-Solana:2011ule,Mehtar-Tani:2011hma,Mehtar-Tani:2011vlz,Mehtar-Tani:2011lic,Mehtar-Tani:2012mfa} a colour dipole interacts coherently as long as its transverse size is smaller than the resolution power of the background medium. Once its size has increased so much that the two partons are resolved separately both partons interact independently and angular ordering is lost. In \textsc{Jewel}, which simulates individual re-scatterings off medium constituents, it is possible to check for each re-scattering of a colour dipole whether or not it resolves the dipole structure, albeit with some approximations and simplifications that will be discussed below. 

Colour dipoles are produced by splitting processes in the parton shower\footnote{When a parton splits into two the daughters are colour connected and form a dipole. This is the dipole I am referring to. There is also a colour connection between the daughter(s) of the splitting and the splitting mother parton, but these dipoles are not relevant in this context.}. The dipole only exists until the first re-scattering that is hard enough to resolve the two partons, because this changes the colour of one of the partons. After that the two partons continue to interact independently. The first step in the colour coherence implementation is thus to find the first resolved scattering after a splitting. Due to limitations imposed by the code structure only the first resolved re-scattering of one of the partons is found. The possibility that the other parton could experience a resolved scattering earlier is thus disregarded. For each re-scattering it is then checked whether the inverse transverse momentum transfer is larger than the separation between the parton and its colour partner. If this is the case the resolved scattering is found, otherwise the re-scattering is rejected and the next one is found. As shown in~\cite{Pablos:2024muu} this procedure is justified even when the recoiling thermal parton is included in the consideration. It is possible that there is no resolved scattering before the next splitting. In this case the splitting has to follow angular ordering, while a splitting after a resolved re-scattering is not angular ordered. 

During the time until the first resolved re-scattering or the first splitting of a parton the dipole can scatter coherently. To simulate this the splitting that produced the dipole is undone, i.e. the two partons are combined into one that is effectively the mother of that splitting. This effective parton is now allowed to re-scatter subject to the constraint that the re-scatterings have to be soft enough not to resolve the two partons. This is again achieved by rejecting re-scatterings with too high transverse momentum transfer. Therefore, as the size of the dipole increases the cross section for resolved re-scattering increases while the cross section for coherent re-scattering decreases. There is thus a smooth transition from a coherent regime with large effects from colour coherence to a decoherent regime where colour coherence effects are small. After the time available for coherent scattering the effective parton is split up again with the same energy sharing and azimuthal angle as the original splitting. Therefore, also re-scatterings that would make this re-splitting kinematically impossible are rejected. The deflections that the effective parton experiences due to re-scattering are passed on to the daughters' momenta and positions during the re-splitting. In this way coherent states consisting of two partons can be considered. In principle there should be coherent states consisting of larger number of partons, but these are not considered here because the larger size of these states implies that the colour coherence is less important. A further constraint imposed by practical limitations is that while resolved scatterings can induce additional radiation according to the standard \textsc{Jewel} procedure, coherent scatterings can only be elastic. In principle it is conceivable that a small dipole could have a coherent re-scattering that is both soft enough not to resolve the dipole and hard enough to induce additional radiation, but this is expected to be rare. An estimate obtained from the simulations shown in this paper is that the fraction of coherent dipoles that experience re-scatterings hard enough to induce radiation is roughly $5\cdot 10^{-4}$. However, the probability for hard coherent scattering is strongly dependent on the dipole size, and therefore the energy loss of very small dipoles is underestimated in the current implementation. 

\smallskip

\begin{figure}
	\centering
	\resizebox{0.45\linewidth}{!}{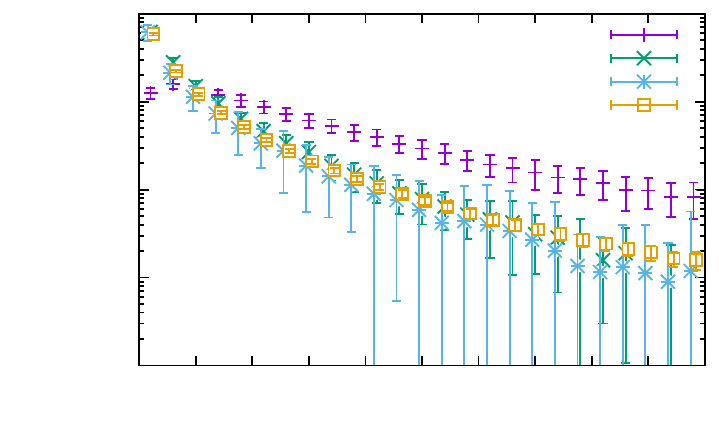}
	\resizebox{0.45\linewidth}{!}{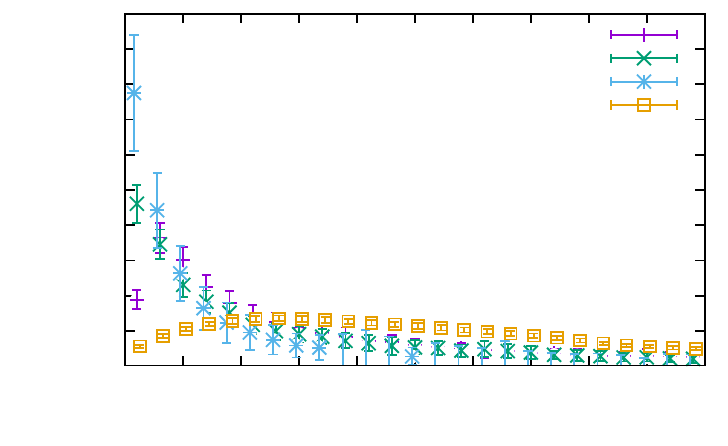}
	\caption{Four-momentum transfer $|\hat t|$ (left) and lab frame time (right) for resolved scatterings of partons that are a part of a colour dipole (purple), scatterings of partons in dipoles that got rejected because they did not resolve the colour charges (green), coherent scatterings of dipoles (blue), and all scatterings including scatterings of individual partons that are not part of a colour dipole (orange). No cuts have been imposed on the final state, i.e.\ all re-scatterings in all generated events are included. The phase space in which the events were generated is defined by the parameter settings given in  \appref{sec:jewelsettings}.}
\label{fig:thats}
\end{figure}

As detailed above, during the evolution \textsc{Jewel} considers scattering of dipoles, where colour coherence has to be respected, and individual partons, where there is no colour coherence. Dipoles are produced in splitting processes, and after a resolved re-scattering turn into two individual partons. To fully take into account the effects of colour coherence the simulation with colour coherence should be run with the dynamical \textit{kinematics first} option for angular ordering, such that splittings of partons that are part of dipoles have to follow angular ordering, while splittings of individual partons don't. This is thus the default choice in \textsc{Jewel}\,2.6.0 and all results presented in \secref{sec:results} and \appref{sec:moreplotsCC} have been generated using it. Colour coherence plays an important role in the re-scattering of dipoles. With the parameter settings used to generate the results shown in \secref{sec:results} (see \appref{sec:jewelsettings} for details) \unit[54]{\%} of the re-scatterings of partons forming dipoles were rejected because they did not resolve the dipole's structure. On the other hand, these are only \unit[5]{\%} of re-scatterings of all partons.\footnote{These numbers were extracted from all generated events, i.e.\ there are no requirements on the final state. The obtained values thus depend on the phase space in which events are generated.} Colour coherence thus affects only a small number of all partons, but for those that are affected it is an important effect. The left panel of \figref{fig:thats} shows the squared four-momentum transfers $\hat t$ for different types of re-scatterings. As expected, resolved re-scatterings of partons forming dipoles are significantly harder than the average represented by the sample of all re-scatterings (which are dominated by re-scatterings of individual partons). Re-scatterings that got rejected because they did not resolve the dipole have a $\hat t$ distribution similar to the average and coherent re-scatterings of dipoles are very slightly softer. Colour coherence is more important at early times when the colour charges are still close together and the parton shower evolves rapidly producing new dipoles, although the splitting angles in the parton shower tend to be larger at early times. This is clearly seen in the right panel of \figref{fig:thats} showing the time in the lab frame at which the different types of re-scatterings appear. Re-scattering of dipoles and partons forming dipoles happen predominantly within the first few \unit{fm/c} while the distribution of all re-scatterings is suppressed at early times and develops a broad maximum later. As expected, resolved re-scatterings are less likely to happen at very early times than coherent re-scattering. 

\begin{table}
	\centering
	\begin{tabular}{|l|c|c|c|}
	   \hline 
 	   & $\langle N_\text{split} \rangle$ & $\langle f_\text{AO} \rangle$ \ [\unit{\%}] & $\langle N_\text{scat}^\text{(per\ parton)} \rangle$  \\ \hline
	   p+p                                  & $15.03 \pm 0.03$  & 1 & 0 \\
	   no colour coherence                  & $26.93 \pm 0.05$ & $15.51 \pm 0.05$ & $20.90 \pm 0.02$ \\
	   only rejections                      & $23.89 \pm 0.04$ & $23.28 \pm 0.06$ & $17.62 \pm 0.02$ \\
	   rejections + coherent scattering     & $23.72 \pm 0.04$ & $23.23 \pm 0.06$ & $20.49 \pm 0.02$ \\
	   no col. coh. lower cross section     & $24.72 \pm 0.04$ & $20.44 \pm 0.06$ & $10.50 \pm 0.01$ \\
	   no col. coh., no induced radiation   & $19.82 \pm 0.04$ & $19.60 \pm 0.06$ & $17.71 \pm 0.01$ \\
	   with col. coh., no induced radiation & $18.39 \pm 0.03$ & $29.10 \pm 0.07$ & $17.91 \pm 0.02$ \\ \hline
    \end{tabular}
	\caption{Characteristics of \textsc{Jewel}'s final state parton shower in 0-10\% central Pb+Pb events in different set-ups with and without colour coherence. Included are events producing at least two jets with $p_\perp > \unit[100]{GeV}$ within $|\eta| < 2.8$. Shown are the mean number of splittings per event generated by \textsc{Jewel}'s final state parton shower (including initial state emissions associated with hard re-scatterings), average fraction of splittings that have to satisfy angular ordering (denoted $\langle f_\text{AO} \rangle$), and average number of scatterings per parton. The first splitting is included in the average number of splittings, while in the fraction of splittings required to satisfy angular ordering the first splitting is not included, because angular ordering for the first splitting is treated differently from subsequent splittings. The number of scatterings per partons is determined by tracing back each parton present at the end of the partonic evolution to either the matrix element or an emission from the initial state parton shower continuing with the parent parton when a splitting vertex is reached. The number of splittings in p+p collisions from Table~\ref{tab:nsplit_incoh} is shown for comparison.}
	\label{tab:nsplit_coh}
\end{table}

A similar picture is conveyed in Table~\ref{tab:nsplit_coh}, which shows the average number of splittings and scatterings per parton and the fraction of splittings that are required to satisfy angular ordering $\langle f_\text{AO} \rangle$ for different setups. As in~\cite{Roux:2024fpv} the number of scatterings is counted by tracing back the partons present at the end of the partonic evolution and counting scatterings along the way. When a splitting is reached the algorithm continues with the splitting parton until the hard scattering matrix element or an initial state parton is reached. In this way early scatterings are counted several times, because they affect all offsprings of the early parton. When turning on the rejection of unresolved scatterings the number of splittings drops because there are fewer induced emissions. At the same time $\langle f_\text{AO} \rangle$ increases because there is an increased probability that a parton does not scatter between two splittings. Adding coherent scatterings does not affect the number of splittings, as these can only be elastic and don't disrupt the colour coherence of a dipole, but the number of scatterings per parton increases again and ends up close to the number without colour coherence. Also shown are the values for a scenario without colour coherence where the scattering cross section has been artificially reduced by a factor 2 (see discussion below). The number of splittings is reduced also in this case because less scattering leads to fewer induced emissions, but the reduction is not as large as with colour coherence. Comparing the number of scatterings and $\langle f_\text{AO} \rangle$ it becomes evident that with colour coherence there are more scatterings per parton, but $\langle f_\text{AO} \rangle$ is higher because colour coherence reduces the scattering of some partons dramatically while leaving others unaffected. In contrast to this reducing the scattering cross section globally affects all partons in the same way, which results in a higher value of $\langle f_\text{AO} \rangle$ although the average number of scatterings per parton is lower. When induced emissions are turned off, colour coherence does not affect the number of splittings and scatterings as much, but leads to a clear increase in $\langle f_\text{AO} \rangle$. 

\begin{figure}
	\centering
	\resizebox{0.45\linewidth}{!}{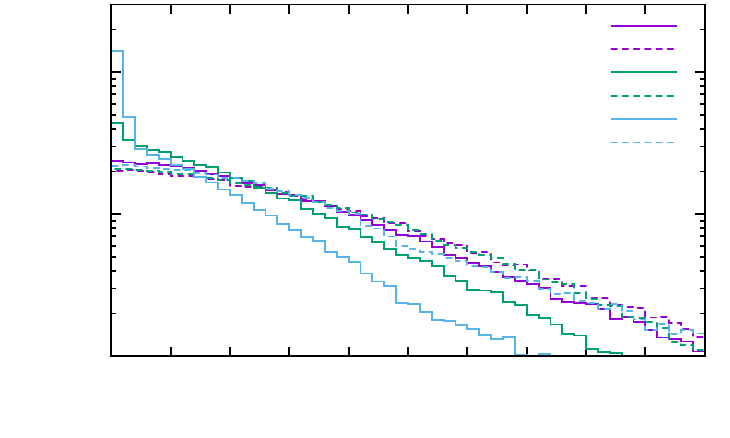}
	\resizebox{0.45\linewidth}{!}{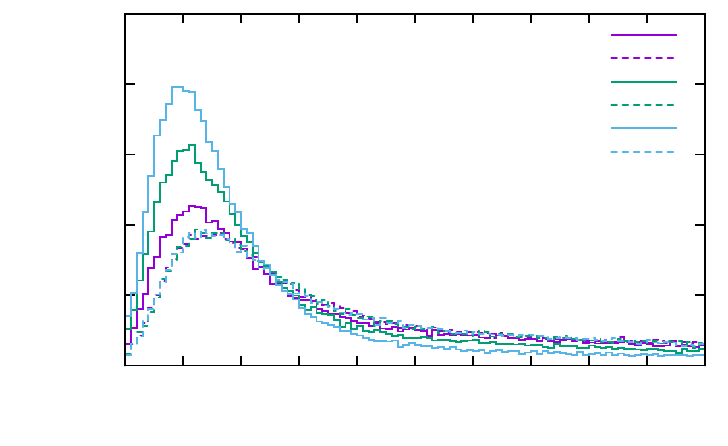}\\
	\resizebox{0.45\linewidth}{!}{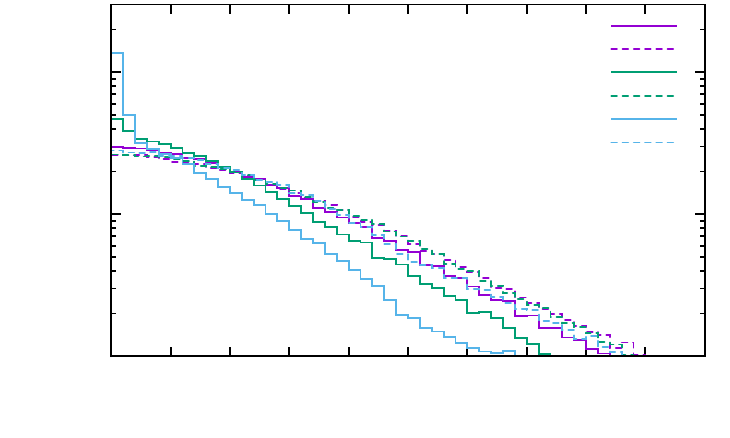}
	\resizebox{0.45\linewidth}{!}{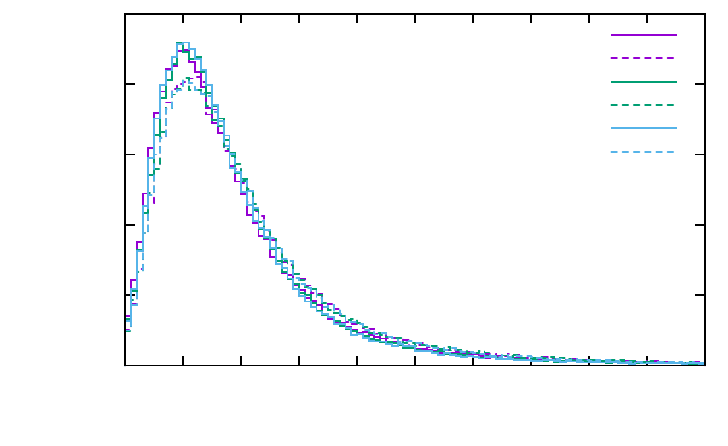}\\
	\caption{Change in opening angle $\theta$ (left) and energy loss (right) of a gluon dipole produced in the centre of a central Pb+Pb collision. In the top panels induced radiation in included, while in the bottom ones it is not. The legs of the dipole each have $\unit[100]{GeV}$ energy and the initial opening angle is $\theta_0$. The dipole legs are on-shell so that there is no vacuum radiation. The results are shown without errorbars for better readability, the statistical uncertainty can be inferred from the fluctuations.}
	\label{fig:dipole}
\end{figure}

To further illustrate the effects of colour coherence for a colour dipole \figref{fig:dipole} shows results from a simplified setup consisting of a single dipole made up of two gluons in a colour octet state and with \unit[100]{GeV} energy each. The gluons are on-shell, so that there is no radiation associated with the production process (commonly referred to as vacuum radiation). The dipole is produced at $t=0$ in the center of a head-on ($b=0$) Pb+Pb collision and at midrapidity ($y=0$). The left panels of \figref{fig:dipole} show the change in opening angle $\theta$ of the dipole due to medium interactions for different initial opening angles $\theta_0$. In the top panels induced radiation is included, while in the bottom panels there is no induced radiation and therefore all scatterings are purely elastic. Induced radiation adds a modest amount of broadening, but leaves the qualitative features of the $\Delta \theta$ distribution unchanged. As expected, when colour coherence is ignored (incoherent scenario) the change in opening angle $\Delta \theta$ is independent of the initial opening angle (except for very small differences coming from the fact that $\theta$ cannot become negative). When colour coherence is taken into account the distribution becomes increasingly peaked at zero as $\theta_0$ is reduced, because the dipole stays coherent for some time. During the coherence time the opening angle doesn't change. For the largest initial opening angle considered here ($\theta_0 = 0.5$), the $\Delta \theta$ distribution resembles the incoherent case, because when the opening angle is large the dipole loses coherence very quickly. The right panels of \figref{fig:dipole} show the energy loss of the dipole. Again, in the incoherent case the energy loss is independent of $\theta_0$, while in the coherent case the average energy loss becomes smaller. For purely elastic energy loss shown in the bottom right panel the energy loss is reduced only slightly due to colour coherence. The coherent dipole has a larger momentum and therefore a larger cross section, which partially cancels against the smaller colur charge. When induced radiation is included (top right panel) the energy loss in the coherent case depends on the initial opening angle of the dipole. This is largely an artifact of the restriction that coherent scatterings can only be elastic. The effect looks worryingly large in this example, but I would like to stress that kinematics in this toy setup is extreme. It is designed to showcase the effects of colour coherence, which requires the dipole to stay coherent for a long time, and is by no means representative of a realistic parton shower configuration. In particular, the importance of medium induced radiation is exaggerated here since the presence of vacuum radiation suppresses medium induced radiation.

The conclusion so far is thus that turning colour coherence on does not affect the number of scatterings per parton much, but it has important consequences for the radiation since it suppresses both medium induced and hard vacuum-like emissions. The former is probably to some extent model dependent, since, as discussed earlier, induced emissions off a coherent dipole are conceivable but cannot be included in \textsc{Jewel} in its current form. The latter, however, is a consequence of restored angular ordering that is consistent with what is seen in the analytical calculations and not a feature of this specific implementation.

\smallskip

The implementation of colour coherence in \textsc{Jewel} is in some aspects similar to the one in the Hybrid Model. In both cases partons don't interact independently after a splitting until they are resolved as individual objects by the background medium. During the coherence time the two partons instead interact coherently, i.e.\ as if the splitting had not yet taken place. One difference is that \textsc{Jewel} constructs coherent states of at most two partons, while in the Hybrid Model a larger number of partons can stay coherent. Other differences are founded in the different physical pictures of medium modifications of jets. As the Hybrid Model (without Moli\`ere scattering, which is the version considered here) assumes a strong coupling scenario the partons originating from a hard scattering do not undergo individual re-scatterings and the medium resolution scale is defined solely by the temperature. In \textsc{Jewel}, on the other hand, hard partons experience individual scattering processes and the resolution power is determined for each re-scattering individually from the momentum transfer. The medium temperature sets the screening mass and thus steers how hard re-scatterings are on average. The effects of colour coherence on the radiation pattern, which are here found to play a dominant role, are absent in the Hybrid Model. At strong coupling there is no distinction between elastic and radiative energy loss, therefore the Hybrid Model first runs the parton shower as a standard vacuum parton shower before the medium modifications are added. The effect of restoration of angular ordering, which in \textsc{Jewel} is treated dynamically on a splitting-by-splitting basis, therefore does not exist in the Hybrid Model. These differences in the consequences of colour coherence explain to a large extent why the two conceptually similar implementations of colour coherence have widely different phenomenological implications.

\section{Effects of colour coherence in jet quenching}
\label{sec:results}

\begin{figure}
	\includegraphics[width=0.5\textwidth]{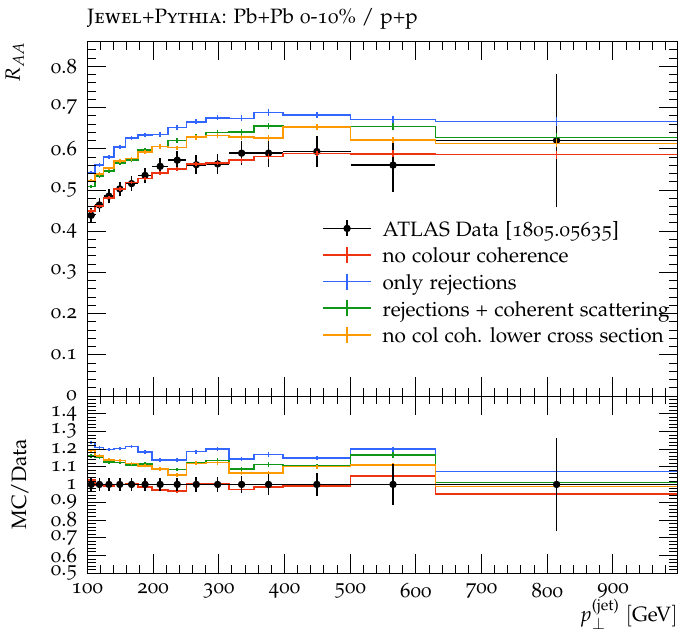}
	\includegraphics[width=0.5\textwidth]{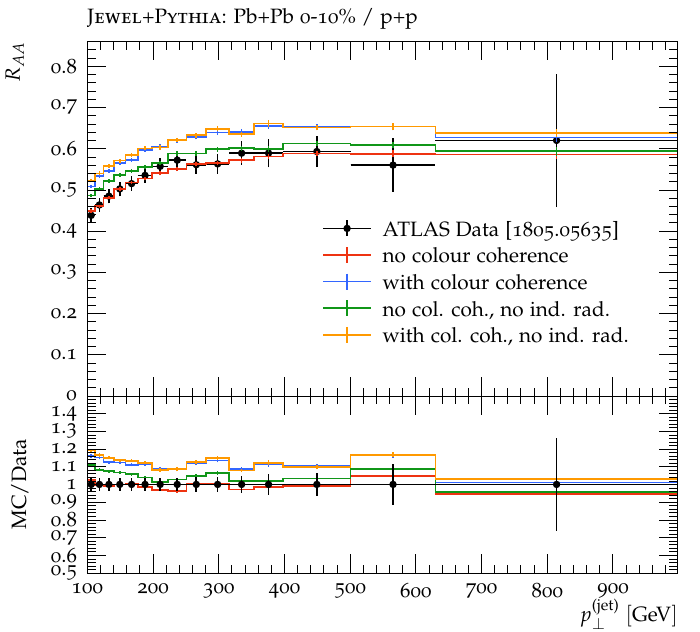}\\
	\caption{Nuclear modification factor $R_\text{AA}$ for $0-10\%$ Pb+Pb collisions at $\sqrt{s_{NN}} = \unit[5.02]{TeV}$ measured by \textsc{Atlas}~\cite{ATLAS:2018gwx} for $R=0.4$ jets within $|\eta^\text{(jet)}| < 2.8$ compared to \textsc{Jewel}\,2.6.0+\textsc{Pythia} with different colour coherence settings. \textbf{Left:} Without colour coherence (red), with only rejections of unresolved scatterings (blue), and with rejection of unresolved scatterings and coherent scatterings of dipoles (green). Also shown for comparison is the case without colour coherence where the scattering cross section has been artificially reduced by a factor of two (orange).  \textbf{Right:} Without (red) and with colour coherence (rejection of unresolved scatterings and coherent scatterings of dipoles, shown in blue), and without (green) and with colour coherence (orange) where medium induced emissions have been disabled.}
	\label{fig:RAA_PbPbCC}
\end{figure}

This section focuses on consequences of colour coherence that are visible in the hadronic final state of heavy ion collisions. \Figref{fig:RAA_PbPbCC} shows the nuclear modification factor in central Pb+Pb collisions. Rejecting unresolved scatterings (shown in blue) leads to a drastic increase of $R_\text{AA}$, for which to a large extent the suppression of splittings is responsible. Adding coherent scattering on top (green histogram) leads to a moderate reduction, which is now due to elastic energy loss. I would like to stress that no attempt at tuning the model to data was made, the point here is to demonstrate the effects of colour coherence when nothing else changes. Shown in orange for comparison is the scenario without colour coherence where the scattering cross section is artificially reduced by a factor of two, which yields a similar $R_\text{AA}$ as the colour coherence. This illustrates how effective colour coherence is: by affecting a comparatively small fraction of the scatterings it changes the splitting pattern in a way that corresponds to halving the scattering cross section when it comes to the overall suppression of hard jets. 

Also shown in \figref{fig:RAA_PbPbCC} is the scenario without induced emissions, i.e.\ purely elastic energy loss, with and without colour coherence (shown in green and orange, respectively, in the right panel of \figref{fig:RAA_PbPbCC}). The effect of colour coherence is here largely due to the suppression of splittings due to restored angular ordering (cf. table~\ref{tab:nsplit_coh}), because the number of scatterings per parton is very similar in both cases (this does not automatically imply the same elastic energy loss as the scattering kinematics is different, but still no large differences are expected). Interestingly, the coherent scenarios with and without induced emissions (shown in blue and orange, respectively) yield almost identical nuclear modification factors, although the number of splittings is more than \unit[20]{\%} higher in the simulation with induced radiation (cf.~table~\ref{tab:nsplit_coh}). At a first glance the similarity between the nuclear modification factors with and without induced radiation implies that elastic scattering is more efficient in transporting energy outside of the jet cone and that colour coherence further reduces the importance of induced radiation. At the same time, the number of splittings is significantly different both with and without colour coherence. The interpretation of these findings in thus not straightforward. While this interesting observation deserves further investigation, it requires a careful assessment of the non-trivial interplay between elastic and inelastic scattering. This is beyond the scope of the present work and will therefore be left for future investigation.

\smallskip

In addition to the jet suppression as function of transverse momentum \textsc{Atlas} has also measured the nuclear modification factor as function of the distance between two sub-jets satisfying the Soft Drop criterion~\cite{ATLAS:2022vii}. At first glance this looks like a promising observable for isolating colour coherence effects. Unfortunately, this turns out not to be the case. The main reasons are a formation time effect and several confounding factors that exacerbate the interpretation of the results. Plots and a more detailed discussion can be found in \appref{sec:rg}.

\smallskip

\begin{figure}
	\includegraphics[width=0.5\textwidth]{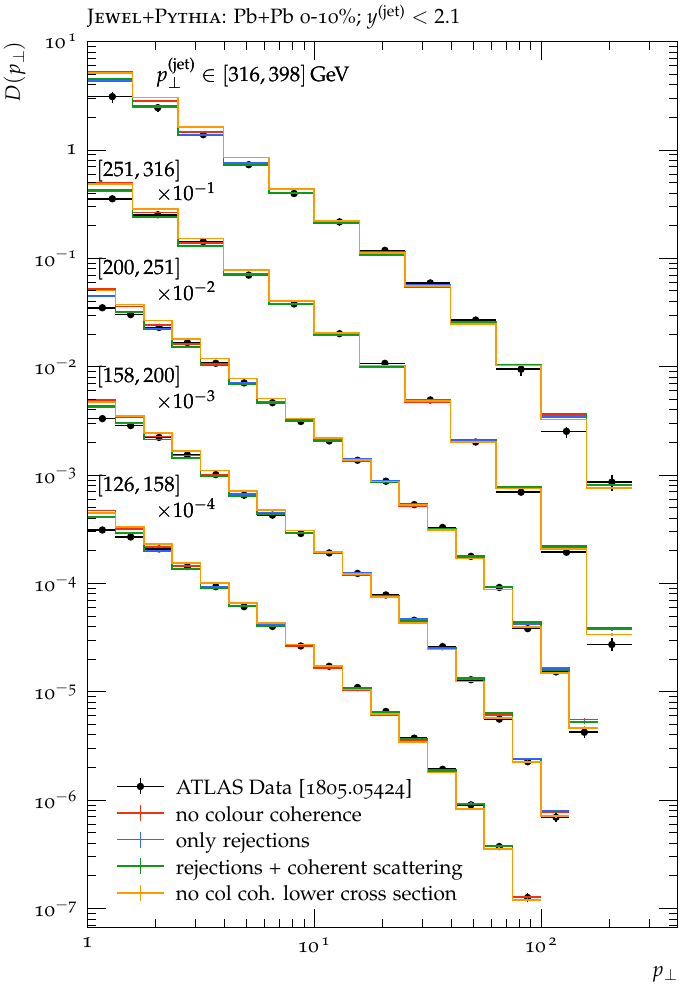}
	\includegraphics[width=0.5\textwidth]{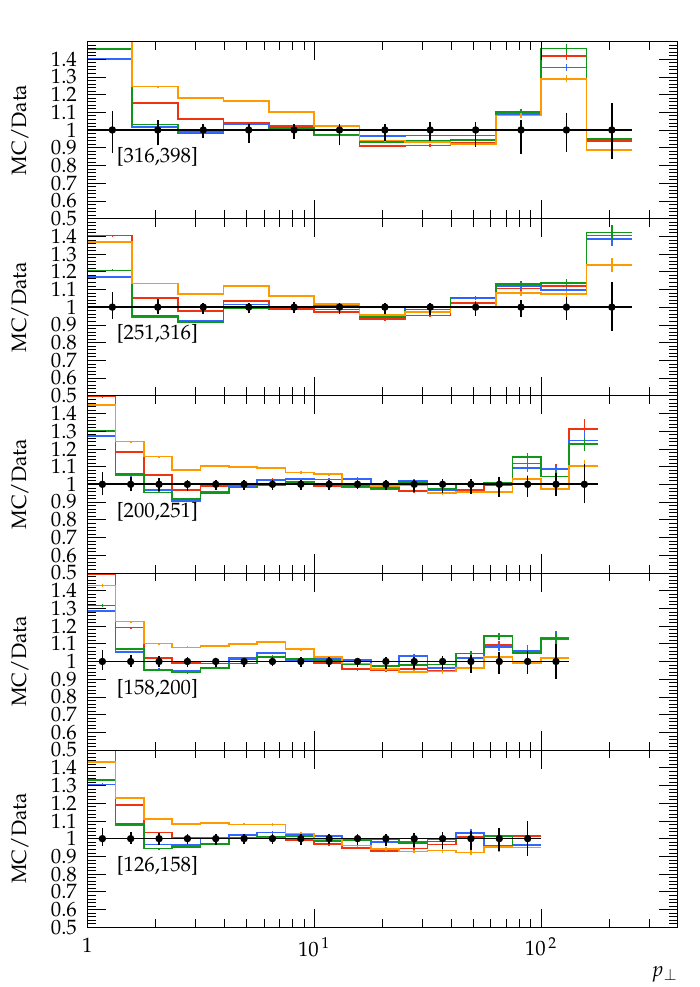}
	\caption{\textsc{Jewel}\,2.6.0+\textsc{Pythia} results with and without colour coherence compared to jet fragmentation functions in $0-10\%$ central Pb+Pb collisions at $\sqrt{s_{NN}} = \unit[5.02]{TeV}$ measured by \textsc{Atlas}~\cite{ATLAS:2018bvp} for $R=0.4$ jets in bins of jet transverse momentum. \textsc{Jewel}\,2.6.0+\textsc{Pythia} results are shown without colour coherence (red), with only rejections of unresolved scatterings (blue), and with rejection of unresolved scatterings and coherent scatterings of dipoles (green). Also shown for comparison is the case without colour coherence where the scattering cross section has been artificially reduced by a factor of two (orange).}
	\label{fig:FF_PbPbCC}
\end{figure}

The jet fragmentation functions (shown in \figref{fig:FF_PbPbCC}) reveal a reduction of soft particles inside the jets due to colour coherence. Medium response is an important source of soft particles. Colour coherence reduces the number of partons by suppressing splittings while leaving the number of scatterings per partons almost unchanged. The total number of scatterings thus goes down, which leads to a corresponding reduction in the medium response component. In addition, the reduced number of splittings translates into a harder fragmentation function, which is seen also in the form of a slight increase in the number of hard particles in \figref{fig:FF_PbPbCC}. One has to keep in mind that the overall suppression is not the same in the different scenarios and the survivor bias hardening the fragmentation function is stronger when the overall suppression is larger. What this means here is that for the scenario with colour coherence the survivor bias is less severe such that these jets can 'afford' a softer fragmentation without failing the $p_\perp$ cut. This counteracts the hardening of the fragmentation pattern.  Interestingly, the scenario without colour coherence and reduced scattering rate, on the other hand, has a softer jet fragmentation function than any of the other cases. This is not caused by medium response, as the total number of scatterings is lower than in the other scenarios, but is due to a softer fragmentation with more splittings than in the case with colour coherence (cf. table~\ref{tab:nsplit_coh}). The jet fragmentation functions thus distinguish between the scenarios with and without colour coherence even when $R_\text{AA}$ is the same, and a better agreement with data is obtained with colour coherence. For completeness, the ratios of the jet fragmentation functions in central Pb+Pb collisions to those in p+p collisions are shown in \appref{sec:ratios}. The jet fragmentation functions for the scenarios without medium induced emissions are shown in \appref{sec:moreplotsCC}  (and the ratios in \appref{sec:ratios}), but here the differences are generally smaller.

\smallskip

\begin{figure}
	\includegraphics[width=0.5\textwidth]{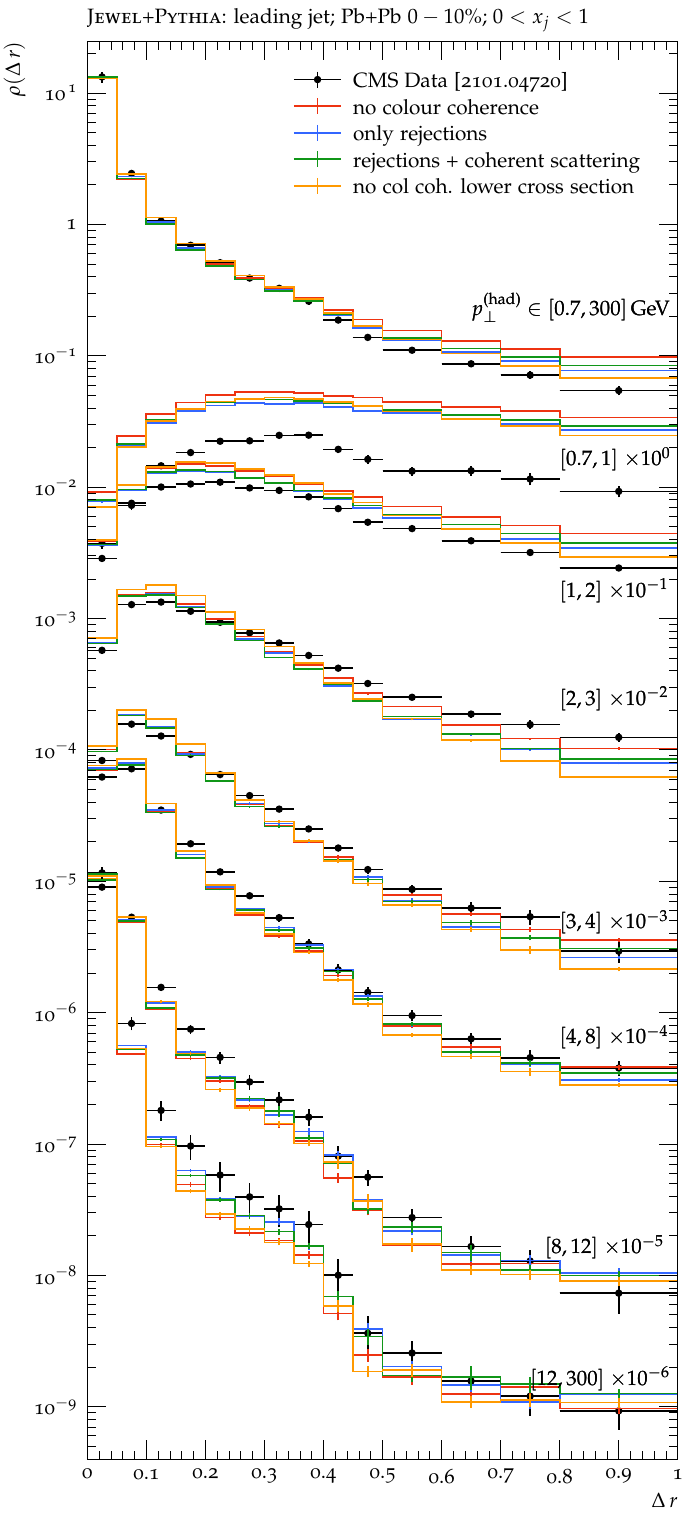}
	\includegraphics[width=0.5\textwidth]{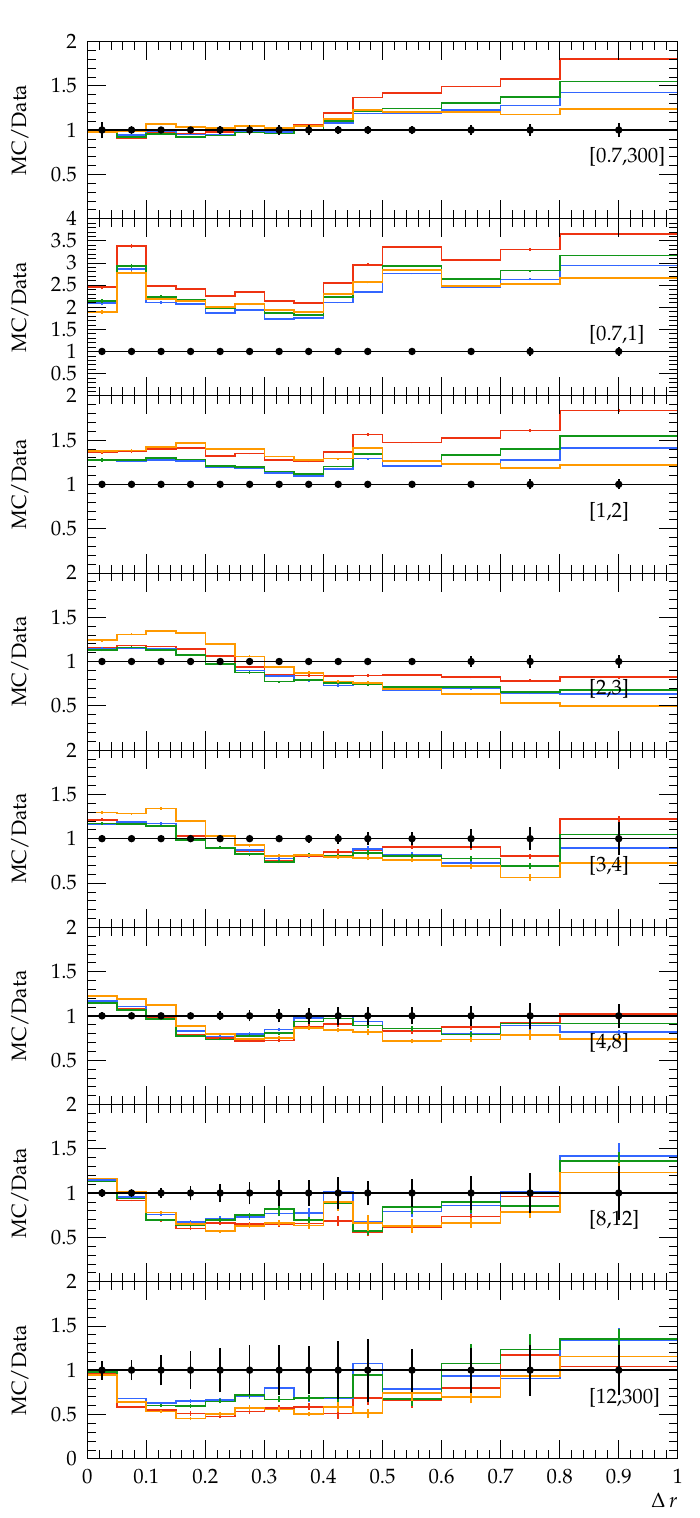}
	\caption{\textsc{Jewel}\,2.6.0+\textsc{Pythia} results with and without colour coherence compared to jet--hadron correlations in $0-10\%$ central Pb+Pb collisions at $\sqrt{s_{NN}} = \unit[5.02]{TeV}$ measured by \textsc{Cms}~\cite{CMS:2021nhn} for $R=0.4$ jets in bins of hadron transverse momentum for $0 < x_\text{j} < 1$. Shown here are the correlations for the leading jet in di-jet events.  The leading jet is required to have $p_\perp^\text{(jet)} > \unit[120]{GeV}$ and the sub-leading jet $p_\perp^\text{(jet)} > \unit[50]{GeV}$, the azimuthal angle between the two jets has to satisfy $\Delta \phi_{jj} > 5\pi/6$ and both jets have to be within $|\eta^\text{(jet)}| < 1.6$. \textsc{Jewel}\,2.6.0+\textsc{Pythia} results are shown without colour coherence (red), with only rejections of unresolved scatterings (blue), and with rejection of unresolved scatterings and coherent scatterings of dipoles (green). Also shown for comparison is the case without colour coherence where the scattering cross section has been artificially reduced by a factor of two (orange).}
	\label{fig:rho_lead_PbPbCC}
\end{figure}

\begin{figure}
	\includegraphics[width=0.5\textwidth]{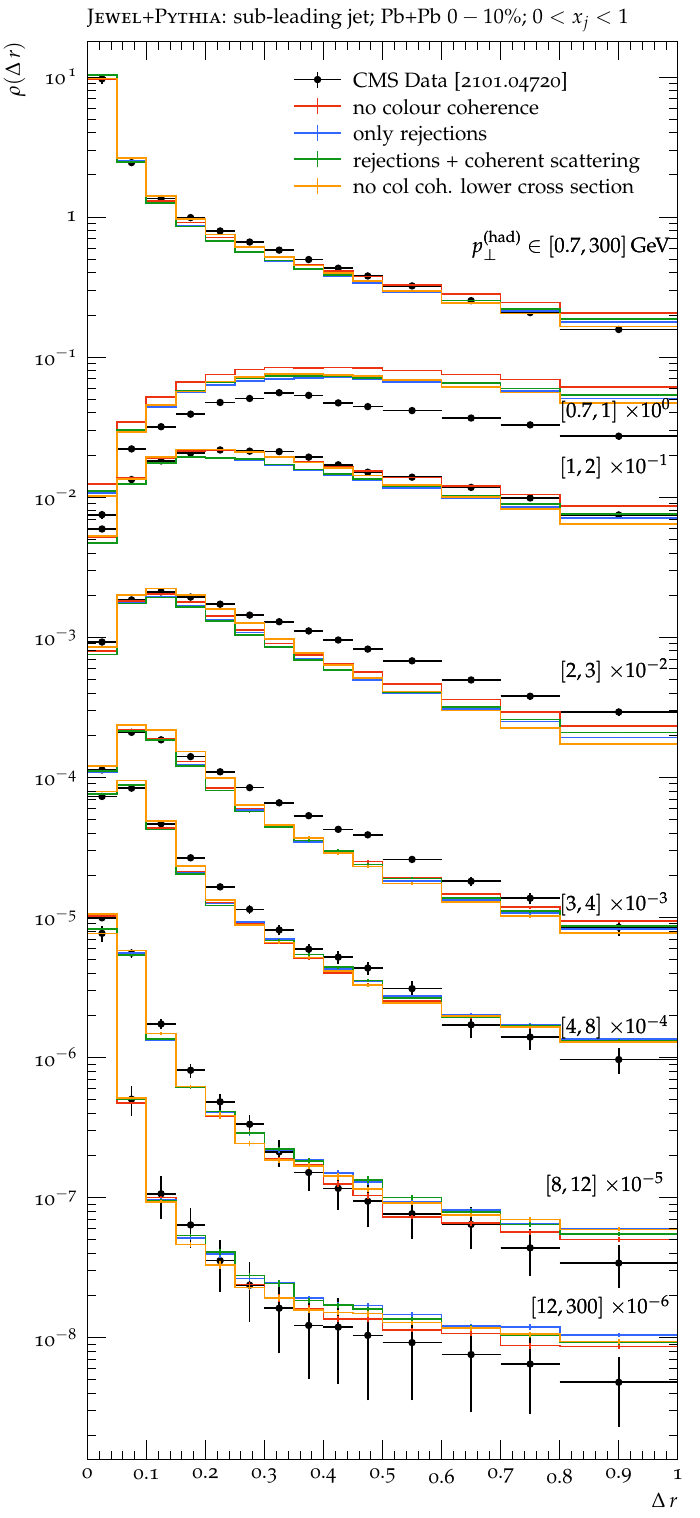}
	\includegraphics[width=0.5\textwidth]{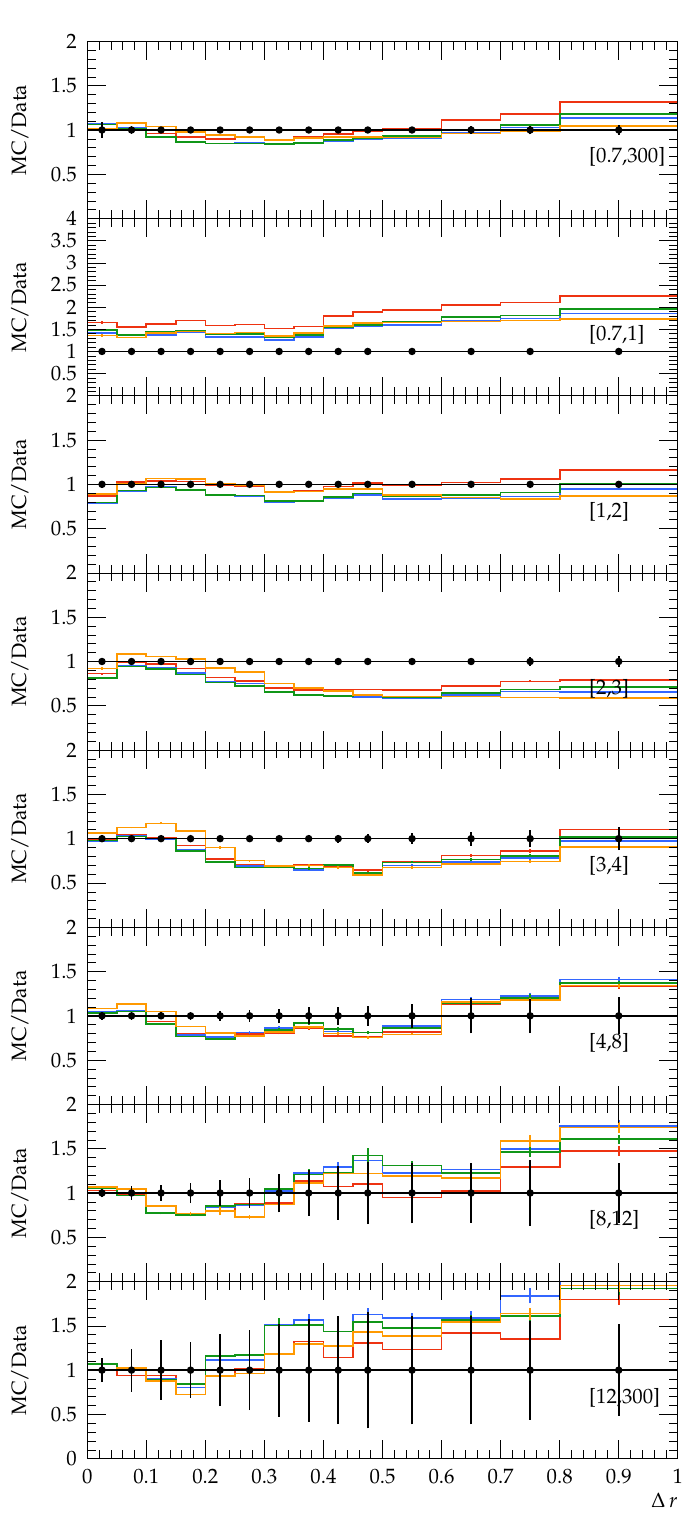}
	\caption{\textsc{Jewel}\,2.6.0+\textsc{Pythia} results with and without colour coherence compared to jet--hadron correlations in $0-10\%$ central Pb+Pb collisions at $\sqrt{s_{NN}} = \unit[5.02]{TeV}$ measured by \textsc{Cms}~\cite{CMS:2021nhn} for $R=0.4$ jets in bins of hadron transverse momentum for $0 < x_\text{j} < 1$. Shown here are the correlations for the sub-leading jet in di-jet events.  The leading jet is required to have $p_\perp^\text{(jet)} > \unit[120]{GeV}$ and the sub-leading jet $p_\perp^\text{(jet)} > \unit[50]{GeV}$, the azimuthal angle between the two jets has to satisfy $\Delta \phi_{jj} > 5\pi/6$ and both jets have to be within $|\eta^\text{(jet)}| < 1.6$. \textsc{Jewel}\,2.6.0+\textsc{Pythia} results are shown without colour coherence (red), with only rejections of unresolved scatterings (blue), and with rejection of unresolved scatterings and coherent scatterings of dipoles (green). Also shown for comparison is the case without colour coherence where the scattering cross section has been artificially reduced by a factor of two (orange).}
	\label{fig:rho_subl_PbPbCC}
\end{figure}

The reduction in medium response due to colour coherence is also visible in the jet--hadron correlations (shown in figures~\ref{fig:rho_lead_PbPbCC} and \ref{fig:rho_subl_PbPbCC}, ratios for the inclusive $p_\perp$-selection can be found in \appref{sec:ratios}) in the form of a reduction of particles at large $\Delta r$ and a general suppression in two lowest hadron $p_\perp$ bins. There is a slight enhancement in the highest two bins related to the hardening also seen in the fragmentation function. The scenario without colour coherence and reduced scattering rate tends to produce more particles close to the jet axis in the intermediate hadron $p_\perp$ range, which is in tension with the \textsc{Cms} data. The jet--hadron correlations for the scenarios without medium induced radiation can be found in \appref{sec:moreplotsCC} (and ratios in \appref{sec:ratios}) and are, again, rather similar to the case with colour coherence. 

\section{Conclusions}
\label{sec:conclusions}

Colour coherence is implemented in \textsc{Jewel} by requiring for each re-scattering of an individual parton that the inverse transverse momentum transfer is smaller than the (transverse) separation between the parton and its colour partner (in case the parton is part of a colour dipole, otherwise there is no restriction). For coherent scatterings of colour dipoles, on the other hand, it is required that the inverse transverse momentum transfer is larger than the transverse size of the dipole. The coherence of a dipole is lost as soon as one of the legs experienced a resolved scattering. In this way the temperature, which determines the screening mass and therefore how hard scatterings are on average, defines the average coherence length. However, there are large fluctuations due to the power-law behaviour of the scattering cross section, which are accounted for in this implementation. Colour coherence affects the radiation pattern even of hard, vacuum like emissions via the angular ordering requirement. In the new \textsc{Jewel} version angular ordering is handled in a much more flexible manner compared to earlier versions. This means that in parton showers evolving in a background medium angular ordering can be turned on or off for each individual splitting depending on whether the splitting parton or its colour partner experienced a resolved scattering. Coherent scatterings of a dipole don't lead to a loss of angular ordering. Presently, the main limitations of the implementation in \textsc{Jewel} are that coherent states cannot consist of more than two partons, and that coherent scatterings of dipoles can only be elastic. 

The general approach to only treat partons as individual objects after their separation is resolved by the medium and to let the coherent state interact until then is the same as in the implementation in the Hybrid Model~\cite{Hulcher:2017cpt}. One difference is due to the different views on the interactions of hard partons in the background medium, namely fluctuations in the momentum transfers are absent in the Hybrid Model due to the strong coupling approach. Other differences are due to the implementations of the model. Most notably, in \textsc{Jewel} coherent states can be composed of only two partons, while in the Hybrid Model they can consist of more than two partons. Also, the splitting pattern in the Hybrid Model is unaffected by medium effects in general and there is thus no loss and restoration of angular ordering.

\smallskip

The more flexible angular ordering implementation allows to compare different options for parton showers evolving in vacuum. In particular, there is freedom in the order in which angular ordering and kinematic constraints are imposed. The option of demanding kinematic constraints to be satisfied before imposing angular ordering yields better agreement with p+p data on jet fragmentation functions and jet--hadron correlations. However, since these choices are beyond the formal accuracy of the parton shower the differences between them should at present rather be regarded as a theoretical uncertainty. The \textit{kinematics first} option is the one that allows to turn angular ordering off after a resolved scattering and is therefore the new default choice. In older \textsc{Jewel} versions, however, \textit{angular ordering first} was the only available option.

\smallskip

Turning colour coherence on also for re-scattering leads to fewer resolved scatterings and therefore the fraction of splittings that have to satisfy angular ordering increases. This leads to a reduction of the number of vacuum-like splittings resulting in a harder fragmentation pattern. The number of medium induced emissions is also reduced. I would like to point out that it is in \textsc{Jewel} not possible to distinguish between vacuum like and medium induced emissions on the basis of individual splittings. The number of medium-induced emissions can only be estimated as an ensemble average by a comparing standard simulation to one where re-scatterings are not allowed to reset the parton shower.

The changes in the fragmentation pattern are to a large extent responsible for the effects of colour coherence observed in the hadronic final state. The number of scatterings per parton does not change significantly, because unresolved scatterings are replaced by coherent scattering. The dipole has a smaller colour charge than the average of the individual partons, but this is partly compensated by the higher momentum of the dipole which increases the scattering cross section. For the parton shower as a whole, however, the number of scatterings is reduced due to colour coherence, since the harder fragmentation produces fewer partons. The harder fragmentation and fewer medium induced splittings are to a large extent responsible for the sizable increase in the nuclear modification factor when colour coherence is turned on. The medium response component, which scales with the number of scatterings, is reduced. This is visible in the jet fragmentation function and jet--hadron correlations. 

\acknowledgments

I would like to thank Guilherme Milhano, Daniel Pablos and Krishna Rajagopal for valuable comments on he manuscript. 
This work is part of a project that has received funding from the European Research Council (ERC) under the European Union's Horizon 2020 research and innovation programme  (Grant agreement No. 803183, collectiveQCD).

\appendix
\section{Changes in \textsc{Jewel}\,2.6 compared to earlier versions}
\label{sec:jewelchanges}

There have been a number of small bug fixes and changes in the implementation (not the model) that are numerically irrelevant and are therefore not listed individually here. Most notably, a bug in the scattering kinematics was fixed, that stalled the event generation. A major restructuring in a central part of the code was necessary to accommodate the new angular ordering options. Again, this is not a change in the model, only in the way it is implemented, but it is a non-trivial check that the new version without the changes listed below reproduces the results obtained with earlier versions.

\subsection{Changes in the model}
\label{sec:modelchanges}

In addition to the new features, there are a few changes in details of the model. These changes were made because the new versions are physically better motivated than the old ones.

\begin{itemize}
	\item The flavour and fragmentation parameters in the hadronisation are taken from the Professor tune to \textsc{Lep} data~\cite{Buckley:2009bj} (instead of the default values).
	\item The energy sharing of the first final state splitting off an initial state emission is obtained from the \textsc{Pythia} parton shower. In older versions a new value is generated by the \textsc{Jewel} parton shower if finding virtual masses for the daughters fails. In version 2.6 there is a general check whether the \textsc{Pythia} value is compatible with the requirement in the \textsc{Jewel} parton shower that a splitting into massless daughters has to be kinematically possible, and a new value is generated when this is not the case. Since a splitting into massive daughters may be possible even when splitting into massless daughters is not possible, this is a stricter requirement.
	\item When a splitting in the \textsc{Jewel} parton shower fails due to kinematic constraints older versions reject the larger of the daughters' masses and generate a new one. Version\ 2.6 randomly picks the daughter mass that gets rejected in case both are non-zero. When one of the daughters is on-shell it is automatically the other one that gets rejected.
\end{itemize}

\begin{figure}[h!]
	\includegraphics[width=0.5\textwidth]{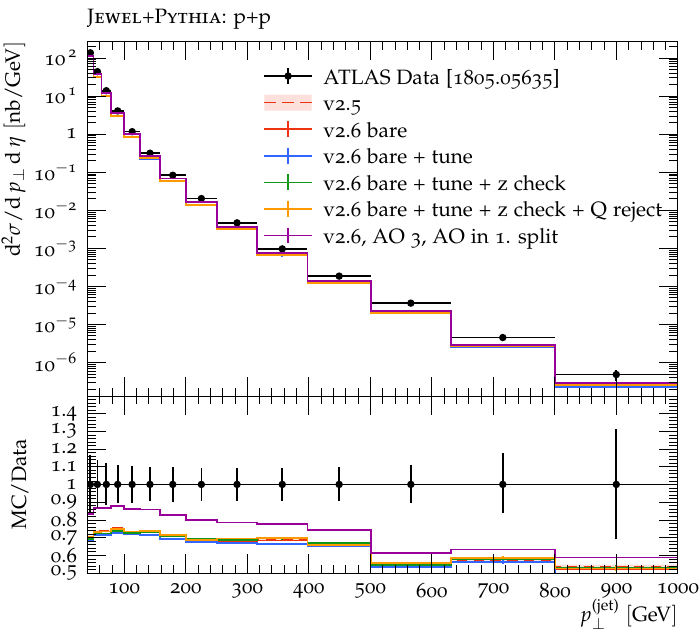}
	\includegraphics[width=0.5\textwidth]{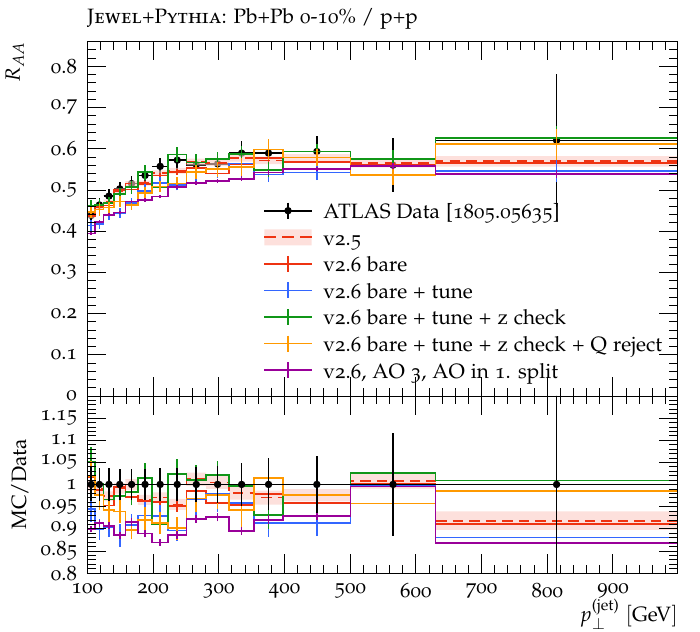}
	\caption{\textsc{Jewel}+\textsc{Pythia} results for the jet $p_\perp$-spectrum for $R=0.4$ jets within $|\eta^\text{(jet)}| < 2.8$ in p+p collisions (left) and the nuclear modification factor in central Pb+Pb collisions (right) compared to \textsc{ATLAS} data~\cite{ATLAS:2018gwx}, both at $\sqrt{s_{NN}} = \unit[5.02]{TeV}$. \textsc{Jewel}+\textsc{Pythia} results are shown for \textsc{Jewel}\,2.5 and different variants of \textsc{Jewel}\,2.6 where the changes described in section~\ref{sec:modelchanges} are successively added. 'v2.6 bare' denotes \textsc{Jewel}\,2.6 without any of the changes listed in section~\ref{sec:modelchanges}, i.e.\ this version corresponds to \textsc{Jewel}\,2.5 albeit with changes in how the model is implemented. 'v2.6 bare + tune' adds the fragmentation tune, 'v2.6 bare + tune + z check' adds the new treatment of the energy sharing in the first splitting off an initial state emission on top of the fragmentation tune. Finally, 'v2.6 bare + tune + z check + Q reject' adds the new treatment of failed splittings, this is thus \textsc{Jewel}\,2.6. In all the variants the angular ordering strategy (cf.~section~\ref{sec:newfeatures}) is \textit{angular ordering first} (\texttt{ANGORD} = 3 in \textsc{Jewel}\,2.6) corresponding to \textsc{Jewel}\,2.5, and there is no angular ordering for the first final state splitting (\texttt{AOFIRSTSPLIT} = false in \textsc{Jewel}\,2.6). In addition, the results for \textsc{Jewel}\,2.6 with the same angular ordering strategy but angular ordering in the first splitting enabled are shown (labeled 'v2.6, AO 3, AO in 1. split').}
	\label{fig:baseline_jetspec}
\end{figure}

\begin{figure}
	\includegraphics[width=0.5\textwidth]{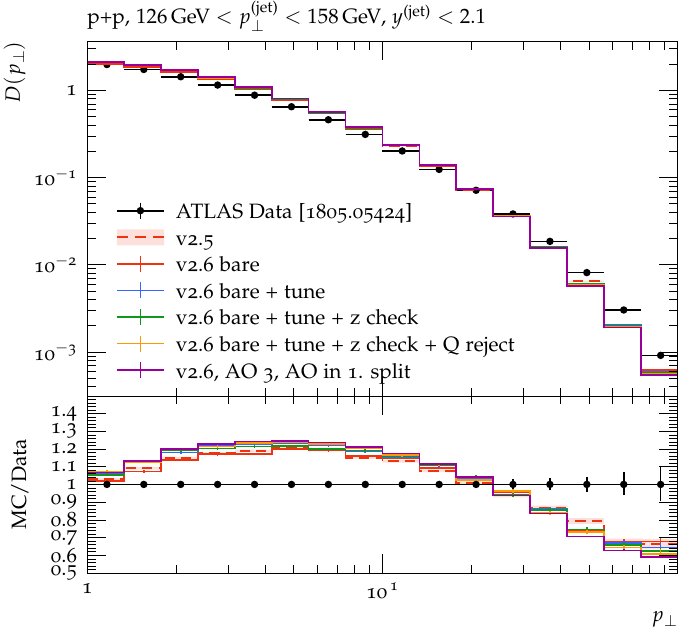}
	\includegraphics[width=0.5\textwidth]{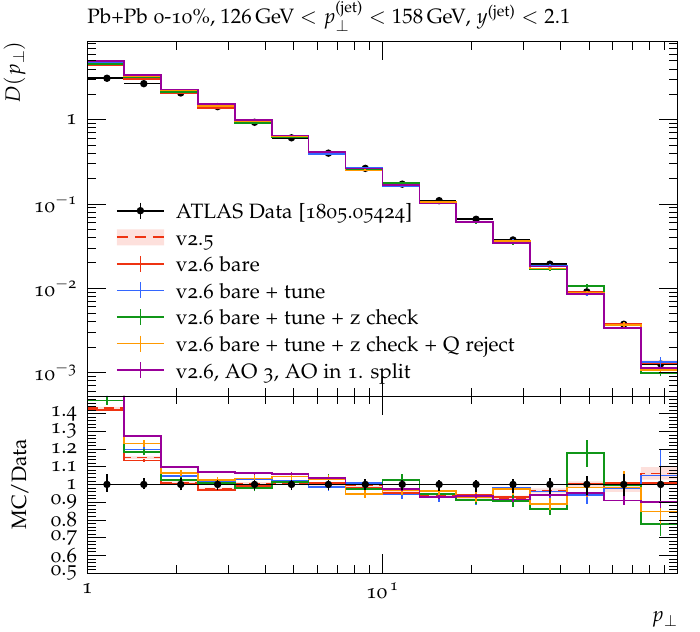}
	\caption{\textsc{Jewel}+\textsc{Pythia} results for the jet fragmentation function in p+p (left) and $0-10\%$ central Pb+Pb collisions (right) at $\sqrt{s_{NN}} = \unit[5.02]{TeV}$ compared to \textsc{Atlas} data~\cite{ATLAS:2018bvp} for $R=0.4$ jets in the jet $p_\perp$ bin from \unit[126]{GeV} to \unit[158]{GeV}. The \textsc{Jewel}+\textsc{Pythia} lines are for the same variants as in figure~\ref{fig:baseline_jetspec}.}
	\label{fig:basline_FF}
\end{figure}

\begin{figure}
	\includegraphics[width=0.5\textwidth]{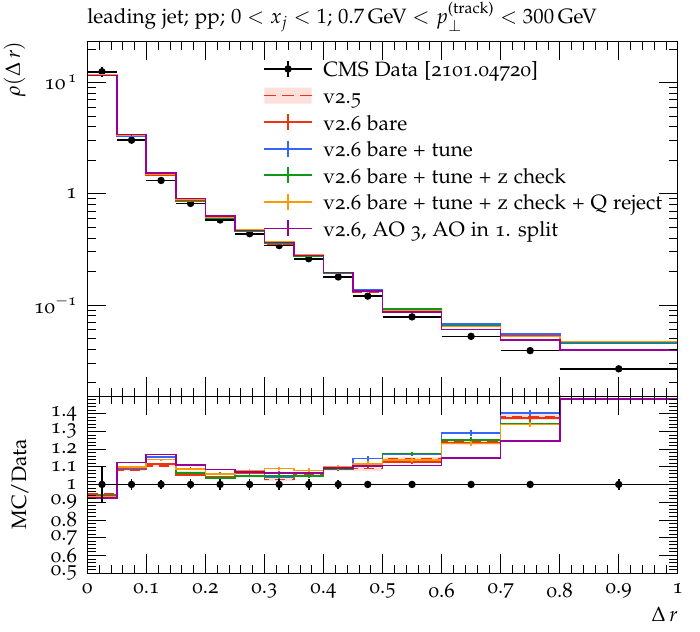}
	\includegraphics[width=0.5\textwidth]{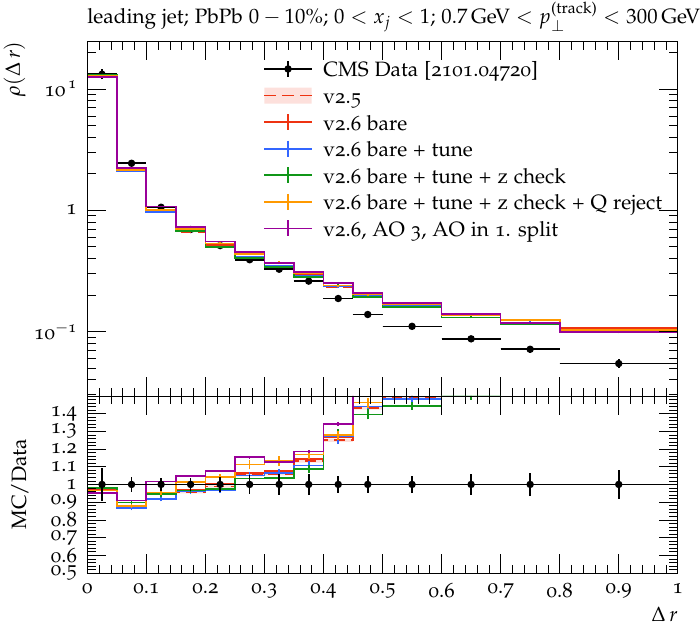}
	\caption{\textsc{Jewel}+\textsc{Pythia} results for the jet--hadron correlations for the leading jet in p+p (left) and $0-10\%$ central Pb+Pb collisions (right) at $\sqrt{s_{NN}} = \unit[5.02]{TeV}$ measured by \textsc{Cms}~\cite{CMS:2021nhn} for $R=0.4$ jets. The leading jet is required to have $p_\perp^\text{(jet)} > \unit[120]{GeV}$ and the sub-leading jet $p_\perp^\text{(jet)} > \unit[50]{GeV}$, the azimuthal angle between the two jets has to satisfy $\Delta \phi_{jj} > 5\pi/6$ and both jets have to be within $|\eta^\text{(jet)}| < 1.6$. The \textsc{Jewel}+\textsc{Pythia} lines are for the same variants as in figure~\ref{fig:baseline_jetspec}.}
	\label{fig:baseline_rholead}
\end{figure}

Figures~\ref{fig:baseline_jetspec}, \ref{fig:basline_FF} and \ref{fig:baseline_rholead} show examples of the effects of these changes on jet observables, namely the jet spectrum, jet fragmentation function, and jet--hadron correlations, respectively. The 'bare' variant of \textsc{Jewel}\,2.6, that contains smaller bug fixes and important changes in algorithms agrees well with \textsc{Jewel}\,2.5. The largest effects due to the model changes are found in the jet $p_\perp$-spectrum in central Pb+Pb collisions, where differences up to \unit[10]{\%} are observed.

\subsection{New features}
\label{sec:newfeatures}

This section briefly summarises the new features in \textsc{Jewel}\,2.6 and lists the corresponding parameters and their default values.

\begin{itemize}
	\item The main new feature is the colour coherence implementation. The corresponding parameters are
	\begin{itemize}
		\item \texttt{COHLENGTHFAC} (default is 1.), which multiplies the transverse distance between the colour and the anti-colour charge, 
		\item \texttt{PCOHREJ} (default is 1.), which is the probability with which an unresolved scattering gets rejected,
		\item \texttt{COHSCAT} (default is \texttt{true}), which is the switch for coherent scattering of unresolved colour dipoles.
	\end{itemize} 
	\item The options for angular ordering are now 
	\begin{itemize} 
		\item \texttt{ANGORD} = 0: angular ordering always disabled,
		\item \texttt{ANGORD} = 1: (default value) kinematic constraints are imposed before angular ordering constraints, when a scattering took place angular ordering is not required for the next splitting,
		\item \texttt{ANGORD} = 2: like option 1, but angular ordering is always required (even when there was a scattering),
		\item \texttt{ANGORD} = 3: angular ordering constraints are imposed before kinematic constraints, when a splitting takes place in a region with $T > T_c$ no angular ordering is required (this was the only option in previous \textsc{Jewel} versions),
		\item \texttt{ANGORD} = 4: like option 3, but angular ordering is always required,
		\item \texttt{ANGORD} = 5: like option 1, but uses the exact value of the splitting angle instead of the approximated one of \eqref{eq:splitangle}.
	\end{itemize}
	\item Angular ordering can be required for the first emission off the final state partons of the matrix element, the switch is \texttt{AOFIRSTSPLIT} (\texttt{false} by default).
	\item The scattering cross section can be artificially scaled by a factor, the corresponding parameter is \texttt{XSECSCALEFAC} (which equals 1. by default).
	\item \textsc{Jewel} can perform the subtraction of thermal momenta for the whole event using the Constituent Subtraction method of~\cite{Milhano:2022kzx} (boolean parameter \texttt{DOSUBTRACTION}, equals \texttt{false} by default). The parameter \texttt{RSUB} sets the maximal distance for a subtraction (default is 1.). The resulting four-momenta are off-shell and there are two methods for putting them on-shell: one of them (parameter \texttt{PUTONSHELL} = 2) does a global re-shuffling of energy and momentum while the other one (\texttt{PUTONSHELL} = 1) keeps the particles' momenta and adjusts the energies. Setting \texttt{PUTONSHELL} to 0, which is the default, leaves the four-momenta unchanged.
	\item A different value of the strong coupling constant $\alpha_s$ can be used in the parton shower and in scattering. The corresponding parameters are \texttt{LAMBDAQCD} (for scattering) and \texttt{LAMBDAQCDPS} (for the parton shower). The default value remains unchanged and is \unit[0.4]{Gev} for both.
\end{itemize}

As seen in figures~\ref{fig:baseline_jetspec}, \ref{fig:basline_FF}, and \ref{fig:baseline_rholead} the main effect of turning on angular ordering of the first splitting (as done throughout this paper) is a substantial increase of the jet $p_\perp$-spectrum, mostly at lower $p_\perp$. The nuclear modification factor, on the other hand, changes much less. It is worth noting that the purple lines in left panels of these figures are the same as the orange lines in figures~\ref{fig:FF_pp}, \ref{fig:rho_lead_pp}, \ref{fig:rho_subl_pp}, and left panel of \figref{fig:RAA_PbPbm01234}. The purple lines in the right panels here are the same as the orange lines in the right panel of \figref{fig:RAA_PbPbm01234} and in figures~\ref{fig:FF_PbPbm01234},  \ref{fig:rho_lead_PbPbm01234}, and \ref{fig:rho_subl_PbPbm01234}.
	
\FloatBarrier

\newpage

\section{\textsc{Jewel}\,2.6 parameter settings}
\label{sec:jewelsettings}

For the \textsc{Jewel} results shown in this paper the parameter settings for \textsc{Jewel} and the simple medium model are, where they depart from the default values, are given in the table below.

\begin{center}
\begin{tabular}{|l|c|}
	\hline
	\textbf{\textsc{Jewel}\,2.6 parameter} & \textbf{value(s)} \\
	\hline
	\texttt{PTMIN} & 3. \\
	\texttt{PTMAX} & 1200. \\
	\texttt{ETAMAX} & 4. \\
	\texttt{ISOCHANNEL} & PP \\
	\texttt{SQRTS} & 5020. \\
	\texttt{PDFSET} & 13100 (p+p) and 901300 (Pb+Pb) \\
	\texttt{WEXPO} & 4.5 \\
	\texttt{ANGORD} & 0, 1, 2 ,3 ,4 ,5 \\
	\texttt{AOFIRSTSPLIT} & T, F \\
	\texttt{KEEPRECOILS} & T \\
	\texttt{DOSUBTRACTION} & T \\
	\texttt{PUTONSHELL} & 1 \\
	\texttt{PCOHREJ} & 0., 1. \\
	\texttt{COHSCAT} & T, F \\
	\hline \hline
	\textbf{medium parameter} & \textbf{value} \\
	\hline
	\texttt{TAUI} & 0.4 \\
    \texttt{TI} & 0.59 \\
    \texttt{MDSCALEFAC} & 1. \\
    \hline
\end{tabular}
\end{center}
	
\FloatBarrier

\newpage

\section{Ratios of jet fragmentation functions and jet--hadron correlations in central Pb+Pb collisions}
\label{sec:ratios}

\begin{figure}[h!]
	\includegraphics[width=0.5\textwidth]{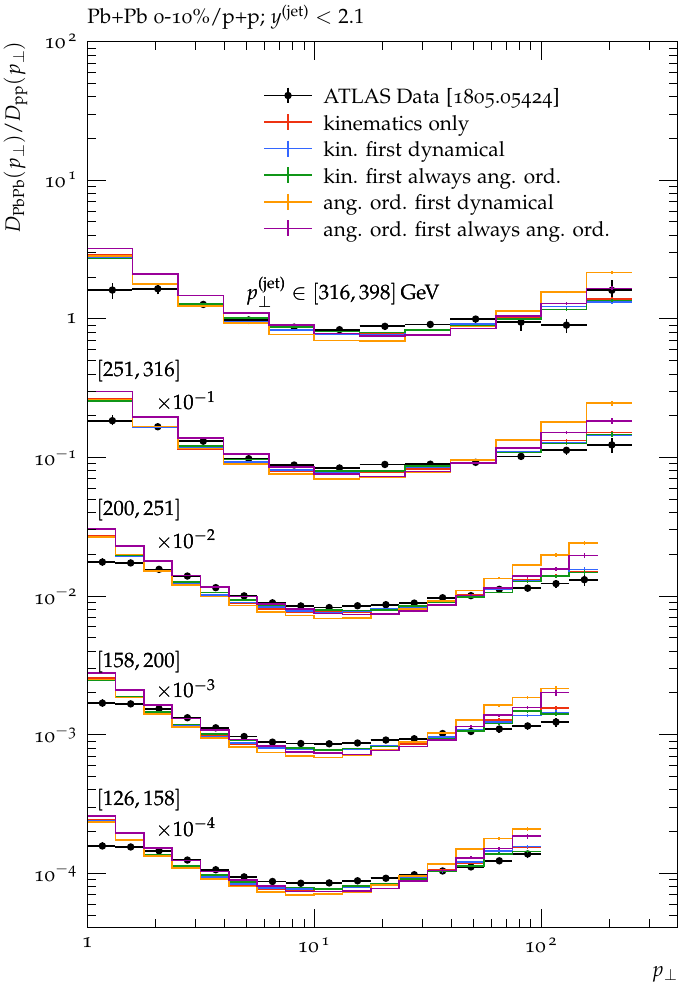}
	\includegraphics[width=0.5\textwidth]{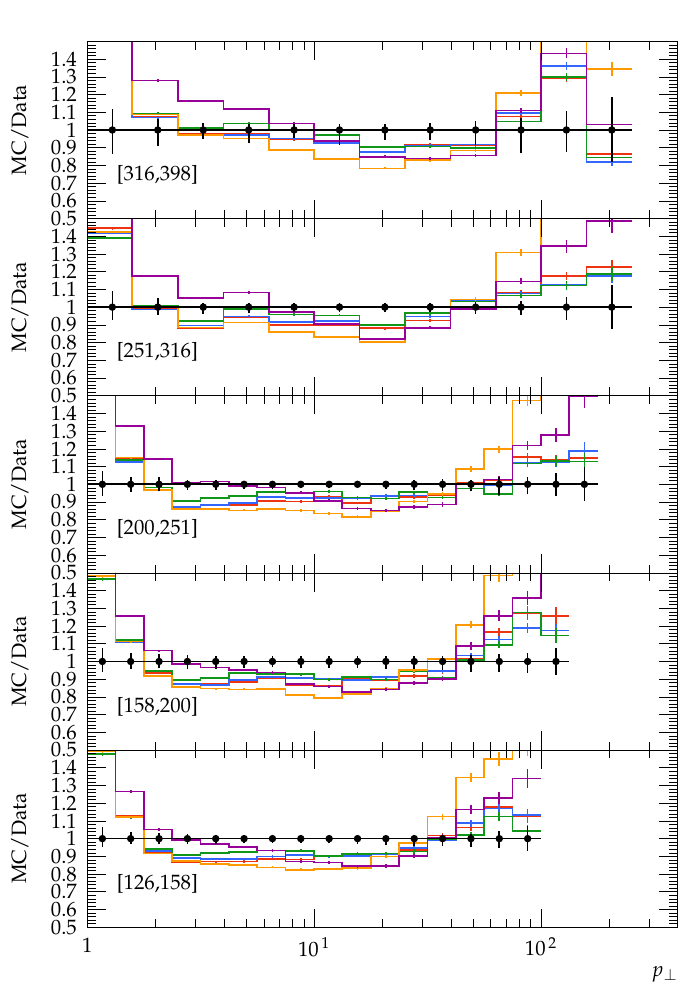}
	\caption{\textsc{Jewel}\,2.6.0+\textsc{Pythia} results with different angular ordering options (without colour coherence) compared to ratios of jet fragmentation functions in $0-10\%$ central Pb+Pb collisions at $\sqrt{s_{NN}} = \unit[5.02]{TeV}$ to p+p collisions measured by \textsc{Atlas}~\cite{ATLAS:2018bvp} for $R=0.4$ jets in bins of jet transverse momentum. }
	\label{fig:FFratios_PbPbm01234}
\end{figure}

\begin{figure}[h!]
	\includegraphics[width=0.5\textwidth]{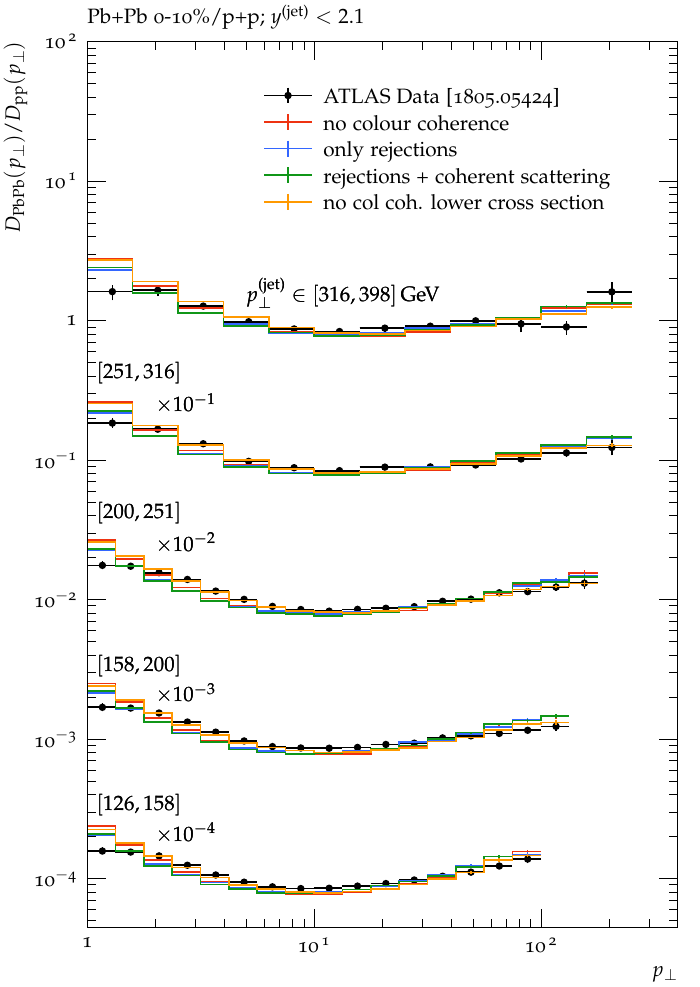}
	\includegraphics[width=0.5\textwidth]{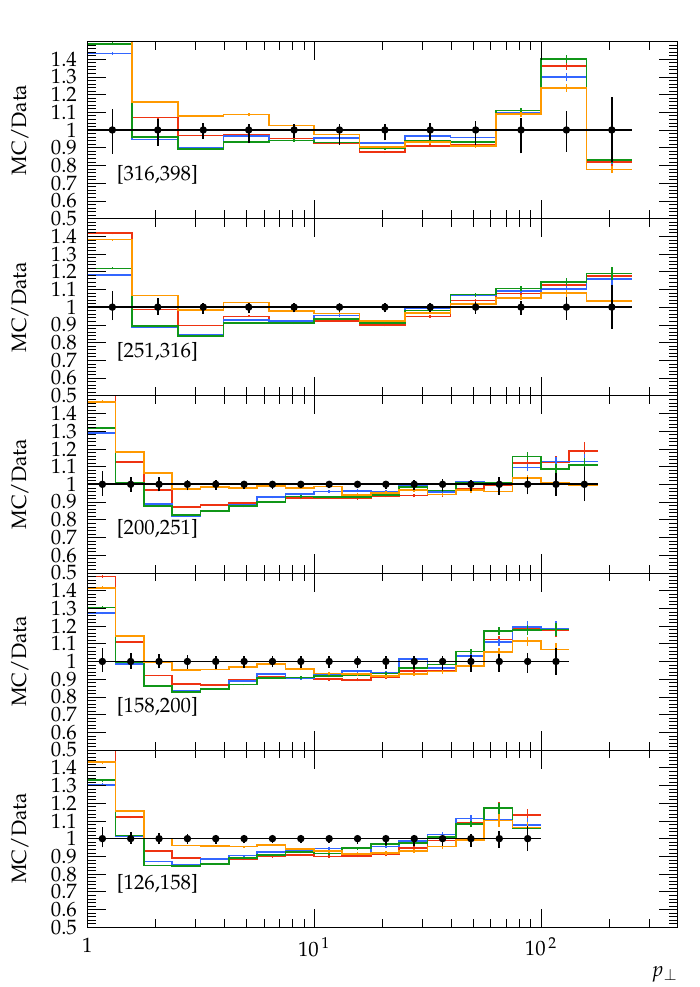}
	\caption{\textsc{Jewel}\,2.6.0+\textsc{Pythia} results with and without colour coherence compared to ratios of jet fragmentation functions in $0-10\%$ central Pb+Pb collisions at $\sqrt{s_{NN}} = \unit[5.02]{TeV}$ to p+p collisions measured by \textsc{Atlas}~\cite{ATLAS:2018bvp} for $R=0.4$ jets in bins of jet transverse momentum. \textsc{Jewel}\,2.6.0+\textsc{Pythia} results are shown without colour coherence (red), with only rejections of unresolved scatterings (blue), and with rejection of unresolved scatterings and coherent scatterings of dipoles (green). Also shown for comparison is the case without colour coherence where the scattering cross section has been artificially reduced by a factor of two (orange).}
	\label{fig:FFratios_PbPbCC}
\end{figure}

\begin{figure}[h!]
	\includegraphics[width=0.5\textwidth]{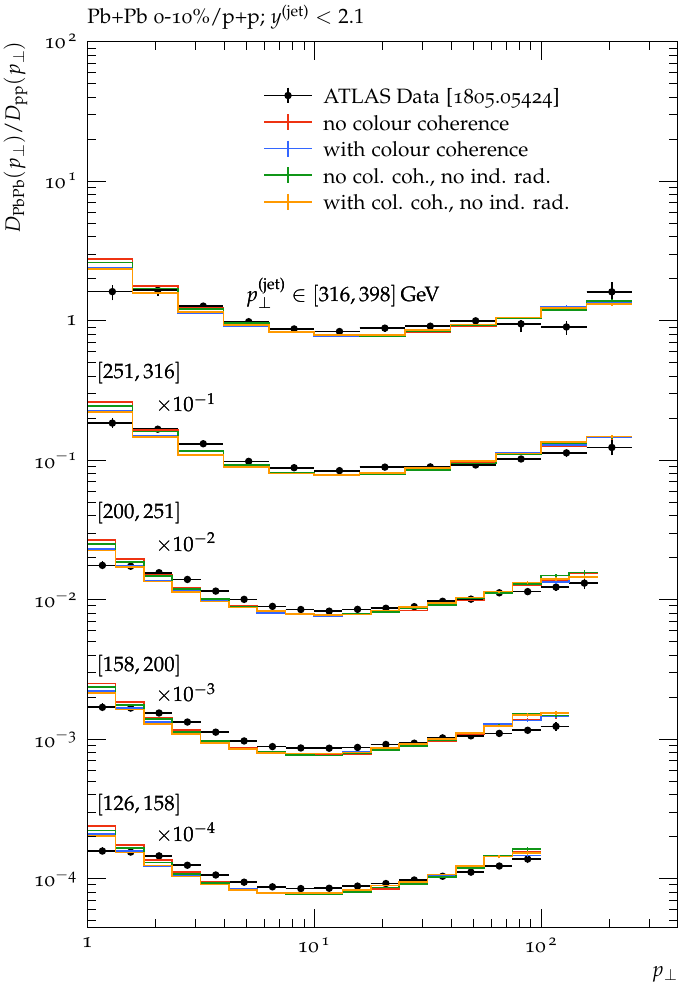}
	\includegraphics[width=0.5\textwidth]{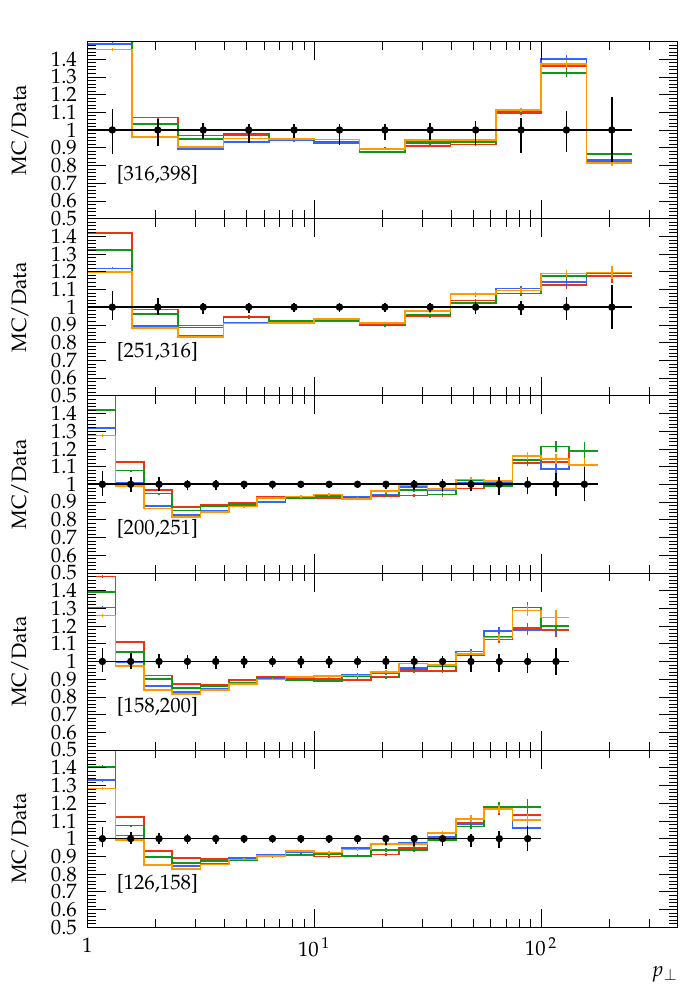}
	\caption{\textsc{Jewel}\,2.6.0+\textsc{Pythia} results with and without colour coherence compared to ratios of jet fragmentation functions in $0-10\%$ central Pb+Pb collisions at $\sqrt{s_{NN}} = \unit[5.02]{TeV}$ to p+p collisions measured by \textsc{Atlas}~\cite{ATLAS:2018bvp} for $R=0.4$ jets in bins of jet transverse momentum. \textsc{Jewel}\,2.6.0+\textsc{Pythia} results are shown without (red) and with colour coherence (rejection of unresolved scatterings and coherent scatterings of dipoles, shown in blue), and without (green) and with colour coherence (orange) where medium induced emissions have been disabled.}
	\label{fig:FFratios_PbPbnoindrad}
\end{figure}

\begin{figure}[h!]
	\includegraphics[width=0.5\textwidth]{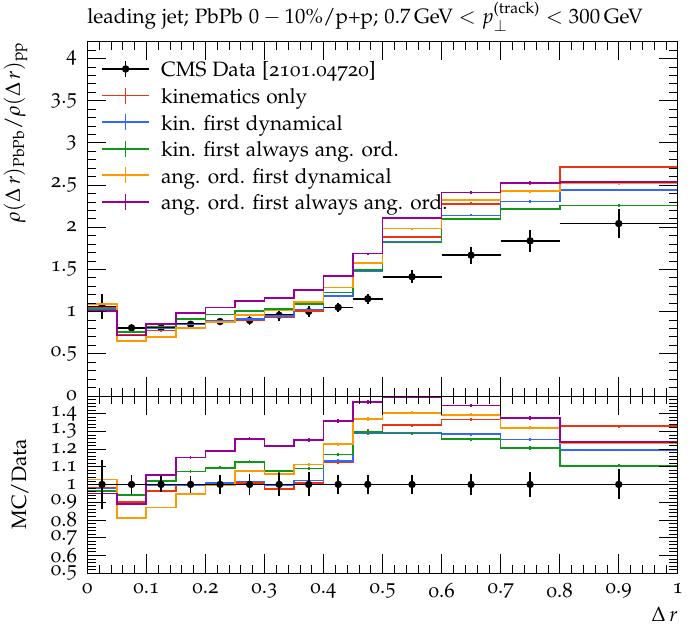}
	\includegraphics[width=0.5\textwidth]{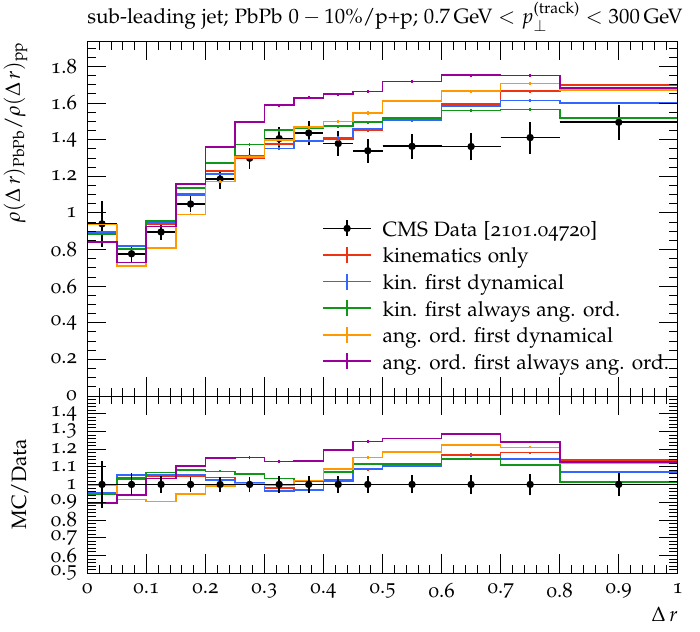}
	\caption{\textsc{Jewel}\,2.6.0+\textsc{Pythia} results with different angular ordering options (without colour coherence) compared to ratios of jet--hadron correlations in $0-10\%$ central Pb+Pb collisions at $\sqrt{s_{NN}} = \unit[5.02]{TeV}$ to p+p collisions measured by \textsc{Cms}~\cite{CMS:2021nhn} for $R=0.4$ jets inclusive in hadron transverse momentum for $0 < x_\text{j} < 1$. Shown here are the correlations for the leading (left) and sub-leading jet (right) in di-jet events. The leading jet is required to have $p_\perp^\text{(jet)} > \unit[120]{GeV}$ and the sub-leading jet $p_\perp^\text{(jet)} > \unit[50]{GeV}$, the azimuthal angle between the two jets has to satisfy $\Delta \phi_{jj} > 5\pi/6$ and both jets have to be within $|\eta^\text{(jet)}| < 1.6$.}
	\label{fig:rhoratios_PbPbm01234}
\end{figure}

\begin{figure}[h!]
	\includegraphics[width=0.5\textwidth]{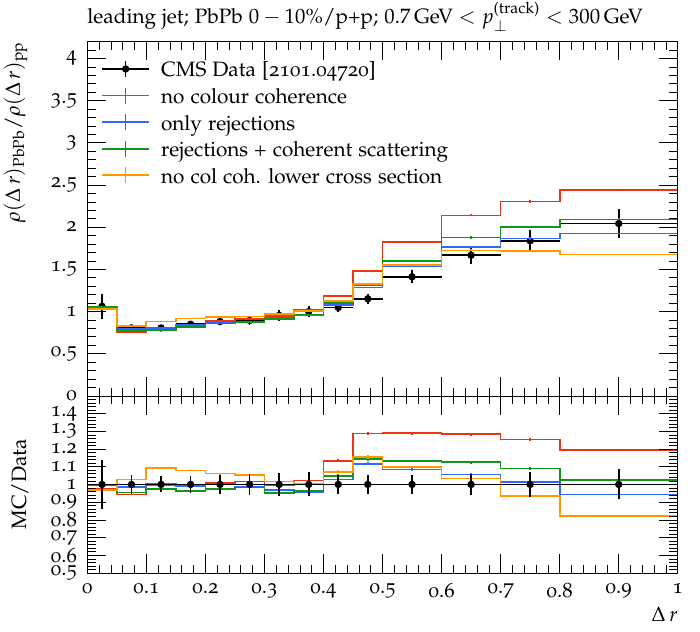}
	\includegraphics[width=0.5\textwidth]{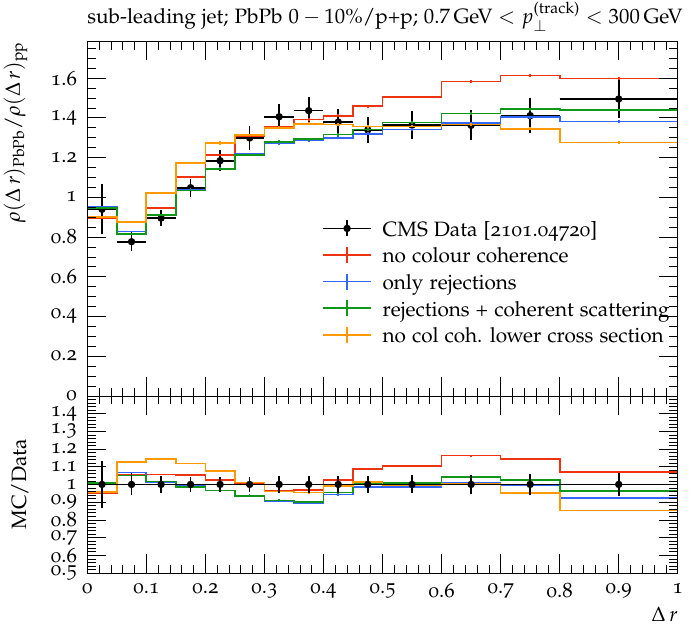}
	\caption{\textsc{Jewel}\,2.6.0+\textsc{Pythia} results with and without colour coherence compared to ratios of jet--hadron correlations in $0-10\%$ central Pb+Pb collisions at $\sqrt{s_{NN}} = \unit[5.02]{TeV}$ to p+p collisions measured by \textsc{Cms}~\cite{CMS:2021nhn} for $R=0.4$ jets inclusive in hadron transverse momentum for $0 < x_\text{j} < 1$. Shown here are the correlations for the leading (left) and sub-leading jet (right) in di-jet events. The leading jet is required to have $p_\perp^\text{(jet)} > \unit[120]{GeV}$ and the sub-leading jet $p_\perp^\text{(jet)} > \unit[50]{GeV}$, the azimuthal angle between the two jets has to satisfy $\Delta \phi_{jj} > 5\pi/6$ and both jets have to be within $|\eta^\text{(jet)}| < 1.6$.  \textsc{Jewel}\,2.6.0+\textsc{Pythia} results are shown without colour coherence (red), with only rejections of unresolved scatterings (blue), and with rejection of unresolved scatterings and coherent scatterings of dipoles (green). Also shown for comparison is the case without colour coherence where the scattering cross section has been artificially reduced by a factor of two (orange)}
	\label{fig:rhoratios_PbPbCC}
\end{figure}

\begin{figure}[h!]
	\includegraphics[width=0.5\textwidth]{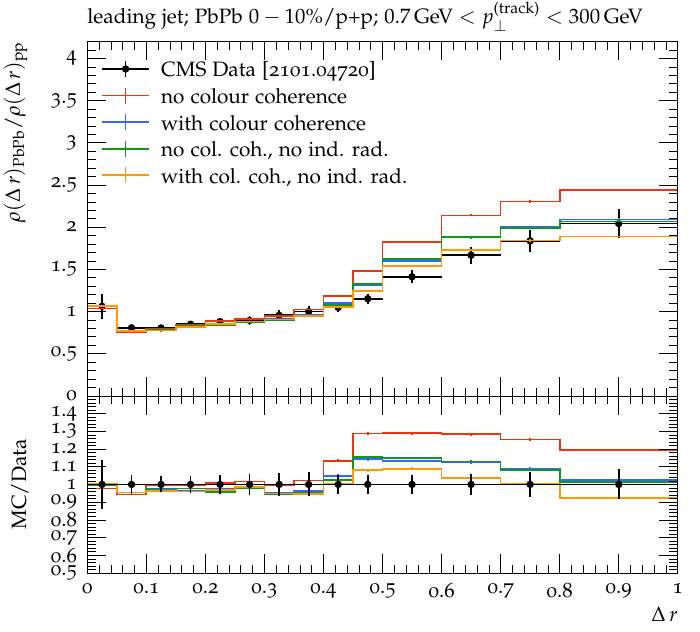}
	\includegraphics[width=0.5\textwidth]{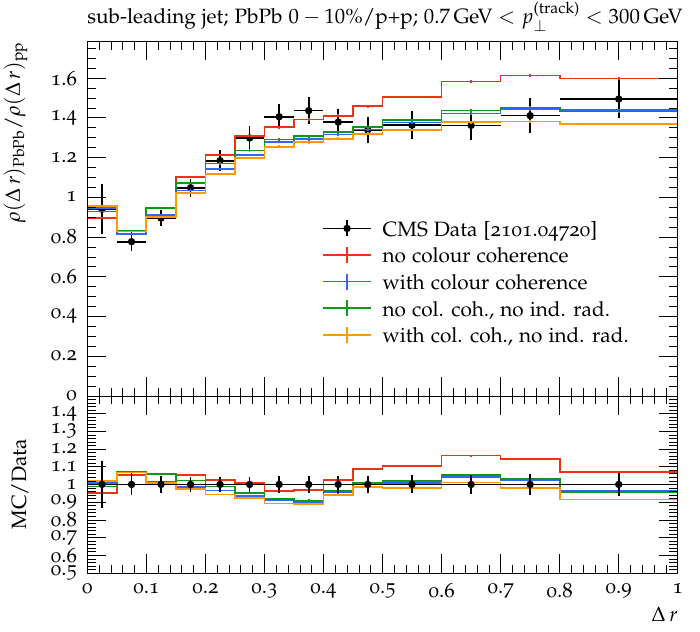}
	\caption{\textsc{Jewel}\,2.6.0+\textsc{Pythia} results with and without colour coherence compared to ratios of jet--hadron correlations in $0-10\%$ central Pb+Pb collisions at $\sqrt{s_{NN}} = \unit[5.02]{TeV}$ to p+p collisions measured by \textsc{Cms}~\cite{CMS:2021nhn} for $R=0.4$ jets inclusive in hadron transverse momentum for $0 < x_\text{j} < 1$. Shown here are the correlations for the leading (left) and sub-leading jet (right) in di-jet events. The leading jet is required to have $p_\perp^\text{(jet)} > \unit[120]{GeV}$ and the sub-leading jet $p_\perp^\text{(jet)} > \unit[50]{GeV}$, the azimuthal angle between the two jets has to satisfy $\Delta \phi_{jj} > 5\pi/6$ and both jets have to be within $|\eta^\text{(jet)}| < 1.6$.  \textsc{Jewel}\,2.6.0+\textsc{Pythia} results are shown without (red) and with colour coherence (rejection of unresolved scatterings and coherent scatterings of dipoles, shown in blue), and without (green) and with colour coherence (orange) where medium induced emissions have been disabled.}
	\label{fig:rhoratios_PbPbnoindrad}
\end{figure}

\FloatBarrier

\newpage

\section{Jet fragmentation functions and jet--hadron correlations in central Pb+Pb collisions with colour coherence but without induced emissions}
\label{sec:moreplotsCC}

\begin{figure}[h!]
	\includegraphics[width=0.5\textwidth]{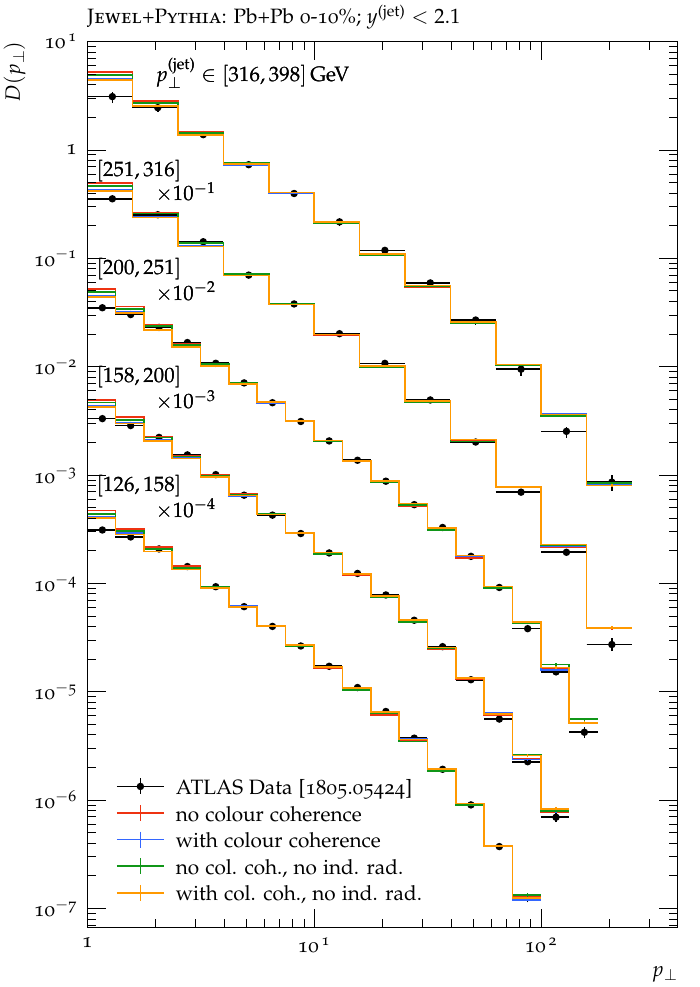}
	\includegraphics[width=0.5\textwidth]{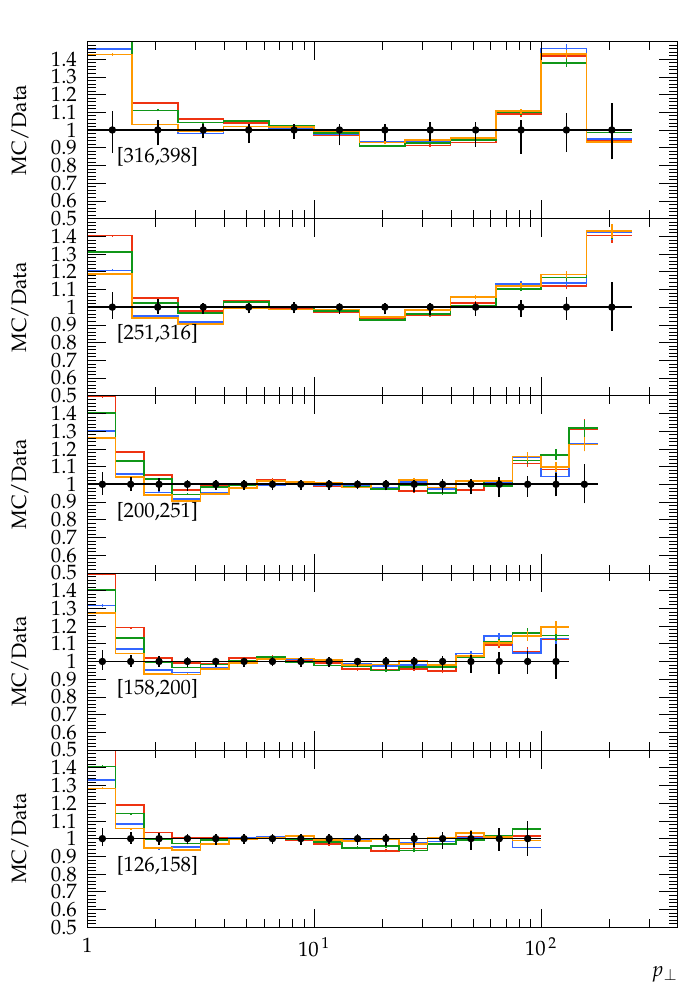}
	\caption{\textsc{Jewel}\,2.6.0+\textsc{Pythia} results with and without colour coherence compared to jet fragmentation functions in $0-10\%$ central Pb+Pb collisions at $\sqrt{s_{NN}} = \unit[5.02]{TeV}$ measured by \textsc{Atlas}~\cite{ATLAS:2018bvp} for $R=0.4$ jets in bins of jet transverse momentum. \textsc{Jewel}\,2.6.0+\textsc{Pythia} results are shown without (red) and with colour coherence (rejection of unresolved scatterings and coherent scatterings of dipoles, shown in blue), and without (green) and with colour coherence (orange) where medium induced emissions have been disabled.}
	\label{fig:FF_PbPbnoindrad}
\end{figure}

\begin{figure}[h!]
	\includegraphics[width=0.5\textwidth]{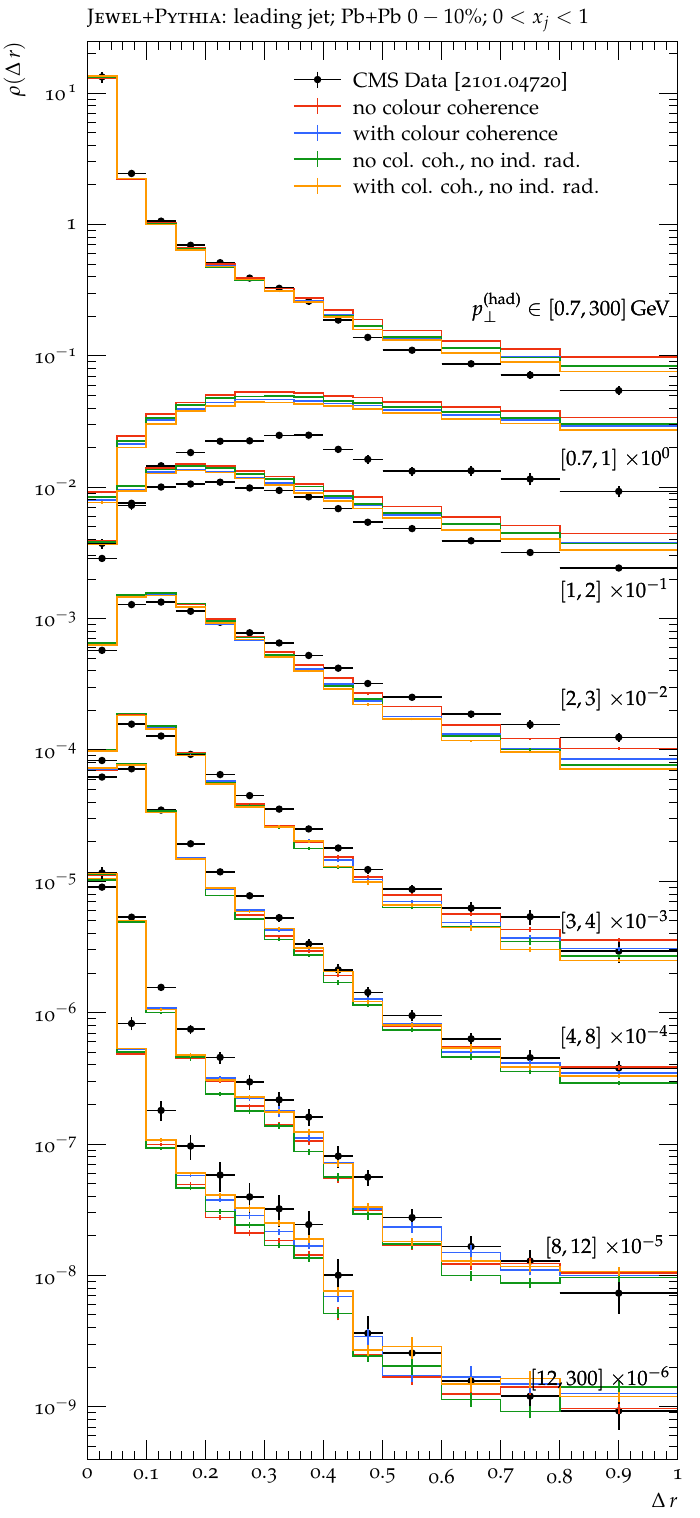}
	\includegraphics[width=0.5\textwidth]{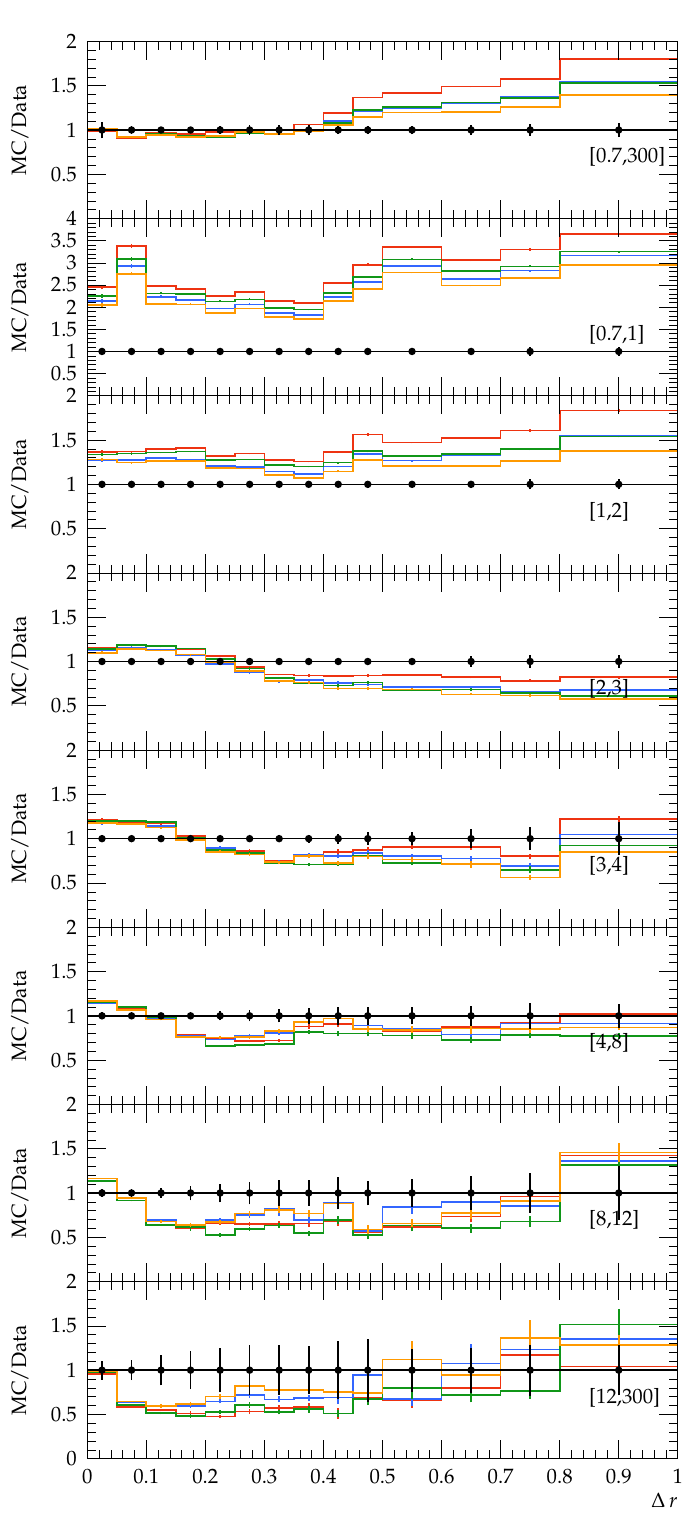}
	\caption{\textsc{Jewel}\,2.6.0+\textsc{Pythia} results with and without colour coherence compared to jet--hadron correlations in $0-10\%$ central Pb+Pb collisions at $\sqrt{s_{NN}} = \unit[5.02]{TeV}$ measured by \textsc{Cms}~\cite{CMS:2021nhn} for $R=0.4$ jets in bins of hadron transverse momentum for $0 < x_\text{j} < 1$. Shown here are the correlations for the leading jet in di-jet events. The leading jet is required to have $p_\perp^\text{(jet)} > \unit[120]{GeV}$ and the sub-leading jet $p_\perp^\text{(jet)} > \unit[50]{GeV}$, the azimuthal angle between the two jets has to satisfy $\Delta \phi_{jj} > 5\pi/6$ and both jets have to be within $|\eta^\text{(jet)}| < 1.6$. \textsc{Jewel}\,2.6.0+\textsc{Pythia} results are shown without (red) and with colour coherence (rejection of unresolved scatterings and coherent scatterings of dipoles, shown in blue), and without (green) and with colour coherence (orange) where medium induced emissions have been disabled.}
	\label{fig:rho_lead_PbPbnoindrad}
\end{figure}

\begin{figure}[h!]
	\includegraphics[width=0.5\textwidth]{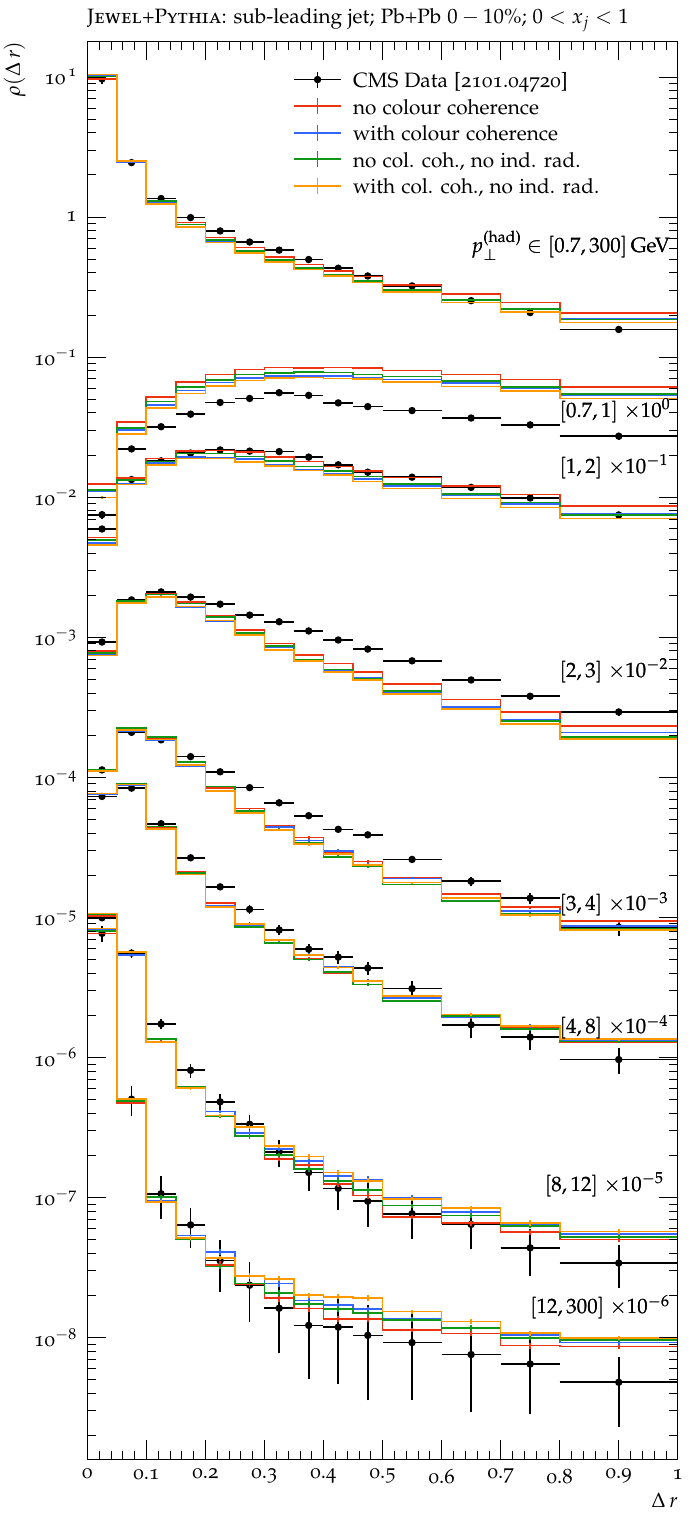}
	\includegraphics[width=0.5\textwidth]{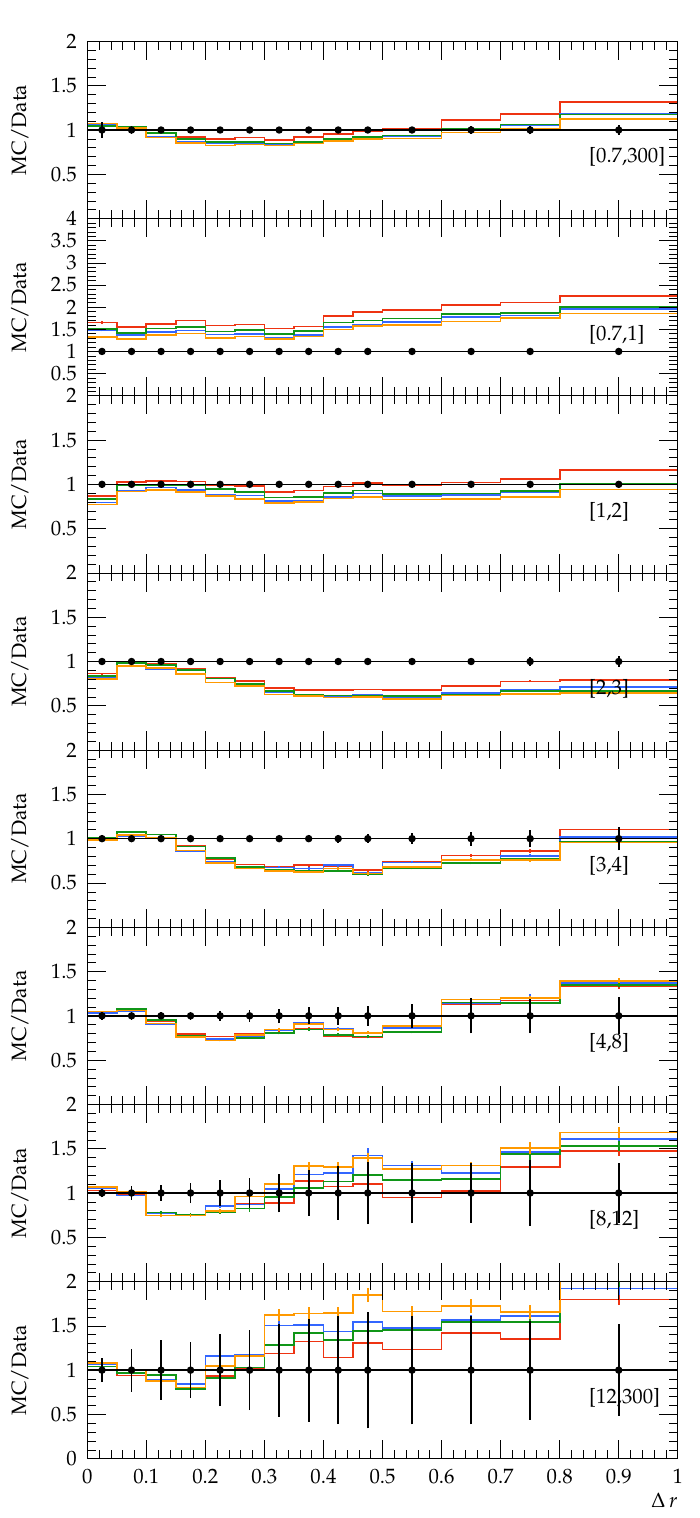}
	\caption{\textsc{Jewel}\,2.6.0+\textsc{Pythia} results with and without colour coherence compared to jet--hadron correlations in $0-10\%$ central Pb+Pb collisions at $\sqrt{s_{NN}} = \unit[5.02]{TeV}$ measured by \textsc{Cms}~\cite{CMS:2021nhn} for $R=0.4$ jets in bins of hadron transverse momentum for $0 < x_\text{j} < 1$. Shown here are the correlations for the sub-leading jet in di-jet events.  The leading jet is required to have $p_\perp^\text{(jet)} > \unit[120]{GeV}$ and the sub-leading jet $p_\perp^\text{(jet)} > \unit[50]{GeV}$, the azimuthal angle between the two jets has to satisfy $\Delta \phi_{jj} > 5\pi/6$ and both jets have to be within $|\eta^\text{(jet)}| < 1.6$. \textsc{Jewel}\,2.6.0+\textsc{Pythia} results are shown without (red) and with colour coherence (rejection of unresolved scatterings and coherent scatterings of dipoles, shown in blue), and without (green) and with colour coherence (orange) where medium induced emissions have been disabled.}
	\label{fig:rho_subl_PbPbnoindrad}
\end{figure}

\FloatBarrier

\section{Jet suppression as function of Soft Drop angle}
\label{sec:rg}

\textsc{Atlas} has measured the suppression of jets as function of the Soft Drop angle between the sub-jets~\cite{ATLAS:2022vii}. The analysis applies Soft Drop grooming with $z_\text{cut} = 0.2$ and $\beta = 0$ to $R=0.4$ jets with $p_\perp^\text{(jet)} > \unit[158]{GeV}$ and $|\eta^\text{(jet)}| < 2.1$. The distance $\Delta R_{12} = \sqrt{(\Delta y_{12})^2 + (\Delta \phi_{12})^2}$ in rapidity and azimuthal angle between the two sub-jets satisfying the Soft Drop criterion is denoted by $r_g$ and is taken as proxy for the opening angle. One would expect colour coherence effects to be prominent in this observable. 

\begin{figure}
	\includegraphics[width=0.5\textwidth]{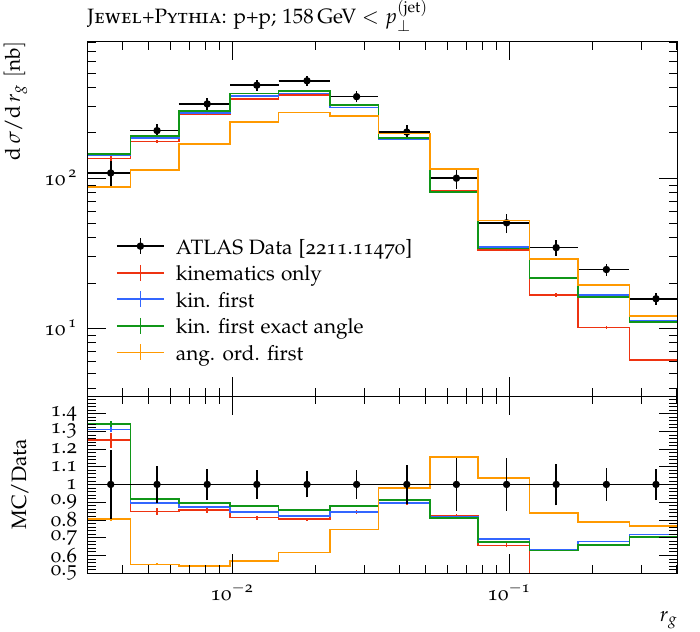}
	\includegraphics[width=0.5\textwidth]{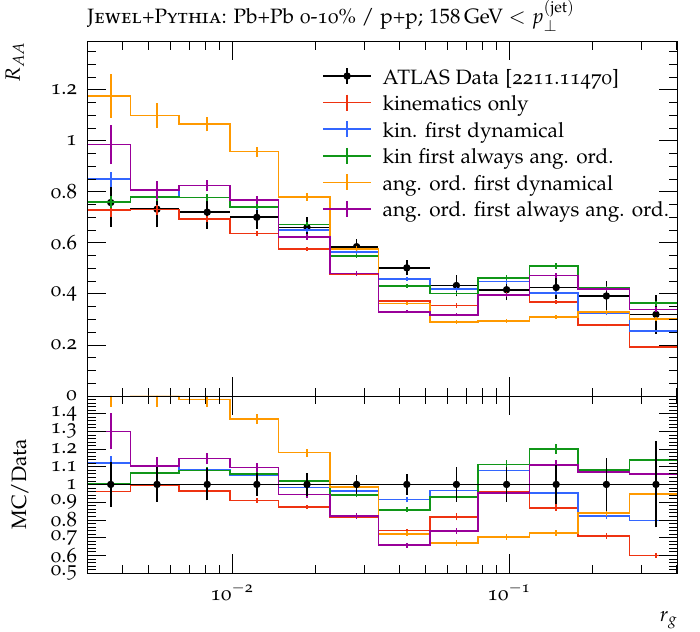}
	\caption{\textbf{Left:} Differential cross section of $R=0.4$ jets with $p_\perp^\text{(jet)} > \unit[158]{GeV}$ and $|\eta^\text{(jet)}| < 2.1$ as function of Soft Drop opening angle $r_g$ in p+p collisions at $\sqrt{s_{NN}} = \unit[5.02]{TeV}$ measured by \textsc{Atlas}~\cite{ATLAS:2022vii} compared to \textsc{Jewel}\,2.6.0+\textsc{Pythia} with different angular ordering options. The Soft Drop grooming parameters are $z_\text{cut} = 0.2$ and $\beta = 0$ and the opening angle $r_g$ is taken to be the distance in rapidity and azimuthal angle between the two sub-jets satisfying the Soft Drop criterion, $r_g = \Delta R_{12} = \sqrt{(\Delta y_{12})^2 + (\Delta \phi_{12})^2}$. \textbf{Right:} Nuclear modification factor $R_\text{AA}$ as function of $r_g$ for $0-10\%$ Pb+Pb collisions at $\sqrt{s_{NN}} = \unit[5.02]{TeV}$ measured by \textsc{Atlas}~\cite{ATLAS:2022vii} compared to \textsc{Jewel}\,2.6.0+\textsc{Pythia} with different angular ordering options. The p+p reference for the different angular ordering options in medium is the corresponding angular ordering option in p+p, with the exception of the \textit{kinematics only} setting, for which a p+p reference with angular ordering (\textit{kinematics first}) was used.}
	\label{fig:rgRAA_PbPbm01234}
\end{figure}

\Figref{fig:rgRAA_PbPbm01234} shows the differential jet cross section in p+p collisions and the nuclear modification factor $R_\text{AA}$ as function of $r_g$. The \textsc{Atlas} measurement is compared to \textsc{Jewel}\,2.6.0+\textsc{Pythia} simulations without colour coherence for the different angular ordering schemes. In p+p collisions (left panel of \figref{fig:rgRAA_PbPbm01234}) the distribution without angular ordering has fewer large $r_g$ jets than those with angular ordering. This unexpected behaviour is a consequence of the selection bias. As seen in \figref{fig:RAA_PbPbm01234} the jet $p_\perp$ spectrum is harder without angular ordering. The average $p_\perp$ of jets entering the distribution in \figref{fig:rgRAA_PbPbm01234} is thus higher without angular ordering, and on average $r_g$ decreases with increasing $p_\perp$. 

The three different angular ordering options have virtually the same $p_\perp$ spectra (cf.~\figref{fig:RAA_PbPbm01234}), but the $r_g$ distribution is markedly different for the \textit{angular ordering first} option. The \textit{kinematics first} options have a lower average $r_g$ than the data, while \textit{angular ordering first} undershoots the data both at high and at low $r_g$. The \textit{kinematics first} options also show a better agreement with the measurement for $R_\text{AA}$ (right panel of \figref{fig:rgRAA_PbPbm01234}).

\begin{figure}
	\includegraphics[width=0.5\textwidth]{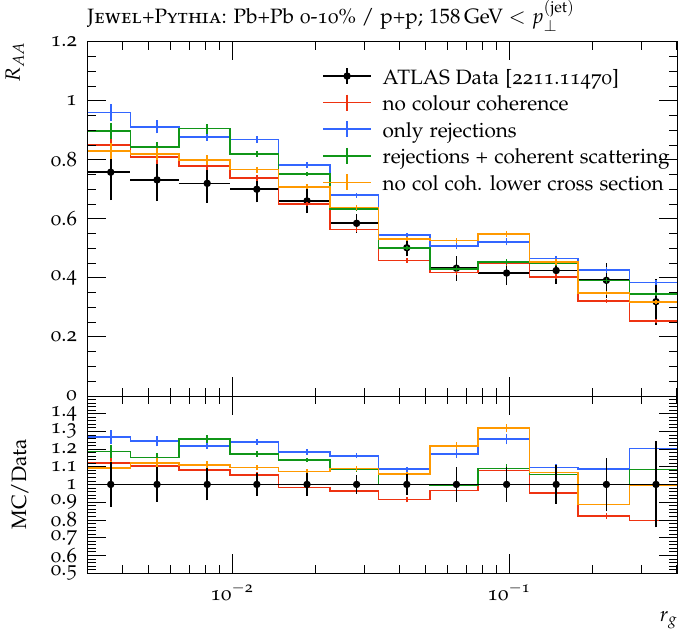}
	\includegraphics[width=0.5\textwidth]{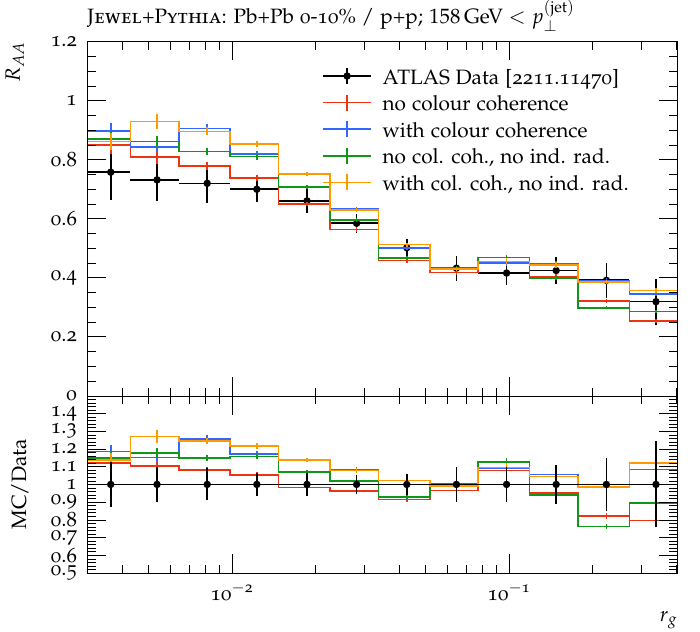}\\
	\caption{Nuclear modification factor $R_\text{AA}$ of $R=0.4$ jets with $p_\perp^\text{(jet)} > \unit[158]{GeV}$ and $|\eta^\text{(jet)}| < 2.1$ as function of Soft Drop opening angle $r_g$ in p+p collisions at $\sqrt{s_{NN}} = \unit[5.02]{TeV}$ measured by \textsc{Atlas}~\cite{ATLAS:2022vii} compared to \textsc{Jewel}\,2.6.0+\textsc{Pythia} with different colour coherence settings. The Soft Drop grooming parameters are $z_\text{cut} = 0.2$ and $\beta = 0$ and the opening angle $r_g$ is taken to be the distance in rapidity and azimuthal angle between the two sub-jets satisfying the Soft Drop criterion, $r_g = \Delta R_{12} = \sqrt{(\Delta y_{12})^2 + (\Delta \phi_{12})^2}$. \textbf{Left:} Without colour coherence (red), with only rejections of unresolved scatterings (blue), and with rejection of unresolved scatterings and coherent scatterings of dipoles (green). Also shown for comparison is the case without colour coherence where the scattering cross section has been artificially reduced by a factor of two (orange).  \textbf{Right:} Without (red) and with colour coherence (rejection of unresolved scatterings and coherent scatterings of dipoles, shown in blue), and without (green) and with colour coherence (orange) where medium induced emissions have been disabled.}
	\label{fig:rgRAA_PbPbCC}
\end{figure}

Surprisingly, including colour coherence (\figref{fig:rgRAA_PbPbCC}) does not lead to less suppression of small $r_g$ jets, as one would naively expect. Instead, one observes a generally somewhat larger $R_\text{AA}$, as seen already in $R_\text{AA}$ versus $p_\perp$ (cf.~\figref{fig:RAA_PbPbCC}), with large $r_g$ jets being somewhat less suppressed. As seen already in the p+p case, the selection bias is a confounding factor in the interpretation of these results. A larger $R_\text{AA}$ means the jets have a smaller initial $p_\perp$ (before medium interactions) and therefore a larger $r_g$. Compared to the incoherent scenario with reduced scattering rate to match the overall level of suppression (shown in orange) the coherent scenario actually has a slightly steeper $r_g$ distribution. Furthermore, medium response and colour coherence induced changes in the hard fragmentation pattern as discussed in \secref{sec:implementation} also affect the shape of the $r_g$ distribution. But there is an additional effect here that is worth noticing: there is a strong correlation between $r_g$ and formation time, with small (large) $r_g$ splittings having a long (short) formation time. Estimating the formation time from the reconstructed kinematics of the sub-jets yields values of up to \unit[10]{fm} at the left and of the $r_g$ plots. Jets with small $r_g$ thus consist of just a single parton for a large part of the evolution irrespective of whether colour coherence is included or not. By the time the parton splits most of the re-scattering has already happened and there is thus no large difference between the coherent and the incoherent scenario. The large $r_g$ splittings happen early, but decohere very quickly due to the large opening angle. Therefore, also in this regime there is no large difference between the two scenarios.

The conclusion is thus that unfortunately, contrary to naive expectations, colour coherence does not leave a clean signature in this observable. This is consistent with the findings in~\cite{Apolinario:2026hff}.

\FloatBarrier

\bibliographystyle{JHEP}  
\bibliography{../bib/jetquenching.bib}

\end{document}